\documentclass[prc,letterpaper,twocolumn,showpacs,showkeys,lengthcheck,tightenlines,
               nofootinbib,preprintnumbers,superscriptaddress]{revtex4-1}

\usepackage[utf8]{inputenc}

\usepackage{titlesec}
\usepackage{dcolumn}
\usepackage{bm}
\usepackage{amsmath,amssymb,amsfonts}

\usepackage{graphicx}
\usepackage{subfigure}
\usepackage{color}
\usepackage{bbold}

\usepackage{slashed}
\usepackage[svgnames]{xcolor}
\usepackage[english]{babel}
\usepackage{blindtext}
\usepackage{microtype}
\usepackage{tikz}

\usepackage[colorlinks=true,urlcolor=blue,citecolor=blue]{hyperref}
\usepackage{ifpdf}
\usepackage{float}

\graphicspath{{.}{figures/}}

\titlespacing{\section}{5pt}{12pt plus 4pt minus 2pt}{8pt plus 2pt minus 2pt}
\titlespacing{\subsection}{0pt}{12pt plus 4pt minus 2pt}{8pt plus 2pt minus 2pt}

\begin{document}

\title{Regularization dependence of pion generalised parton distributions}

\author{Jin-Li Zhang}
\email[]{jlzhang@njnu.edu.cn}
\affiliation{Department of Physics, Nanjing Normal University, Nanjing 210023, China }

\author{Guang-Zhen Kang}
\email[]{gzkang@nju.edu.cn}
\affiliation{School of Science, Yangzhou Polytechnic Institute, Yangzhou 225127, China }

\author{Jia-Lun Ping}
\email[]{jlping@njnu.edu.cn}
\affiliation{Department of Physics, Nanjing Normal University, Nanjing 210023, China }

\begin{abstract}
Pion generalised parton distributions are calculated within the framework of the Nambu--Jona-Lasinio model using different regularization schemes, including the proper time regularization scheme, the three dimensional momentum cutoff scheme, the four dimensional momentum cutoff scheme, and the Pauli-Villars regularization scheme. Furthermore, we check the theoretical constraints of pion generalised parton distributions required by the symmetries of quantum chromodynamics in different regularization schemes. The diagrams of pion parton distribution functions are plotted, in addition, we evaluate the Mellin moments of generalised parton distributions, which related to the electromagnetic and gravitational form factors of pion. Pion generalised parton distributions are continuous but not differential at $x=\pm \,\xi$, when considering the effect of D-term, generalised parton distributions become not continuous at $x=\pm \,\xi$ in all the four regularization schemes. Generalised parton distributions in impact parameter space are considered, the width distribution of $u$ quark in the pion and the mean-squared $\langle \bm{b}_{\bot}^2\rangle_{\pi}^u$ are calculated. The light-front transverse-spin distributions are studied, when quark polarized in the light-front-transverse $+\,x$ direction, the transverse-spin density is no longer symmetric around $(b_x=0,b_y=0)$, the peaks shift to $(b_x=0,b_y>0)$, we compare the average transverse shift $\langle b_{\bot}^y\rangle_1^u$ and $\langle b_{\bot}^y\rangle_2^u$ in different regularization schemes. The light-cone energy radius $r_{E,LC}$ and the light-cone charge radius $r_{c,LC}$ are also evaluated,  we find that in the proper time regularization scheme the values of these quantities are the largest, in the three dimensional momentum cutoff scheme they are the smallest. 
\end{abstract}


\maketitle
\section{Introduction}
Generalised parton distributions (GPDs)~\cite{Mueller:1998fv,Ji:1996nm,Radyushkin:1997ki,Ji:1998pc,Diehl:2003ny} include a wealth of new information on three dimensional structure of hadrons. They provide a consistent unified framework containing the elastic form factors (FFs)~\cite{Rodriguez-Quintero:2019yec,Xu:2019ilh,Cui:2020rmu,Chen:2021guo} and usual parton distribution functions (PDFs)~\cite{Rodriguez-Quintero:2019fyc,Cui:2020dlm,Cui:2020tdf}, and offering even more information. For example, GPDs have relationship with quantum chromodynamics (QCD) energy-momentum tensor (EMT), which would otherwise probably only be accessible by means of graviton scattering, that's another infusive feature of GPDs. The relationship between the EMT and GPDs can be used to obtain the message about the mechanical properties of partonic systems, like pressure distributions inside the nucleon. The Mellin moments of GPDs contain the information of electromagnetic and gravitational properties of hadrons. The gravitational properties yield additional insights on the way that angular momentum and mass are distributed among the quarks and gluons inside the hadron, thus addressing deep issues about the spin~\cite{Leader:2013jra} and mass~\cite{Lorce:2018egm,Hatta:2018sqd} decompositions of hadrons straightforwardly. Through the two-dimensional Fourier transformation, GPDs become spatial light cone distributions of partons in impact parameter space. GPDs emerge in the calculation of hard exclusive process such as deeply virtual Compton scattering (DVCS), deeply virtual meson production (DVMP) and time like Compton scattering (TCS). Factorization allows us to express the amplitudes of these reactions as the convolution of a hard scattering kernel and a soft matrix element of quark and/or gluon fields which consists in the GPDs.

We are interested in pion GPDs~\cite{Roberts:2021nhw,Zhang:2021mtn,Zhang:2021shm,Raya:2021zrz}, although there are little chances of measuring them directly in experiment, the pion GPDs are amenable to indirect experimental determination as well as studies both on transverse~\cite{Burkardt:2001jg} as well as Euclidean~\cite{Hagler:2007hu} lattices.


At present, it is impossible to determine the GPDs from QCD directly, effective models have been applied to supply estimates which should serve to direct future experiments. However, up to now, it is still a challenge to modelling GPDs and various methods have been developed. The calculation of GPD models should satisfy the theoretical constraints required by the symmetries of QCD.

In this paper, we will use the Nambu--Jona-Lasinio (NJL)~\cite{Nambu:1961tp,Nambu:1961fr,Klevansky:1992qe,Buballa:2003qv} model to investigate the pion GPDs using different regularization schemes. NJL model is widely used in many fields, it has an effective Lagrangian of relativistic fermions interacting through local fermion-fermion couplings. Besides, it keeps the fundamental symmetries of QCD, of which the most important one is chiral symmetry. A fly in the ointment is that NJL is non-renormalizable, so we need a regularization scheme to totally define the model. Several regularization methods are commonly used, such as the three dimensional (3D) momentum non-covariant cutoff scheme, the four dimensional (4D) momentum cutoff scheme~\cite{Harada:1994wy,Ishii:1995bu,Florkowski:1996wf,Jafarov:2004zs}, proper time (PT) regularization scheme~\cite{Florkowski:1996wf,Ebert:1996vx,Hellstern:1997nv,Zhang:2016zto,Endrodi:2019whh,Zhang:2020ecj,Zhang:2021shm,Zhang:2021mtn,Zhang:2021tnr}, the Pauli-Villars (PV) scheme~\cite{Davidson:1994uv,Davidson:2001cc}. The latter three have the attractive character of being Lorentz invariant. The 3D cutoff scheme is the most popular method in this model and a lot of works have been done in this way~\cite{Andersen:2007qv}, when the model is applied to thermodynamics, most authors prefer to regularize the integrals by a (sharp or smooth) 3-momentum cutoff~\cite{Buballa:2003qv,Kashiwa:2006rc,Kashiwa:2007hw,Coppola:2018vkw,Li:2018ltg}. The 4D cutoff method preserves the Lorentz symmetry in which space and time are treated on equal footing. The Pauli-Villars regularization is based on the subtraction of the amplitude considering the virtually heavy particle to suppress the nonphysical high energy contribution coming from the loop integrals. The proper time regularization makes integrals finite through the exponentially dumping factor, in this regularization scheme we can introduce the infrared cutoff $\Lambda_{\text{IR}}$ to mimic confinement.

We use the four regularization schemes in this paper, each of them have its advantages and disadvantages, we want to see the difference of pion GPDs in different regularization schemes.


This paper is organized as follows: In Sect.~\ref{nice}, we give a brief introduction to NJL model and then demonstrate the definition and calculation of the pion GPDs. In Sect.~\ref{well1}, results of the pion GPDs using different regularization schemes are given. A brief summary and conclusion is given in Sect.~\ref{excellent}.


\section{Basics properties of pion GPDs}\label{nice}

\subsection{NJL model}\label{good}
The SU(2) flavor NJL Lagrangian is
\begin{align}\label{1}
\mathcal{L}&=\bar{\psi }\left(i\gamma ^{\mu }\partial _{\mu }-\hat{m}\right)\psi\nonumber\\
&+\frac{1}{2} G_{\pi }\left[\left(\bar{\psi }\psi\right)^2-\left( \bar{\psi }\gamma _5 \vec{\tau }\psi \right)^2\right]-\frac{1}{2}G_{\omega}\left(\bar{\psi }\gamma _{\mu}\psi\right)^2\nonumber\\
&-\frac{1}{2}G_{\rho}\left[\left(\bar{\psi }\gamma _{\mu} \vec{\tau } \psi\right)^2+\left( \bar{\psi }\gamma _{\mu}\gamma _5 \vec{\tau } \psi \right)^2\right],
\end{align}
where $\vec{\tau}$ are the Pauli matrices represent isospin and the current quark mass matrix $\hat{m}=\text{diag}\left(m_u,m_d\right)$. In the isospin symmetry, $m_u = m_d =m$. $G_{\pi}$ , $G_{\omega}$, and $G_{\rho}$ are the four-fermion coupling constants in each chiral channel.

The fundamental quark-antiquark interaction kernel is defined as~\cite{Cloet:2014rja}
\begin{align}\label{bc1}
\mathcal{K}_{\alpha\beta,\gamma\delta}&=\sum_{\Omega}\text{K}_{\Omega}\Omega_{\alpha\beta}\bar{\Omega}_{\gamma\delta}\nonumber\\
&=2iG_{\pi}[(\mathbb{1})_{\alpha\beta}(\mathbb{1})_{\gamma\delta}-(\gamma_5\tau_i)_{\alpha\beta}({\gamma_5\tau_i})_{\gamma\delta}]\nonumber\\
&-2iG_{\rho}[(\gamma_{\mu}\tau_i)_{\alpha\beta}(\gamma_{\mu}\tau_i)_{\gamma\delta}+(\gamma_{\mu}\gamma_5\tau_i)_{\alpha\beta}(\gamma_{\mu}\gamma_5\tau_i)_{\gamma\delta}]\nonumber\\
&-2iG_{\omega}(\gamma_{\mu})_{\alpha\beta}(\gamma_{\mu})_{\gamma\delta},
\end{align}
where the label represent the Dirac, color, and isospin indices.

The dressed-quark propagator is obtained by solving the gap equation
\begin{align}\label{2}
S(k)=\frac{1}{{\not\!k}-M+i \varepsilon},
\end{align}
we obtained a constant dressed-quark mass $M$ which satisfies
\begin{align}\label{nocutoffgap}
M=m+12 i G_{\pi}\int \frac{\mathrm{d}^4l}{(2 \pi )^4}\mathrm{tr}_D[S(l)],
\end{align}
where the trace is over Dirac indices. When the effective coupling strength $G_{\pi}$ is larger than $G_{\mathrm{critical}}$, dynamical chiral symmetry breaking (DCSB) occurs and gives a nontrivial solution $M > 0$.
\begin{figure}
\centering
\includegraphics[width=0.47\textwidth]{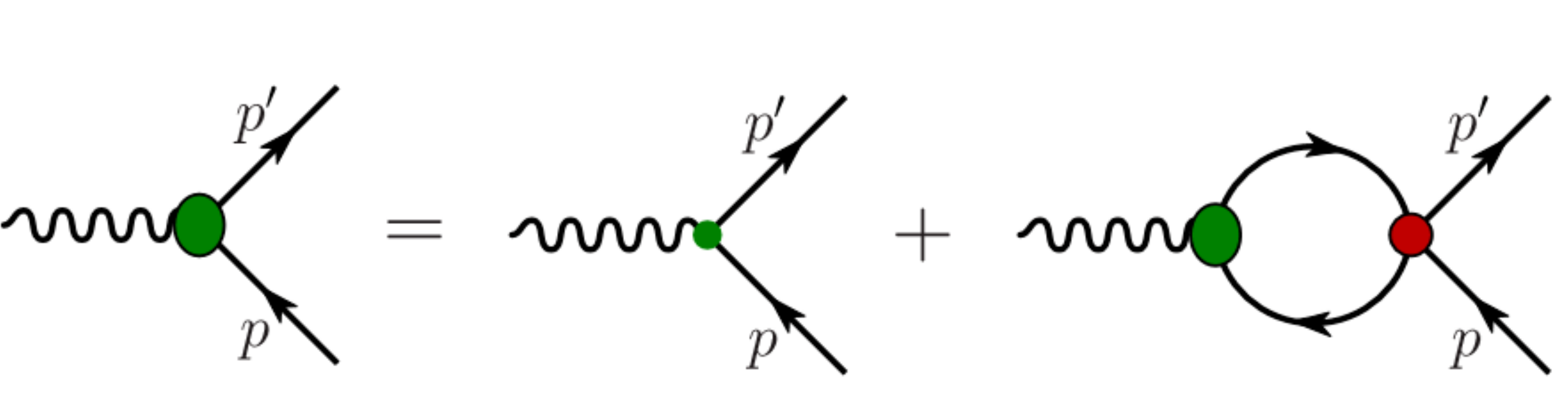}
\caption{Inhomogeneous BSE whose solution gives the quark-photon vertex, represented as the large shaded oval.}\label{qpv}
\end{figure}
The quark charge operator in the NJL model is
\begin{align}\label{bqp}
&\hat{Q}=\left(
\begin{array}{cc}
e_u & 0 \\
0 & e_d \\
\end{array}
\right)=(\frac{1}{6}+\frac{\tau_3}{2}),
\end{align}
where $e_u$ and $e_d$ are the $u$ and $d$ quark charges. This means quark-photon vertex has both an isovector and an isoscalar component, now the dressed quark-photon vertex can be expressed as
\begin{align}\label{dqpver}
&\Lambda_{\gamma Q}^{\mu}(p',p)=\frac{1}{6}\Lambda_{\omega}^{\mu}(p',p)+\frac{\tau_3}{2}\Lambda_{\rho}^{\mu}(p',p),
\end{align}
the effective vertex
\begin{align}\label{dqp}
\Lambda_i^{\mu}(Q^2)&=\gamma^{\mu}F_{1i}(Q^2)+\frac{\sigma^{\mu\nu}q_{\nu}}{2M}F_{2i}(Q^2),
\end{align}
where $i=(\omega,\rho)$. This vertex has the same form as the electromagnetic current for an on-shell spin-$\frac{1}{2}$ fermion. For a point-like quark $F_{1i}(Q^2)=1$ and $F_{2i}(Q^2) =0$. The inhomogeneous Bethe-Salpeter equation (BSE) for the quark-photon vertex is depicted in Fig. \ref{qpv},
\begin{align}\label{dqffff}
&\Lambda_{\gamma Q}^{\mu}(p',p)=\gamma^{\mu}(\frac{1}{6}+\frac{\tau_3}{2})\nonumber\\
&+\sum_{\Omega}\text{K}_{\Omega}\Omega\int \frac{d^4k}{(2\pi)^4}\text{tr}[\bar{\Omega} S(k+q)\Lambda_{\gamma Q}^{\mu}(k+q,k)S(k)],
\end{align}
where $\sum_{\Omega}\text{K}_{\Omega}\Omega_{\alpha\beta} \bar{\Omega}_{\gamma\delta}$ stand for the interaction kernels in Eq. (\ref{bc1}), only the isovector-vector, $-2iG_{\rho}(\gamma_{\mu}\vec{\tau })_{\alpha\beta}(\gamma_{\mu}\vec{\tau })_{\gamma\delta} $, and isoscalar-vector $-2iG_{\omega}(\gamma_{\mu})_{\alpha\beta}(\gamma_{\mu})_{\gamma\delta} $ terms can contribute. 

From the inhomogeneous BSE, the dressed-quark form factors, associated with the electromagnetic current of Eq. (\ref{dqp}), are
\begin{align}\label{bsam}
F_{1i}(Q^2)=\frac{1}{1+2G_i \Pi_{\text{VV}}(Q^2)},\quad \quad F_{2i}(Q^2)=0.
\end{align}
The vector bubble diagram has the form
\begin{align}\label{vbb}
&\quad \Pi_{\text{VV}}(q^2)(g^{\mu\nu}-\frac{q^{\mu}q^{\nu}}{q^2})\delta_{ij}\nonumber\\
&=3i\int \frac{d^4k}{(2\pi)^4}\text{tr}[\gamma^{\mu}\tau_iS(k)\gamma^{\nu}\tau_jS(k+q)].
\end{align}
The pion vertex function, in the light-cone normalization, is given by
\begin{align}\label{6C}
\Gamma_{\pi}^{i}=\sqrt{Z_{\pi}}\gamma_5\tau_i
\end{align}
where $Z_{\pi}$ is the effective meson-quark-quark coupling constant. The pseudoscalar bubble diagram
\begin{align}\label{ab35}
\Pi_{\text{PP}}(q^2)\delta_{ij}=3i\int \frac{\mathrm{d}^4k}{(2 \pi )^4}\text{tr}[\gamma^5\tau_iS(k)\gamma^5\tau_jS(k+q)],
\end{align}
where the traces are over Dirac and isospin indices. Meson masses are then defined by the pole in the two body $t$ matrix, respectively. The normalization factor is determined by
\begin{align}\label{ab35}
Z_{\pi}^{-1}&=-\frac{\partial}{\partial q^2}\Pi_{\text{PP}}(q^2)|_{q^2=m_{\pi}^2}.
\end{align}
The pion decay constant are defined as Ref.~\cite{Klevansky:1992qe}
%
%
%
%
%
%
\begin{align}\label{pdc1}
f_{\pi}&= \int \frac{\mathrm{d}^4k}{(2 \pi )^4}\frac{-4iN_c\sqrt{Z_{\pi}}M}{((k+\frac{p}{2})^2-M^2)((k-\frac{p}{2})^2-M^2)}\nonumber\\
&=\int_0^1 \mathrm{d}x \int \frac{\mathrm{d}^4k}{(2 \pi )^4}\frac{-4iN_c\sqrt{Z_{\pi}}M }{(k^2-x(x-1)m_{\pi}^2-M^2)},
\end{align}
in all these regularization schemes, there are two parameters, the coupling constant $G_{\pi}$ and the cutoff $\Lambda$ are fixed by the pion decay constant $f_{\pi}$ and quark condensate $\langle \bar{u}u\rangle=\langle \bar{d}d\rangle$. $G_{\omega}$ and $G_{\rho}$ are determined by $m_{\omega}=0.782$ GeV, $m_{\rho}=0.770$ GeV through $1+2G_i \Pi_{\text{VV}} (m_i^2)=0$, where $i=(\omega,\rho)$.

We will use the notations and formulas in the appendix in the following.

\subsection{The definition and calculation of pion GPDs}\label{qq}
\begin{figure}
\centering
\includegraphics[width=0.47\textwidth]{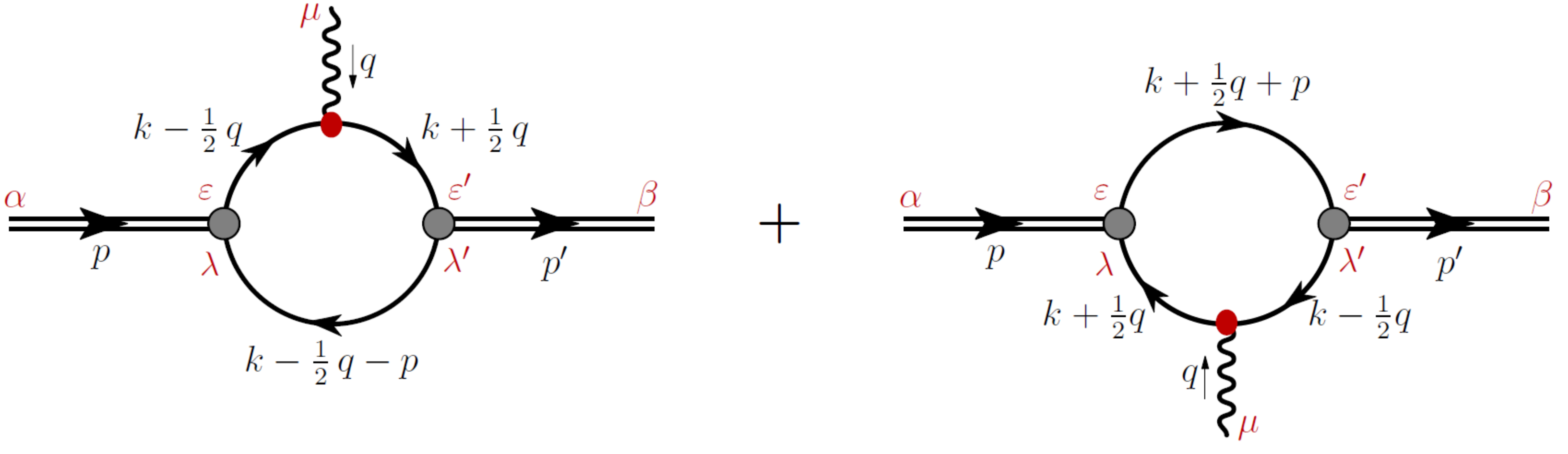}
\caption{Pion GPDs diagrams.}\label{GPD}
\end{figure}

The pion GPDs are given in Fig. \ref{GPD}, where $p$ is the incoming and $p'$ the outgoing pion momentum, here we use the symmetry notation, the kinematics and the associated quantities are defined as
\begin{align}\label{4}
p^{'2}=p^2=m_{\pi}^2, \quad \quad t=q^2=(p'-p)^2=-Q^2,
\end{align}
\begin{align}\label{5}
\xi=\frac{p^+-p'^+}{p^++p'^+},\quad P=\frac{p^++p^{'+}}{2}, \quad n^2=0,
\end{align}
$\xi$ is the skewness parameter, in the light-cone coordinate
\begin{align}\label{4A}
v^{\pm}=(v^0\pm v^3), \quad  \mathbf{v}=(v^1,v^2),
\end{align}
for any four-vector, $n$ is the light-cone four-vector $n=(1,0,0,-1)$, $v^+$ in the light-cone coordinate is defined as
\begin{align}\label{4B}
v^+=v\cdot n.
\end{align}
GPDs are defined as the Fourier transform of a non-perturbative matrix element of a quark operators depending on a light-like distance,
\begin{align}\label{dgpd}
&H(x,\xi,t)=\frac{1}{2}\int \frac{\mathrm{d}z^-}{2\pi}e^{\frac{i}{2}x(p^++p'^+)z^-}\nonumber\\
&\times \langle p'|\bar{q}(-\frac{1}{2}z)\gamma^+q(\frac{1}{2}z)|p\rangle \mid_{z^+=0,\bm{z}=\bm{0}},
\end{align}
\begin{align}\label{dtgpd}
&\frac{P^+q^j-P^jq^+}{P^+m_{\pi}}E(x,\xi,t)=\frac{1}{2}\int \frac{\mathrm{d}z^-}{2\pi}e^{\frac{i}{2}x(p^++p'^+)z^-}\nonumber\\
&\times \langle p'|\bar{q}(-\frac{1}{2}z)i\sigma^{+j}q(\frac{1}{2}z)|p\rangle \mid_{z^+=0,\bm{z}=\bm{0}},
\end{align}
where $x$ is the longitudinal momentum fraction. The first equation corresponds to the vector or no spin flip GPD, the second equation corresponds to tensor or spin flip GPD.

The insert operators in Fig. \ref{GPD} have the form
\begin{subequations}\label{a91}
\begin{align}\label{6A}
\textcolor{red}{\bullet}_1 &=\gamma^+\delta(x-\frac{k^+}{P^+})\,, \\
\textcolor{red}{\bullet}_2 &=i\sigma^{+j} \delta(x-\frac{k^+}{P^+}),
\end{align}
\end{subequations}
$\textcolor{red}{\bullet}_1$ for vector GPD and $\textcolor{red}{\bullet}_2$ for tensor GPD.

In the NJL model, GPDs are defined as
\begin{align}\label{gpddd}
&H^a\left(x,\xi,t\right)
=2i N_c Z_{\pi} \int \frac{\mathrm{d}^4k}{(2 \pi )^4}\delta_n^x (k)\nonumber\\
&\times \text{tr}_{\text{D}}\left[\gamma _5 S \left(k_{+q}\right)\gamma ^+ S\left(k_{-q}\right)\gamma _5 S\left(k-P\right)\right],
\end{align}
\begin{align}\label{tgpddd}
& \frac{P^+q^j-P^jq^+}{P^+m_{\pi}} E\left(x,\xi,t\right)=2i N_c Z_{\pi}\int \frac{\mathrm{d}^4k}{(2 \pi )^4}\delta_n^x (k)\nonumber\\
&\times \text{tr}_{\text{D}}\left[\gamma _5 S \left(k_{+q}\right)i\sigma^{+j}S\left(k_{-q}\right)\gamma _5 S\left(k-P\right)\right],
\end{align}
where $\text{tr}_{\text{D}}$ indicates a trace over spinor indices, $\delta_n^x (k)=\delta (xP^+-k^+)$, $k_{+q}=k+\frac{q}{2}$, $k_{-q}=k-\frac{q}{2}$.

Here we use the following reduce formulas
\begin{subequations}\label{dr}
\begin{align}
p\cdot q&=-\frac{q^2}{2}\,, \\
k\cdot q&=\frac{1}{2} \left(D(k_{+q}^2)-D(k_{-q}^2)\right)\,, \\
k\cdot p&=-\frac{1}{2} \left(D((k-P)^2)-D(k_{-q}^2)-p^2+\frac{q^2}{2}\right)\,, \\
k^2&=\frac{1}{2} \left(D(k_{+q}^2)+D(k_{-q}^2)\right)+M^2-\frac{q^2}{4},
\end{align}
\end{subequations}
where $D(k^2)=k^2-M^2$, applying these relationship to Eqs. (\ref{gpddd}) and (\ref{tgpddd}), cancel each identical numerator and denominator factor; at last use Feynman parametrization to simplify all remaining denominators. With this method, one obtained the final results.

\begin{figure}
\centering
\includegraphics[width=0.33\textwidth]{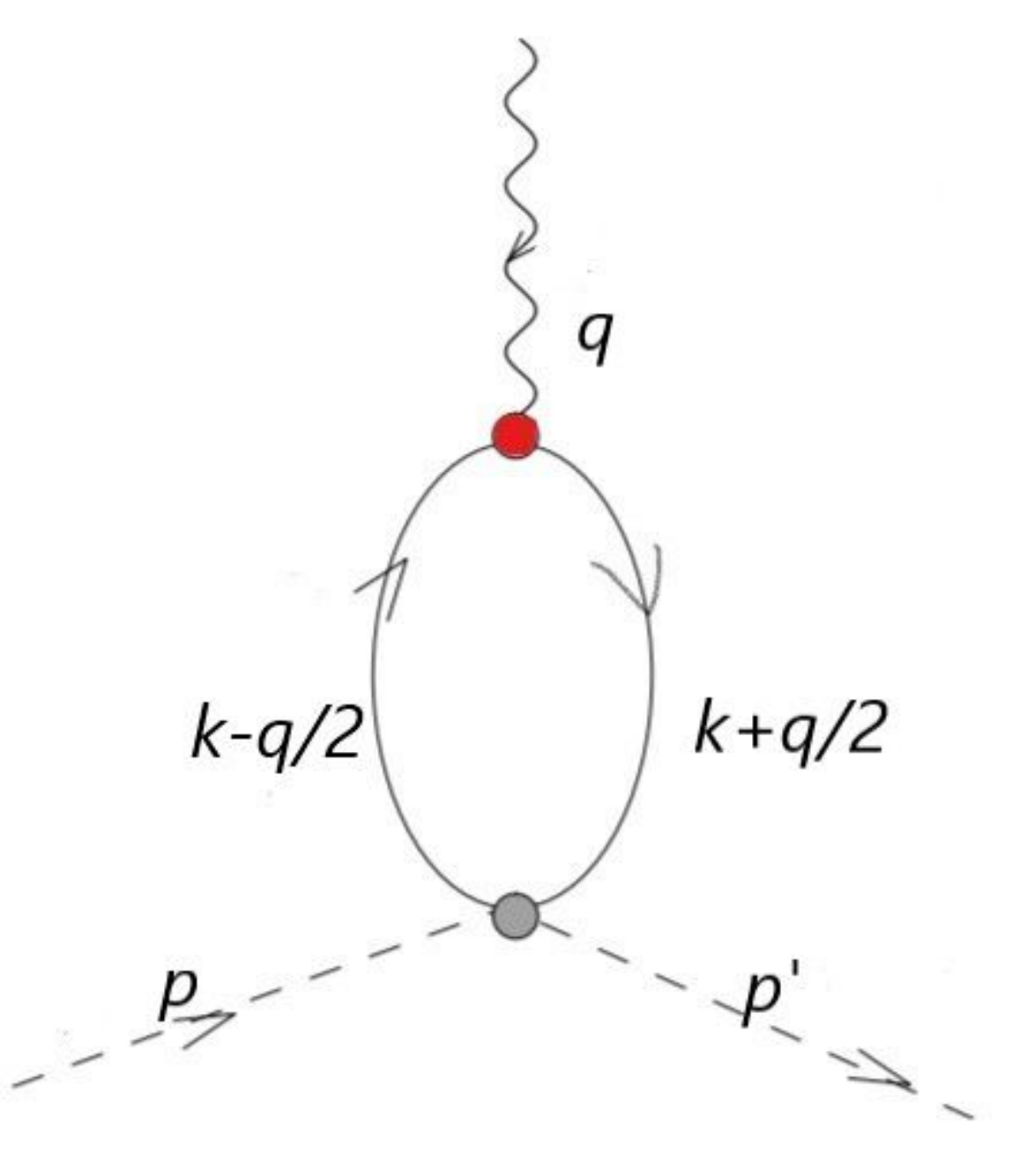}
\caption{The contact contribution responsible for the D-term. }\label{dterm}
\end{figure}
\begin{figure}
\centering
\includegraphics[width=0.40\textwidth]{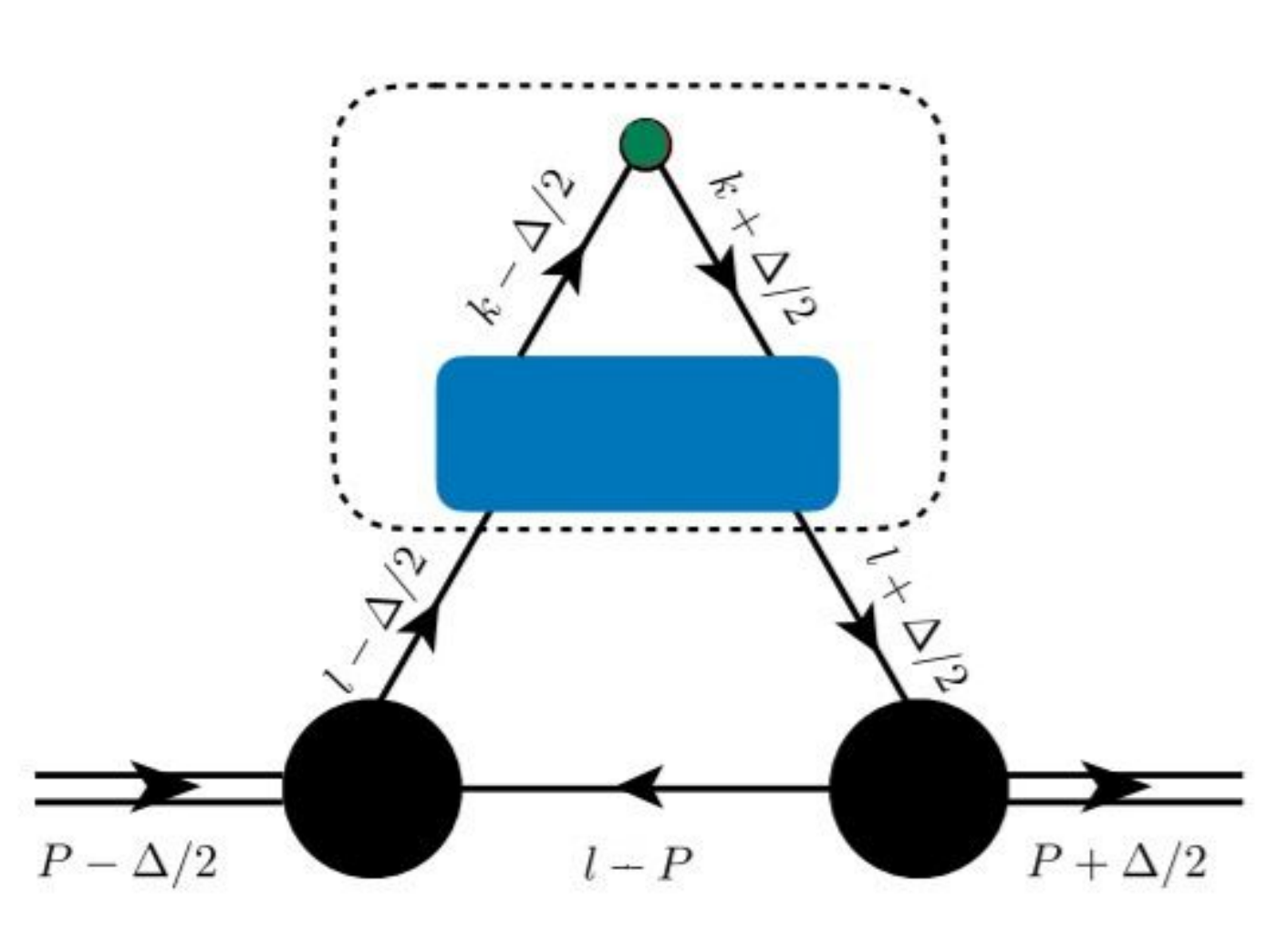}
\caption{Diagram for $H^c(x,\xi,t)$. The dash line boxed area represents the dressed vertex $\Gamma^+$.}\label{dqpvgpd}
\end{figure}
In addition to the GPDs in Fig. \ref{GPD}, we also include the contribution responsible for the D-term as Fig. 2 in Ref.~\cite{Broniowski:2007si}, which is plotted in Fig. \ref{dterm}, the definition of this term is
\begin{align}\label{dgpd}
&H^b\left(x,\xi,t\right)=2i N_c Z_{\pi}\int \frac{\mathrm{d}^4k}{(2 \pi )^4}\delta_n^x (k)\nonumber\\
&\times \text{tr}_{\text{D}}\left[\gamma _5 S \left(k_{+q}\right)\gamma ^+S\left(k_{-q}\right)\gamma _5\right],
\end{align}
what's more, the definition in Eq. (\ref{gpddd}) use the bare vertex, if we use the dressed quark--photon vertex, for the FF, we can just use $\gamma^{\mu}F_{1\rho}(t)$ for the vertex. For vector GPD, we can use the BSE to get the dressed GPD term with not only variable $t$, but also $x$ and $\xi$. Ref.~\cite{Shi:2020pqe} has showed the process in their appendix, the inhomogeneous BSE in Fig. \ref{qpv}, on the right side, the first diagram is the bare vertex corresponds to the bare vector GPD $H^a(x,\xi,t)$, the second term corresponds to the dressed GPD term $H^c(x,\xi,t)$ which is plotted in Fig. \ref{dqpvgpd}, after some calculation, we get the final form of $H^c$
\begin{align}\label{bdgpd}
&H^c\left(x,\xi,t\right)=F^{\alpha}(t)P^{\alpha}(x,\xi,t),
\end{align}
where $F^{\alpha}(t)=(p^{\alpha}+p^{'\alpha})F(t)$, $F(t)$ is the electromagnetic form factor, which is defined as
\begin{align}\label{dgpd}
F^{\alpha}(t)&=2i N_c Z_{\pi} \int \frac{\mathrm{d}^4l}{(2 \pi )^4}\nonumber\\
 &\times \text{tr}_{\text{D}}\left[\gamma _5 S \left(l_{+q}\right)\gamma ^{\alpha} S\left(l_{-q}\right)\gamma _5 S\left(l-P\right)\right],
\end{align}
$P^{\alpha}(x,\xi,t)$ is defined as
\begin{align}\label{dgpd}
P^{\alpha}(x,\xi,t)&=2i G_{\rho}\int \frac{\mathrm{d}^4k}{(2 \pi )^4}\delta_n^x (k)\nonumber\\
 &\times\text{tr}_{\text{D}}\left[\gamma _{\alpha} S \left(k_{+q}\right)\Gamma ^+S\left(k_{-q}\right)\right],
\end{align}
where $\Gamma ^+$ is the big green vertex in the inhomogeneous BSE in Fig. \ref{qpv}, $F_{1\rho}(t)\gamma^+$, $\alpha$ contract at last.

\subsection{The properties of pion GPDs}\label{qq}

\subsubsection{Forward limit}\label{qqQ}
When the initial and final hadrons have the same momentum $p=p'$, $\xi=0$, $t=q^2=0$, GPD should reduce to usual pion PDF,
\begin{align}\label{as}
u(x)&=H(x,0,0),
\end{align}
which should satisfy
\begin{align}\label{as}
\int_0^1u(x) \mathrm{d}x =1.
\end{align}

\subsubsection{Symmetry properties}\label{qqS}
GPDs are neither even nor odd of momentum fraction $x$, the combinations,
\begin{subequations}\label{conjugation}
\begin{align}\label{8}
H^{I=0}(x,\xi,t)&=H(x,\xi,t)-H(-x,\xi,t)\,, \\
H^{I=1}(x,\xi,t)&=H(x,\xi,t)+H(-x,\xi,t),
\end{align}
\end{subequations}
represent the isoscalar (isovector) pion GPDs, respectively, tensor GPD $E(x,\xi,t)$ is resemble.

Time reversal invariance requires,
\begin{subequations}
\begin{align}\label{spgpd}
H^u(x,\xi,t)&=H^u(x,-\xi,t)\,,\\\label{sptgpd}
E^u(x,\xi,t)&=E^u(x,-\xi,t),
\end{align}
\end{subequations}
we will check this property in different regularization separately.

\subsubsection{Polynomiality condition}\label{Bqq}
The Mellin moments of GPDs play an important role, which should satisfy the polynomiality condition
\begin{subequations}\label{pc}
\begin{align}
\int_{-1}^1 x^n H(x,\xi,t) \mathrm{d}x  &= \sum _{i=0}^{[(n+1)/2]}\xi^{2i} A_{n+1,2i}(t)\,, \\
\int_{-1}^1 x^n E(x,\xi,t)\mathrm{d}x  &= \sum _{i=0}^{[(n+1)/2]}\xi^{2i} B_{n+1,2i}(t),
\end{align}
\end{subequations}
where $A_{n+1,2i}(t)$ and $B_{n+1,2i}(t)$ are vector and tensor generalized factor factors (GFFs), respectively.

For $n=0$, the FFs are $\xi$ independent,
\begin{subequations}
\begin{align}\label{ptc}
\int_{-1}^1  H(x,\xi,t) \mathrm{d}x &=  A_{1,0}(t)\,, \\
\int_{-1}^1 E(x,\xi,t)\mathrm{d}x &=B_{1,0}(t),
\end{align}
\end{subequations}
where $ A_{1,0}(t)$ and $B_{1,0}(t)$ are the pion electromagnetic FF and tensor FF. $B_{1,0}(0)$ is the tensor anomalous magnetic moment for $n=0$.

For $n=1$:
\begin{subequations}
\begin{align}\label{ab94}
\int_{-1}^1 x H(x,\xi,t) \mathrm{d}x&= A_{2,0}(t)+\xi^2 A_{2,2}(t) \nonumber\\
&=\theta_2(t)-\xi^2\theta_1(t)\,, \\
\int_{-1}^1 x E(x,\xi,t) \mathrm{d}x&= B_{2,0}(t)+\xi^2 B_{2,2}(t),
\end{align}
\end{subequations}
where $\theta_1$ relates to the quark pressure distribution and $\theta_2$ relates to the quark mass distribution inside the pion. The gravitational form factors satisfy the low-energy theorem $\theta_1(0)-\theta_2(0)=\mathcal{O}(m_{\pi}^2)$~\cite{Donoghue:1991qv}

A light-cone energy radius can be defined in relation to the generalized FF $A_{2,0}(Q^2)$~\cite{Freese:2019bhb}
\begin{subequations}
\begin{eqnarray}\label{a93}
\langle r_{E,LC}^2\rangle&=&-4\frac{\partial A_{2,0}(Q^2)}{\partial Q^2}|_{Q^2=0}\,,\\
\langle r_{c,LC}^2\rangle&=&-4\frac{\partial F_K(Q^2)}{\partial Q^2}|_{Q^2=0},
\end{eqnarray}
\end{subequations}
the second equation is analogous to light-cone charge radius, which is used to make comparison with the light-cone energy radius.

\subsubsection{Impact parameter space PDF}
The impact parameter dependent PDFs are the Fourier transform of GPDs at $\xi=0$
\begin{align}\label{aG9}
q\left(x,\bm{b}_{\perp}^2\right)=\int \frac{\mathrm{d}^2\bm{q}_{\perp}}{(2 \pi )^2}e^{-i\bm{b}_{\perp}\cdot \bm{q}_{\perp}} H\left(x,0,-\bm{q}_{\perp}^2\right).
\end{align}
$H(x,0,-\bm{q}_{\bot}^2)$ should be independent of $-\bm{q}_{\bot}^2$ when $x\rightarrow 1$ as Ref.~\cite{Burkardt:2002hr}.

The width distribution of $u$ quark in the pion is defined as~\cite{Burkardt:2002hr}
\begin{align}\label{tmmf}
\langle \bm{r}_{\bot}^2\rangle_x &=\frac{\int \mathrm{d}^2 \bm{b}_{\bot}\bm{b}_{\bot}^2u(x,\bm{b}_{\bot}^2)}{\int \mathrm{d}^2 \bm{b}_{\bot}u(x,\bm{b}_{\bot}^2)},
\end{align}
when $x\rightarrow 1$, the struck quark becomes closer to the centre of momentum since its weight increases, this average impact parameter should be zero.

The mean-squared $\bm{b}_{\bot}$ is defined as
\begin{align}\label{br1}
\langle \bm{b}_{\bot}^2\rangle &=\int_0^1 \mathrm{d}x \int \mathrm{d}^2 \bm{b}_{\bot}\bm{b}_{\bot}^2q(x,\bm{b}_{\bot}^2),
\end{align}
then we will obtain $\langle \bm{b}_{\bot}^2\rangle$ in different regularization schemes.

\subsubsection{Light-front transverse-spin distributions}
The light-front transverse-spin distributions are defined as~\cite{Brommel:2007xd}
\begin{align}\label{lftsd}
\rho^n\left(\bm{b}_{\bot },\bm{s}_{\perp}\right)
&=\frac{1}{2}\left[A_{n,0}(\bm{b}_{\bot }^2)-\frac{\varepsilon_{\bot}^{ij}\bm{s}_{\perp}^i\bm{b}_{\bot}^j}{m_{H}}B_{n,0}^{'}(\bm{b}_{\bot }^2)\right],
\end{align}
where $m_{H}$ is the hadron mass, $B_{n,0}^{'}(\bm{b}_{\bot }^2)=\partial_{\bm{b}_{\bot }^2}B_{n,0}(\bm{b}_{\bot }^2)$, $A_{n,0}(\bm{b}_{\bot }^2)$ and $B_{n,0}(\bm{b}_{\bot }^2)$ are the two-dimensional Fourier transform of $A_{n,0}(\bm{q}_{\bot }^2)$ and $B_{n,0}(\bm{q}_{\bot }^2)$.

When quark polarized in the light-front-transverse $+\,x$ direction, the transverse-spin density is not symmetric around $\bm{b}_{\bot}=(b_x=0,b_y=0)$ anymore, the peaks shift to $(b_x=0,b_y>0)$~\cite{Zhang:2020ecj,Zhang:2021tnr}.

The average transverse shift
\begin{align}\label{ats}
\langle b_{\bot}^y\rangle_n^u=\frac{\int \mathrm{d}^2\bm{b}_{\bot }b_{\bot }^y \rho_u^n\left(\bm{b}_{\bot },\bm{s}_{\perp}\right)}{\int \mathrm{d}^2\bm{b}_{\bot } \rho_u^n\left(\bm{b}_{\bot },\bm{s}_{\perp}\right)}=\frac{1}{2m_{\pi}}\frac{B_{n,0}^u(0)}{A_{n,0}^u(0)},
\end{align}
in the $y$ direction for a transverse quark spin $s_{\bot} = (1, 0)$ in the $x$ direction.

\section{GPDs in different regularization schemes}\label{well1}

\subsection{PT regularization}
The basic idea of PT regularization is based on~\cite{Ebert:1996vx,Hellstern:1997nv,Bentz:2001vc}.
\begin{align}\label{4}
\frac{1}{X^n}&=\frac{1}{(n-1)!}\int_0^{\infty}\mathrm{d}\tau \,\tau^{n-1}e^{-\tau X}\nonumber\\
& \rightarrow \frac{1}{(n-1)!} \int_{1/\Lambda_{\text{UV}}^2}^{1/\Lambda_{\text{IR}}^2}\mathrm{d}\tau \,\tau^{n-1}e^{-\tau X}
\end{align}
where the $1/\Lambda_{\text{UV}}^2$ induces the dumping factor into the original propagator. Besides the ultraviolet cutoff, we also introduce the infrared cutoff $\Lambda_{\text{IR}}$ to mimic confinement, the infrared cutoff is used to mimic confinement. The parameters used for PT regularization scheme are demonstrated in Table \ref{tbpt}.

Using PT regularization, the gap equation in Eq. (\ref{nocutoffgap}) becomes
\begin{align}\label{ptcutoffgap}
M
&=m+\frac{3G_{\pi}M}{\pi^2}\int_{1/\Lambda_{\text{UV}}^2}^{1/\Lambda_{\text{IR}}^2}\mathrm{d}\tau \frac{1}{\tau^2} e^{-\tau M^2},
\end{align}
the pion decay constant becomes
\begin{align}\label{ptfpi}
f_{\pi}^{\text{PT}}=\frac{N_c\sqrt{Z_{\pi }}M}{4\pi ^2}\int_0^1 \mathrm{d}x\,\bar{\mathcal{C}}_1(\sigma_1),
\end{align}
the vector bubble diagram $\Pi_{\text{VV}}$ is defined as
\begin{align}\label{ptvvbu}
\Pi_{\text{VV}}^{\text{PT}}(t)=-\frac{3 }{\pi ^2} \int_0^1\mathrm{d}x\, x (1-x) t \,\bar{\mathcal{C}}_1(\sigma_2).
\end{align}

\begin{figure*}
\centering
\includegraphics[width=0.47\textwidth]{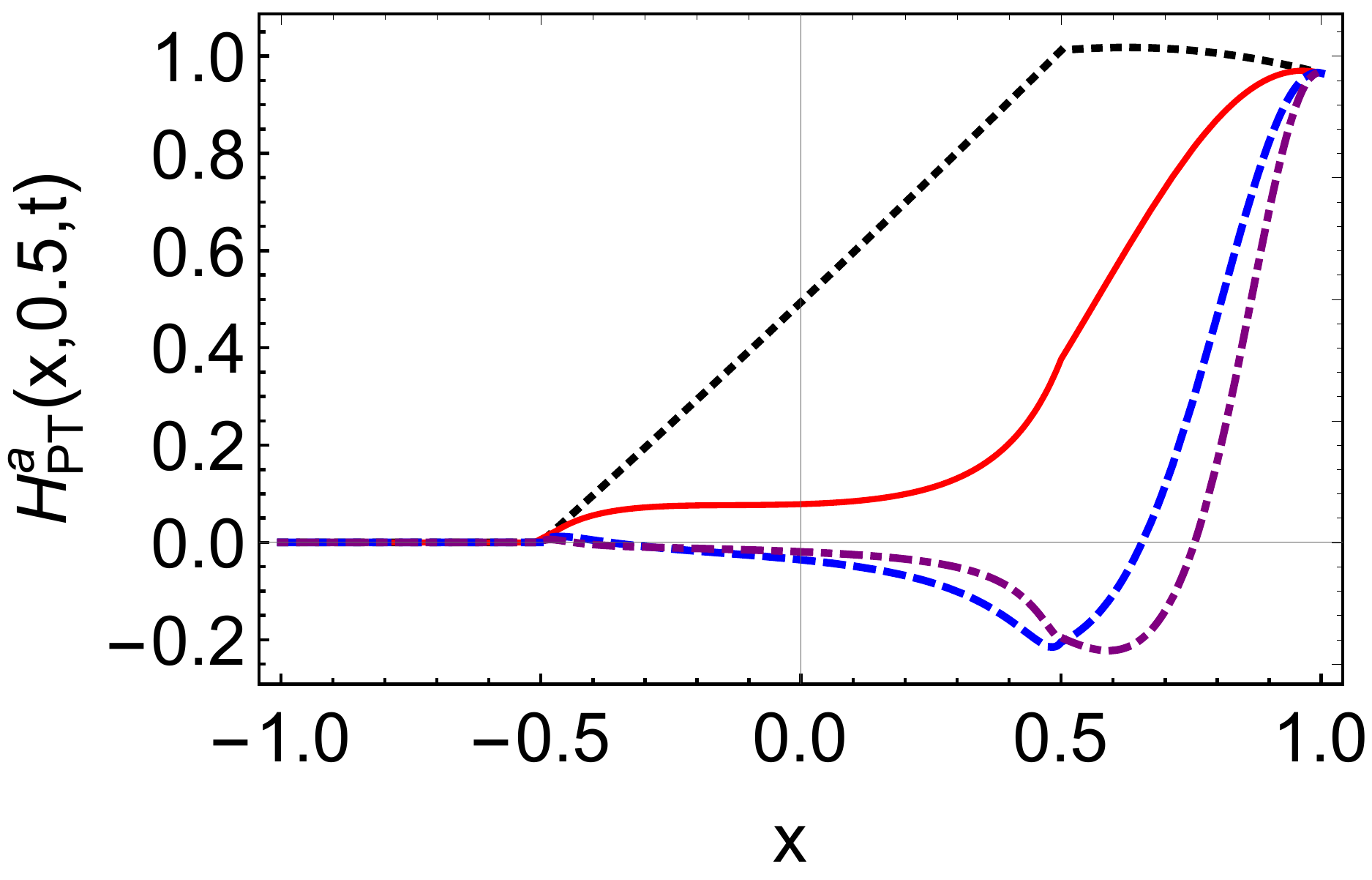}
\qquad
\includegraphics[width=0.47\textwidth]{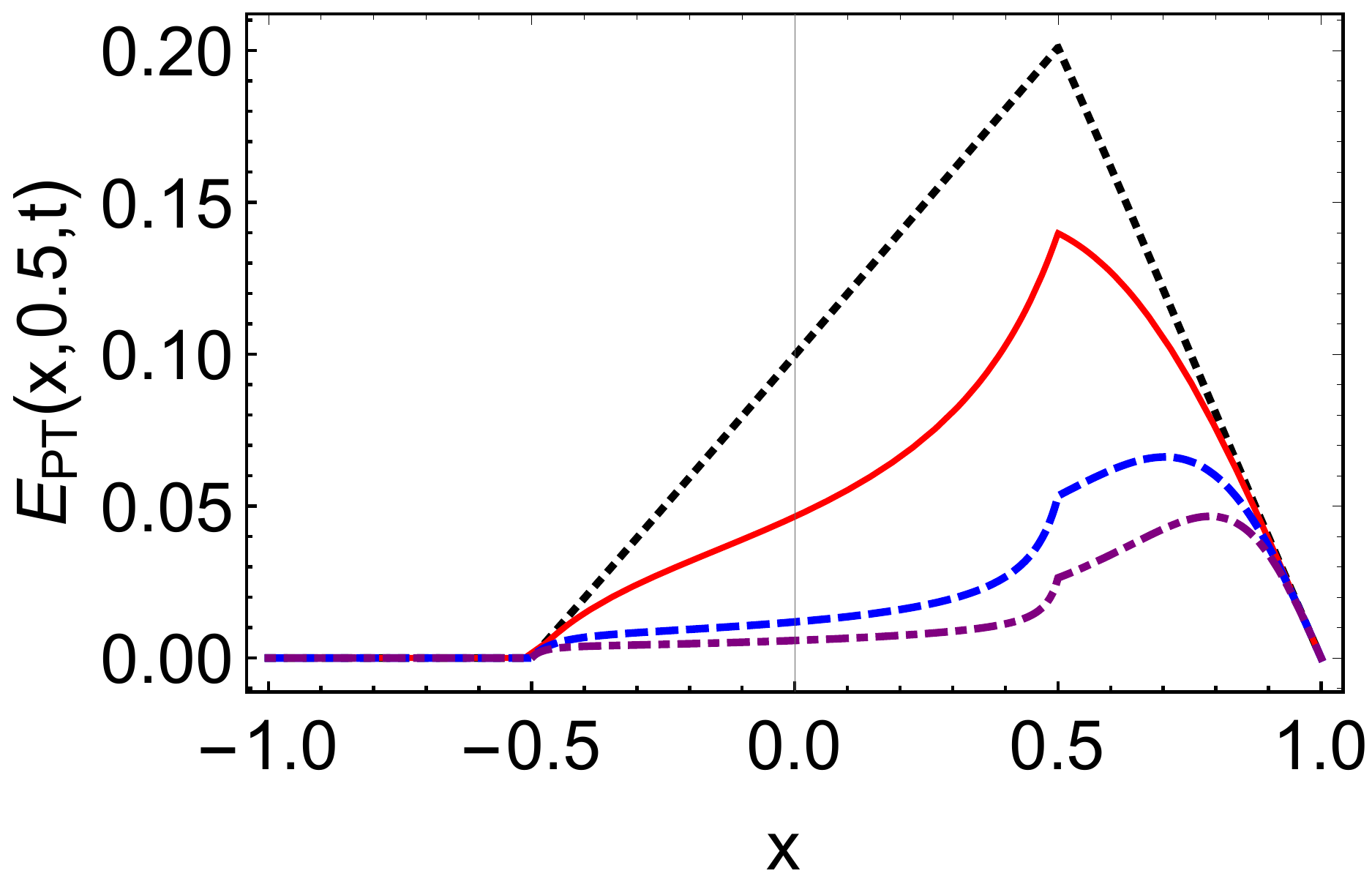}
\caption{Pion GPDs (left panel: vector GPD, right panel: tensor GPD) at $\xi=0.5$ with different $t$ using PT regularization: black dotted line -- $t=0$ GeV$^2$, red line -- $t=-1$ GeV$^2$, blue dashed line -- $t=-5$ GeV$^2$, purple dotdashed line -- $t=-10$ GeV$^2$. }\label{ptgpd}
\end{figure*}

\begin{figure*}
\centering
\includegraphics[width=0.47\textwidth]{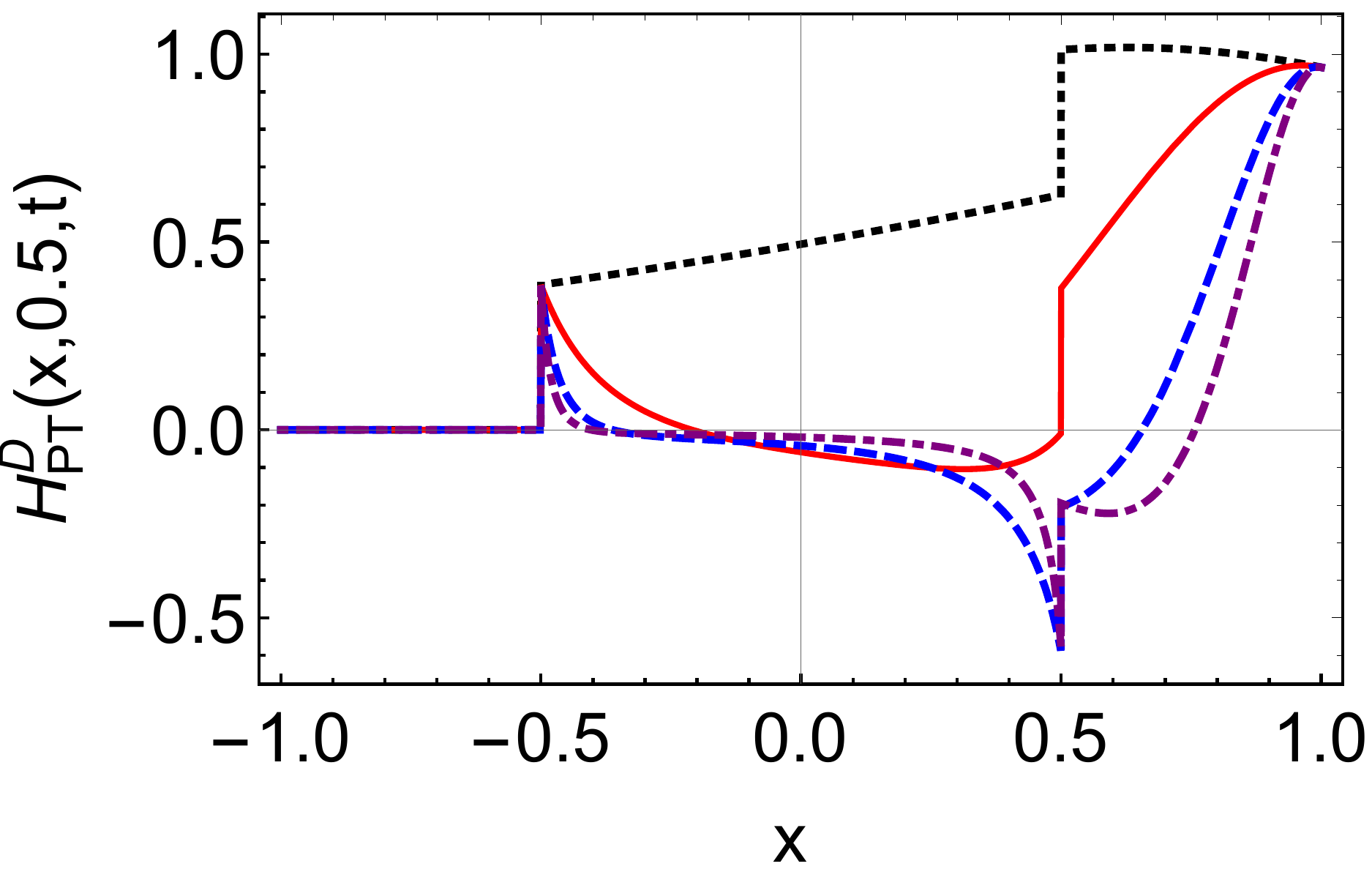}
\qquad
\includegraphics[width=0.47\textwidth]{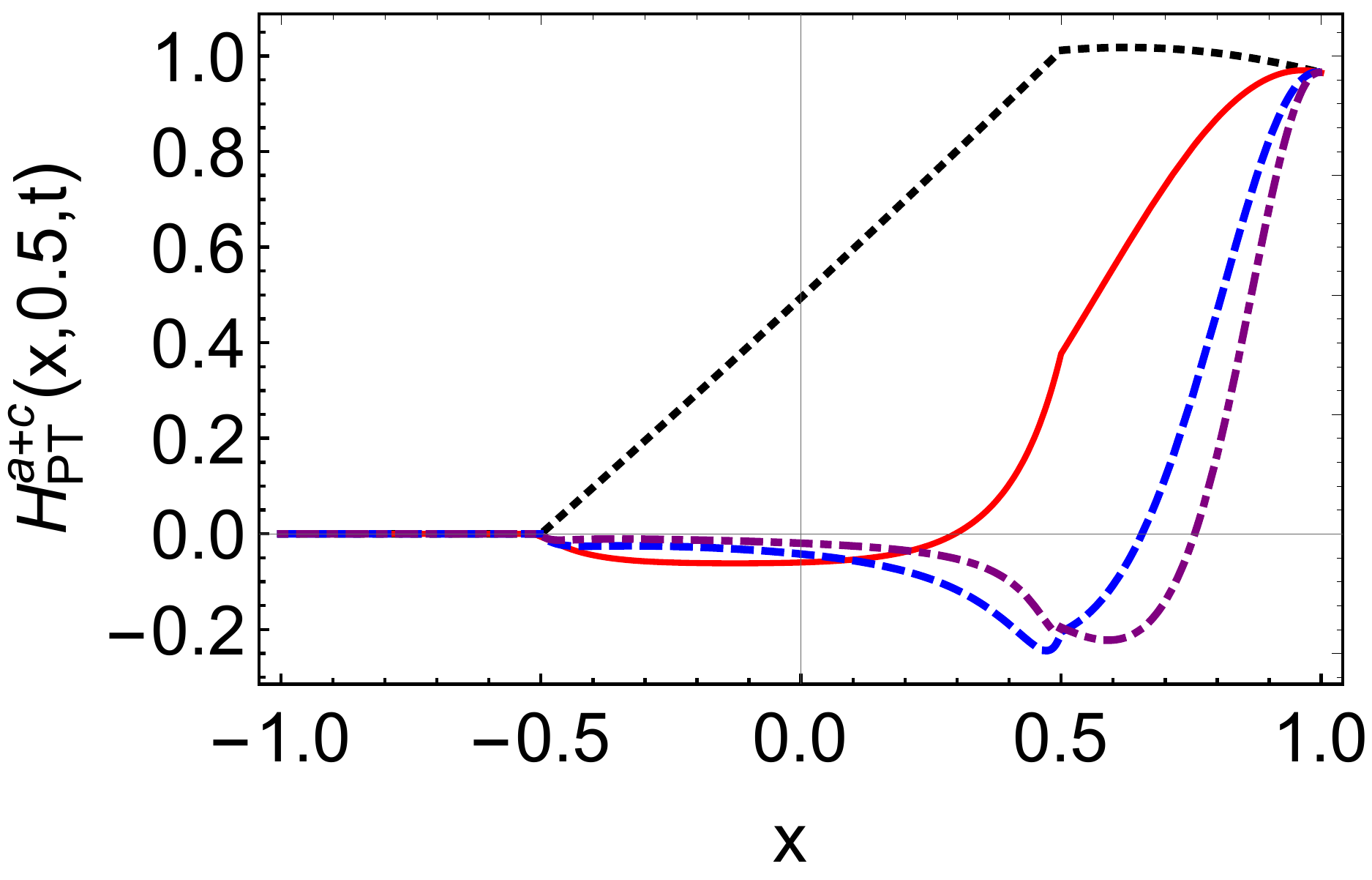}
\caption{Pion GPDs (left panel: $H_{\text{PT}}^D$, right panel: $H_{\text{PT}}^a+H_{\text{PT}}^c$) at $\xi=0.5$ with different $t$ using PT regularization: black dotted line -- $t=0$ GeV$^2$, red line -- $t=-1$ GeV$^2$, blue dashed line -- $t=-5$ GeV$^2$, purple dotdashed line -- $t=-10$ GeV$^2$. }\label{ptgpd1}
\end{figure*}

\begin{center}
\begin{table*}
\caption{Parameter set used for PT regularization. The dressed quark mass and regularization parameters are in units of GeV, while coupling constant are in units of GeV$^{-2}$.}\label{tbpt}
\begin{tabular}{p{1.0cm} p{1.0cm} p{1.0cm} p{1.0cm}p{1.0cm}p{1.0cm}p{1.0cm}p{1.0cm}p{1.0cm}p{1.0cm}p{1.4cm}}
\hline\hline
$\Lambda_{\text{IR}}$&$\Lambda _{\text{PT}}$&$M$&$Z_{\pi}$&$G_{\pi}$&$G_{\rho}$&$G_{\omega}$&$m_{\pi}$&$f_{\pi}$&$m$&$\langle\bar{u}u\rangle$\\
\hline
0.240&0.645&0.4&17.847 &18.541&11.042&10.412&0.14&0.93&0.016&--(0.173)$^3$\\
\hline\hline
\end{tabular}
\end{table*}
\end{center}
Using PT regularization scheme, the final results of pion GPDs are
\begin{align}\label{agpdf}
&\quad H_{\text{PT}}^a\left(x,\xi,t\right)\nonumber\\
&=\frac{N_cZ_{\pi }}{8\pi ^2} \left[\theta_{\bar{\xi} 1}\bar{\mathcal{C}}_1(\sigma_3)+ \theta_{\xi 1} \bar{\mathcal{C}}_1(\sigma_4)+\theta_{\bar{\xi} \xi}\frac{x}{\xi }\bar{\mathcal{C}}_1(\sigma_5)\right]\nonumber\\
&+\frac{N_c Z_{\pi } }{8\pi ^2}\int_0^1 \mathrm{d}\alpha \frac{\theta_{\alpha \xi}}{\xi} (2 x m_{\pi}^2+(1-x)t) \frac{1}{\sigma_6}\bar{\mathcal{C}}_2(\sigma_6),
\end{align}
\begin{align}\label{agtpdf}
E_{\text{PT}}\left(x,\xi,t\right)=\frac{N_c Z_{\pi } }{4\pi ^2}\int_0^1 \mathrm{d}\alpha \frac{\theta_{\alpha \xi}}{\xi}m_{\pi}M\frac{1}{\sigma_6}\bar{\mathcal{C}}_2(\sigma_6),
\end{align}
\begin{align}\label{agtpdf}
H_{\text{PT}}^b\left(x,\xi,t\right)=-\frac{N_cZ_{\pi }}{4\pi ^2} \frac{\theta_{\bar{\xi} \xi}}{\xi } M x\, \bar{\mathcal{C}}_1(\sigma_5),
\end{align}
\begin{align}\label{agtpdf}
H_{\text{PT}}^c\left(x,\xi,t\right)=F_{1\rho}^{\text{PT}}(t)A_{1,0}^{\text{PT}}(t)\frac{N_cG_{\rho }}{4\pi ^2} \frac{\theta_{\bar{\xi} \xi}}{\xi }t\left(1-\frac{x^2}{\xi^2}\right)\bar{\mathcal{C}}_1(\sigma_5),
\end{align}
and
\begin{subequations}\label{region1}
\begin{align}
\theta_{\bar{\xi} 1}&=x\in[-\xi, 1]\,, \\
\theta_{\xi 1}&=x\in[\xi, 1]\,, \\
\theta_{\bar{\xi} \xi}&=x\in[-\xi, \xi]\,, \\
\theta_{\alpha \xi}&=x\in[\alpha (\xi +1)-\xi , \alpha  (1-\xi)+\xi ]\cap x\in[-1,1],
\end{align}
\end{subequations}
where $\theta$ is the step function, $x$ only exist in the corresponding region. One can write $\theta_{\bar{\xi} \xi}/\xi=\Theta(1-x^2/\xi^2) $, where $\Theta(x)$ is the Heaviside function, and $\theta_{\alpha \xi}/\xi=\Theta((1-\alpha^2)-(x-\alpha)^2/
\xi^2)\Theta(1-x^2)$. These results are in the region $\xi > 0$, under $\xi \rightarrow -\xi $: $\theta_{\bar{\xi} 1} \leftrightarrow \theta_{\xi 1}$; and $\theta_{\bar{\xi} \xi}/\xi$, $\theta_{\alpha \xi}/\xi$ are invariant. The dressed vector GPDs are
\begin{align}
H_{\text{PT}}^D\left(x,\xi,t\right)=H_{\text{PT}}^a\left(x,\xi,t\right)+H_{\text{PT}}^b\left(x,\xi,t\right)+H_{\text{PT}}^c\left(x,\xi,t\right).
\end{align}
We plot the diagrams of vector and tensor GPDs using PT regularization in Figs. \ref{ptgpd} and \ref{ptgpd1}. From the diagrams we can see, $H^a\left(x,\xi,0\right)=0$ when $x<-\xi$, GPDs are continuous but not differentiable at $x =\pm \, \xi$ except $\xi=0$~\cite{Zhang:2021tnr}, tensor GPD is similar. When considering the dressed quark-photon vertex and D-term, vector GPD becomes not continuous at $x =\pm \, \xi$, from Fig. \ref{ptgpd1} we can see that the discontinuity comes from the D-term.  

%
%
%
%

%

\subsubsection{Forward limit}
\begin{figure}
\centering
\includegraphics[width=0.47\textwidth]{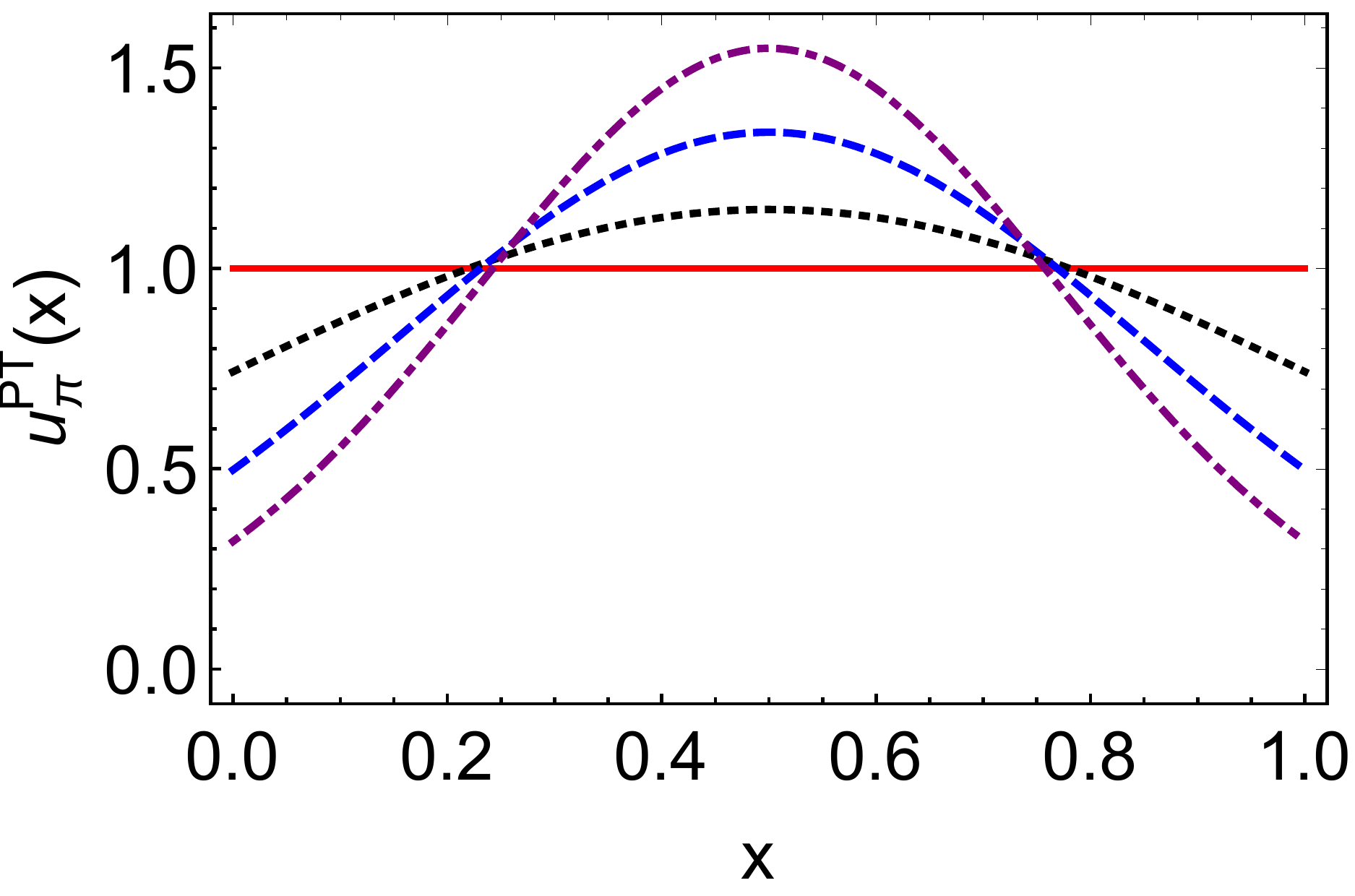}
\caption{Pion $u$ quark PDF with different values of $m_{\pi}$ using PT regularization scheme: red line -- $m_{\pi}=0$ GeV, black dotted line -- $m_{\pi}=0.4$ GeV, blue dashed line -- $m_{\pi}=0.6$ GeV, purple dotdashed line -- $m_{\pi}=0.75$ GeV. }\label{ptpdf}
\end{figure}

In the forward limit
\begin{align}\label{hpdf}
u_{\text{PT}}(x)&=\frac{3Z_{\pi }}{4\pi ^2}\bar{\mathcal{C}}_1(\sigma_1) +\frac{3Z_{\pi }}{2\pi ^2} x(1-x) m_{\pi}^2 \frac{\bar{\mathcal{C}}_2(\sigma_1)}{\sigma_1},
\end{align}
which satisfies
\begin{align}\label{h1pdf}
\int_0^1  u(x)\mathrm{d}x=1.
\end{align}
This PDF, varies around $1$, only changes a little with $x$. In the chiral limit $m_{\pi}=0$, pion PDF $q(x)=1$, Refs.~\cite{Noguera:2015iia,Theussl:2002xp,Zhang:2021shm}. When $m_{\pi}\neq 0$, our results show pion $q(x)\approx 1$, it fluctuates around $1$. Ref.~\cite{Theussl:2002xp} plotted pion PDF with different values of $m_{\pi}$. For comparison, we plot the diagrams of $u$ quark PDFs of pion with different values of $m_{\pi}$ in Fig. \ref{ptpdf}.

\subsubsection{Form factors}
The dressed form factors
\begin{align}
A_{1,0}^{D,\text{PT}}(t)&= \int_{-1}^1  H_{\text{PT}}^D\left(x,\xi,t\right)\mathrm{d}x\nonumber\\
&=\int_{-1}^1 H_{\text{PT}}^a\left(x,\xi,t\right)\mathrm{d}x\nonumber\\
&+F_{1\rho}^{\text{PT}}(t)A_{1,0}^{\text{PT}}(t)\int_{-1}^1  P\left(x,\xi,t\right)\mathrm{d}x\nonumber\\
&=A_{1,0}^{\text{PT}}(t)-F_{1\rho}^{\text{PT}}(t)A_{1,0}^{\text{PT}}(t)2G_{\rho}\Pi_{\text{VV}}(t)\nonumber\\
&=A_{1,0}^{\text{PT}}(t)F_{1\rho}^{\text{PT}}(t),
\end{align}
where $H_{\text{PT}}^b(x,\xi,t)$ is odd function of $x$, so
\begin{align}
\int_{-1}^1 H_{\text{PT}}^b(x,\xi,t)\mathrm{d}x=0,
\end{align}
but
\begin{align}
&\int_{-1}^1 x H_{\text{PT}}^b(x,\xi,t)\mathrm{d}x\nonumber\\
=&-\frac{N_c Z_{\pi } }{\pi ^2}  \int_0^1 \mathrm{d}x \,M x (1-2x) \bar{\mathcal{C}}_1(\sigma_2),
\end{align}
so it supplies $A_{2,2}$, the final form of quark pressure distribution is $A_{2,2}^D=A_{2,2}+A_{2,2}^b$.
\begin{align}\label{aF3}
A_{1,0}^{\text{PT}}(t)&=\frac{N_cZ_{\pi } }{4\pi^2}\int_0^1 \mathrm{d}x\,\bar{\mathcal{C}}_1(\sigma_1)\nonumber\\
&+\frac{N_cZ_{\pi }}{4\pi^2} \int_0^1 \mathrm{d}x \int_0^{1-x}\mathrm{d}y   \nonumber\\
&\times (2m_{\pi }^2-(x+y)(2m_{\pi }^2-t)) \frac{1}{\sigma_7}\bar{\mathcal{C}}_2(\sigma_7),
\end{align}
\begin{align}\label{a1F3}
B_{1,0}^{\text{PT}}(t)
&=\frac{ N_c Z_{\pi } }{2\pi ^2}  \int _0^1\mathrm{d}x \int _0^{1-x}\mathrm{d}y \frac{M m_{\pi}}{\sigma_7}\bar{\mathcal{C}}_2(\sigma_7).
\end{align}
\begin{align}\label{a93}
A_{2,0}^{\text{PT}}(t)&=\frac{N_cZ_{\pi }}{8\pi ^2}\int_0^1 \mathrm{d}x\,\bar{\mathcal{C}}_1(\sigma_1) \nonumber\\
&+\frac{N_c Z_{\pi }}{4\pi ^2}  \int_0^1 \mathrm{d}x \int_0^{1-x} \mathrm{d}y\frac{1}{\sigma_7}\bar{\mathcal{C}}_2(\sigma_7) \nonumber\\
&\times (2 m_{\pi }^2 (1-x-y)^2+t (1-x-y) (x+y)),
\end{align}
\begin{align}\label{a93}
A_{2,2}^{\text{PT}}(t)&=-\frac{N_cZ_{\pi }}{4\pi ^2}\int_0^1 \mathrm{d}x\,(1-x)\bar{\mathcal{C}}_1(\sigma_1)\nonumber\\
&-\frac{N_c Z_{\pi } }{2\pi ^2}  \int_0^1 \mathrm{d}x \,x (1-2x) \bar{\mathcal{C}}_1(\sigma_2)\nonumber\\
&-\frac{N_c Z_{\pi } }{4\pi^2} \int_0^1  \mathrm{d}x  \frac{ (1-x ) \left(2 m_{\pi }^2-t\right)}{t}\bar{\mathcal{C}}_1(\sigma_1)\nonumber\\
&+\frac{N_c Z_{\pi }}{4\pi^2}\int_0^1 \mathrm{d}x \int_0^{1-x} \mathrm{d}y \frac{\left(2 m_{\pi }^2-t\right)}{t }\bar{\mathcal{C}}_1(\sigma_7),
\end{align}
\begin{eqnarray}\label{a93}
B_{2,0}^{\text{PT}}(t)&=&\frac{N_c Z_{\pi}  }{2\pi ^2} \int_0^1 \mathrm{d}x \int_0^{1-x} \mathrm{d}y  \nonumber\\
&\times & m_{\pi}M (1-x-y)\frac{1}{\sigma_7}\bar{\mathcal{C}}_2(\sigma_7),
\end{eqnarray}
$B_{2,2}(t)$ is zero.

\subsubsection{Impact parameter space PDFs}
\begin{figure}
\centering
\includegraphics[width=0.47\textwidth]{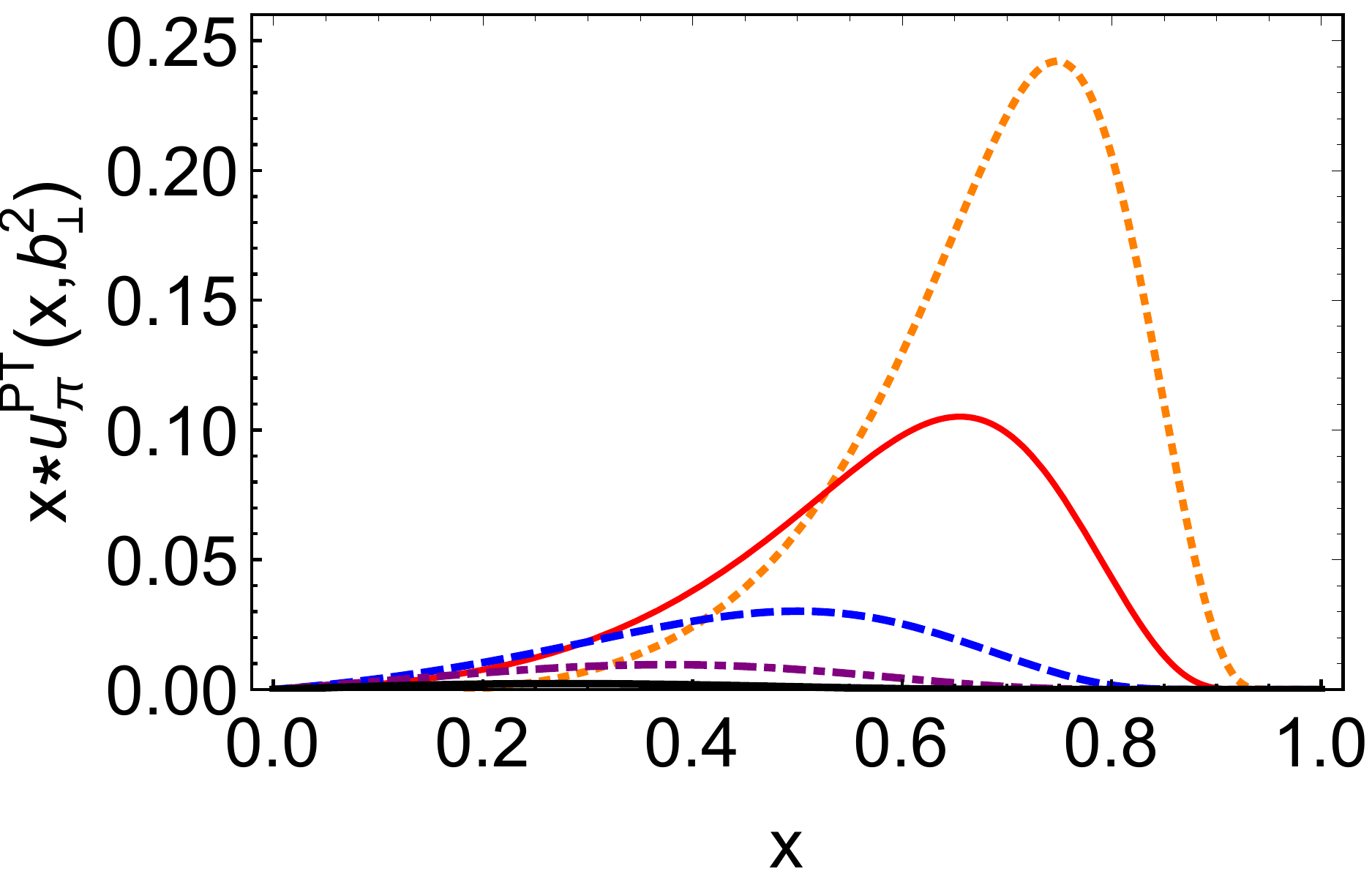}
\qquad
\includegraphics[width=0.47\textwidth]{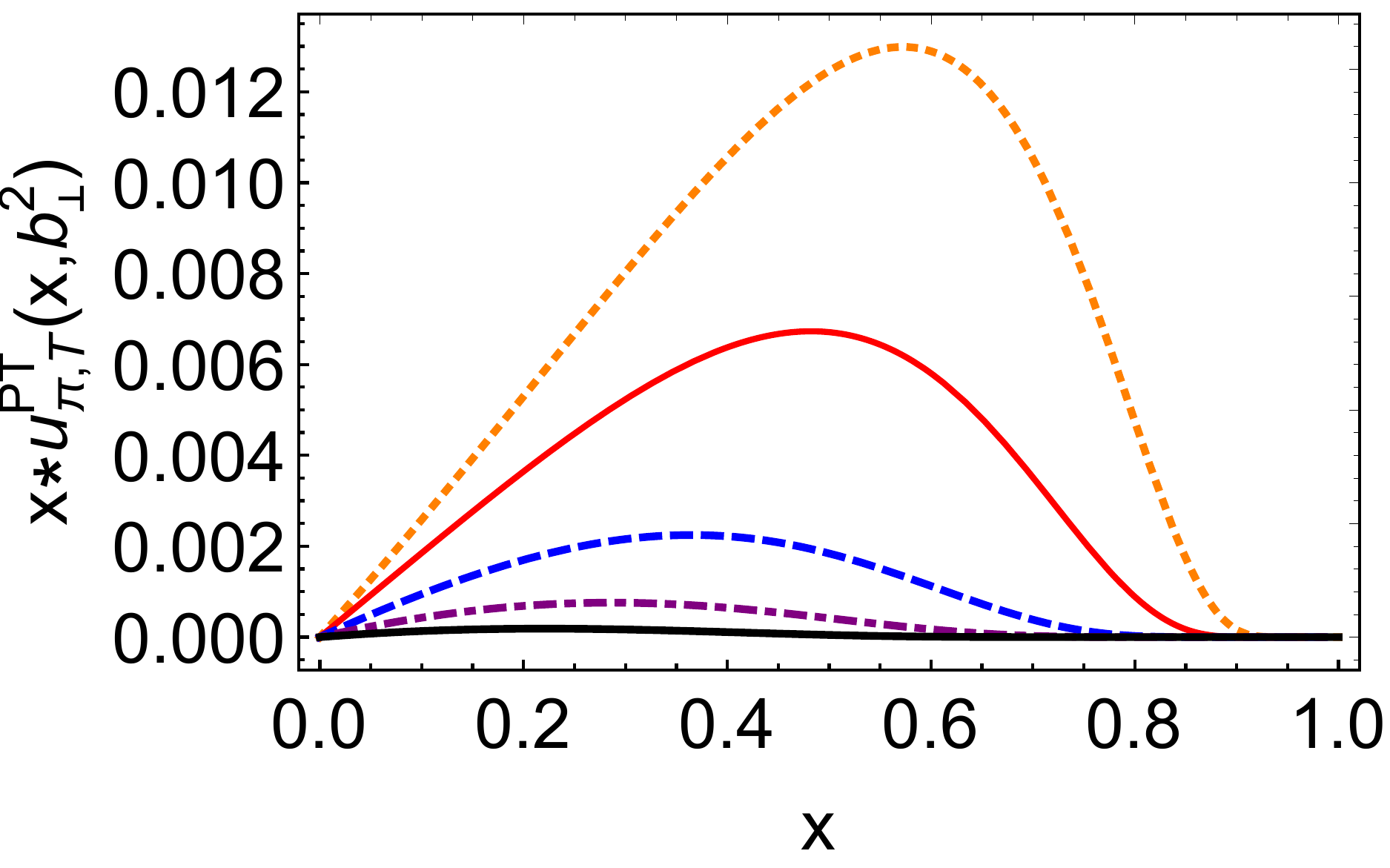}
\caption{Impact parameter space PDFs using PT regularization scheme: left panel -- $x*u\left(x,\bm{b}_{\perp}^2\right)$, the $\delta^2(\bm{b}_{\perp})$ component first line of Eq. (\ref{1ipspdf}) - is suppressed in the image, and right panel -- $x *u_T\left(x,\bm{b}_{\perp}^2\right)$ both panels with $\bm{b}_{\perp}^2=0.5$ GeV$^{-2}$ --- orange dotted curve, $\bm{b}_{\perp}^2=1$ GeV$^{-2}$ --- red solid curve, $\bm{b}_{\perp}^2=2.5$ GeV$^{-2}$ --- blue dashed curve, $\bm{b}_{\perp}^2=5$ GeV$^{-2}$ --- purple dot-dashed curve, $\bm{b}_{\perp}^2=10$ GeV$^{-2}$ --- thick black solid curve.}\label{ptqxb}
\end{figure}

In the PT regularization scheme
\begin{align}\label{1ipspdf}
&\quad u^{\text{PT}}\left(x,\bm{b}_{\perp}^2\right)\nonumber\\
&=\frac{N_cZ_{\pi }}{4\pi ^2} \int \frac{\mathrm{d}^2\bm{q}_{\perp}}{(2 \pi )^2}e^{-i\bm{b}_{\perp}\cdot \bm{q}_{\perp}} \bar{\mathcal{C}}_1(\sigma_1) \nonumber\\
&+\frac{N_cZ_{\pi }}{32\pi ^3}\int_0^{1-x} \mathrm{d}\alpha \int \mathrm{d}\tau  \nonumber\\
&\times \frac{(x-1) \left(4-\frac{\bm{b}_{\perp}^2}{\alpha (1-\alpha -x)\tau }\right)+8 \alpha   x (1-\alpha-x)\tau m_{\pi }^2 }{4 \alpha ^2 \tau ^2 (1-\alpha-x)^2 } \nonumber\\
&\times e^{ -\tau  \left(M^2-x \left(1-x\right) m_{\pi}^2\right)} e^{-\frac{\bm{b}_{\perp}^2}{4\tau \left(1-\alpha-x\right) \alpha}},
\end{align}
\begin{align}\label{2ipspdf}
u_T^{\text{PT}}\left(x,\bm{b}_{\perp}^2\right)
&=\frac{N_cZ_{\pi }}{16\pi^3}\int_0^{1-x}  \mathrm{d}\alpha \int \mathrm{d}\tau  \frac{m_{\pi}M }{\alpha\left(1-\alpha-x\right)\tau }\nonumber\\
&\times e^{- \frac{\bm{b}_{\perp}^2}{4\tau (1-\alpha-x)\alpha}}e^{-\tau \left(M^2-x \left(1-x\right) m_{\pi}^2\right)},
\end{align}
for $u^{\text{PT}}\left(x,\bm{b}_{\perp}^2\right)$, when integrating $\bm{b}_{\perp}$ one can get PDF $u_{\text{PT}}(x)$ in Eq. (\ref{hpdf}). We plot the diagrams of $x *u^{\text{PT}}\left(x,\bm{b}_{\perp}^2\right)$ and $x *u_{\text{T}}^{\text{PT}}\left(x,\bm{b}_{\perp}^2\right)$ in Fig. \ref{ptqxb}.

\subsection{3D regularization}
The three momentum cutoff is after integrating out the $p_0$, a cutoff imposed on all the left integrals $\mathbf{p}^2<\Lambda^2$, the gap equation in Eq. (\ref{nocutoffgap}) becomes
\begin{align}\label{3dcutoffgap}
M=m-\frac{12G_{\pi}}{\pi^2}\int_0^{\Lambda_{\text{3D}}} \mathrm{d}k\frac{k^2 M}{(k^2+M^2)^{1/2}},
\end{align}
the pion decay constant becomes
\begin{align}\label{3dfpi}
f_{\pi}^{\text{3D}}=\frac{N_c\sqrt{Z_{\pi }}M}{2\pi ^2}\int_0^1 \mathrm{d}x\int_0^{\Lambda_{\text{3D}}}\mathrm{d}k\frac{k^2}{(k^2+\sigma_1)^{3/2}},
\end{align}
the vector bubble diagram $\Pi_{\text{VV}}$ is defined as
\begin{align}\label{3dvvbu}
\Pi_{\text{VV}}^{\text{3D}}(t)=-\frac{6 }{\pi ^2} \int_0^1\mathrm{d}x\int_0^{\Lambda_{\text{3D}}} \mathrm{d}k\,  \frac{x (1-x) tk^2}{(k^2+\sigma_2)^{3/2}}.
\end{align}
The parameters used in 3D momentum cutoff scheme are listed in Table \ref{tb3d}.
%
%
\begin{center}
\begin{table*}
\caption{Parameter set used for 3D regularization. The dressed quark mass and regularization parameters are in units of GeV, while coupling constant are in units of GeV$^{-2}$.}\label{tb3d}
\begin{tabular}{p{1.0cm} p{1.0cm}p{1.0cm} p{1.0cm}p{1.0cm}p{1.0cm}p{1.0cm}p{1.0cm}p{1.0cm}p{1.4cm}}
\hline\hline
$\Lambda _{\text{3D}}$&$M$&$G_{\pi}$&$G_{\rho}$&$G_{\omega}$&$m_{\pi}$&$f_{\pi}$&$m$&$Z_{\pi}$&$\langle\bar{u}u\rangle$\\
\hline
0.595&0.396&6.630&7.968&6.898&0.14&0.93&0.006&17.497&--\,(0.245)$^3$\\
\hline\hline
\end{tabular}
\end{table*}
\end{center}

\begin{figure*}
\centering
\includegraphics[width=0.47\textwidth]{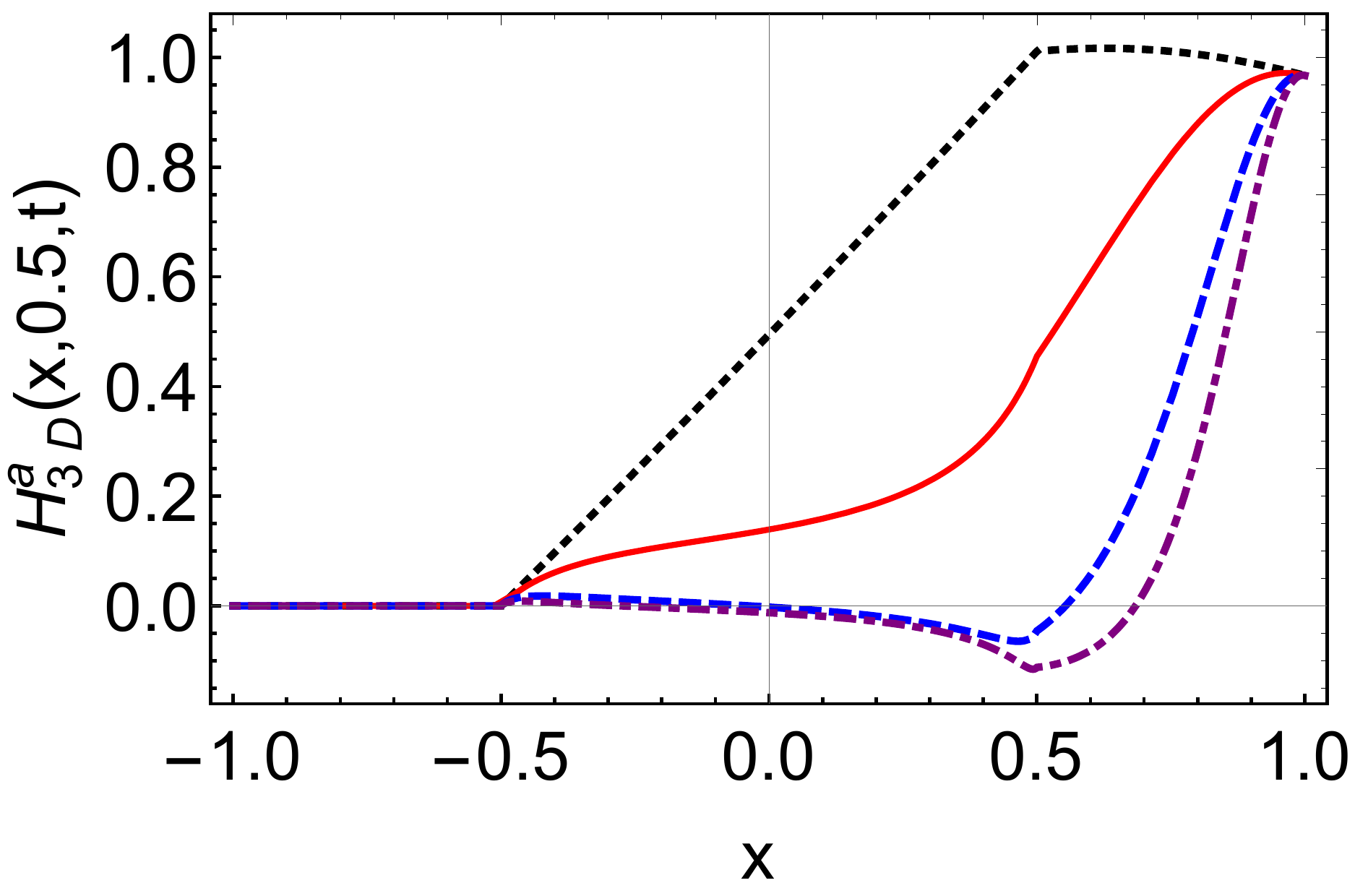}
\qquad
\includegraphics[width=0.47\textwidth]{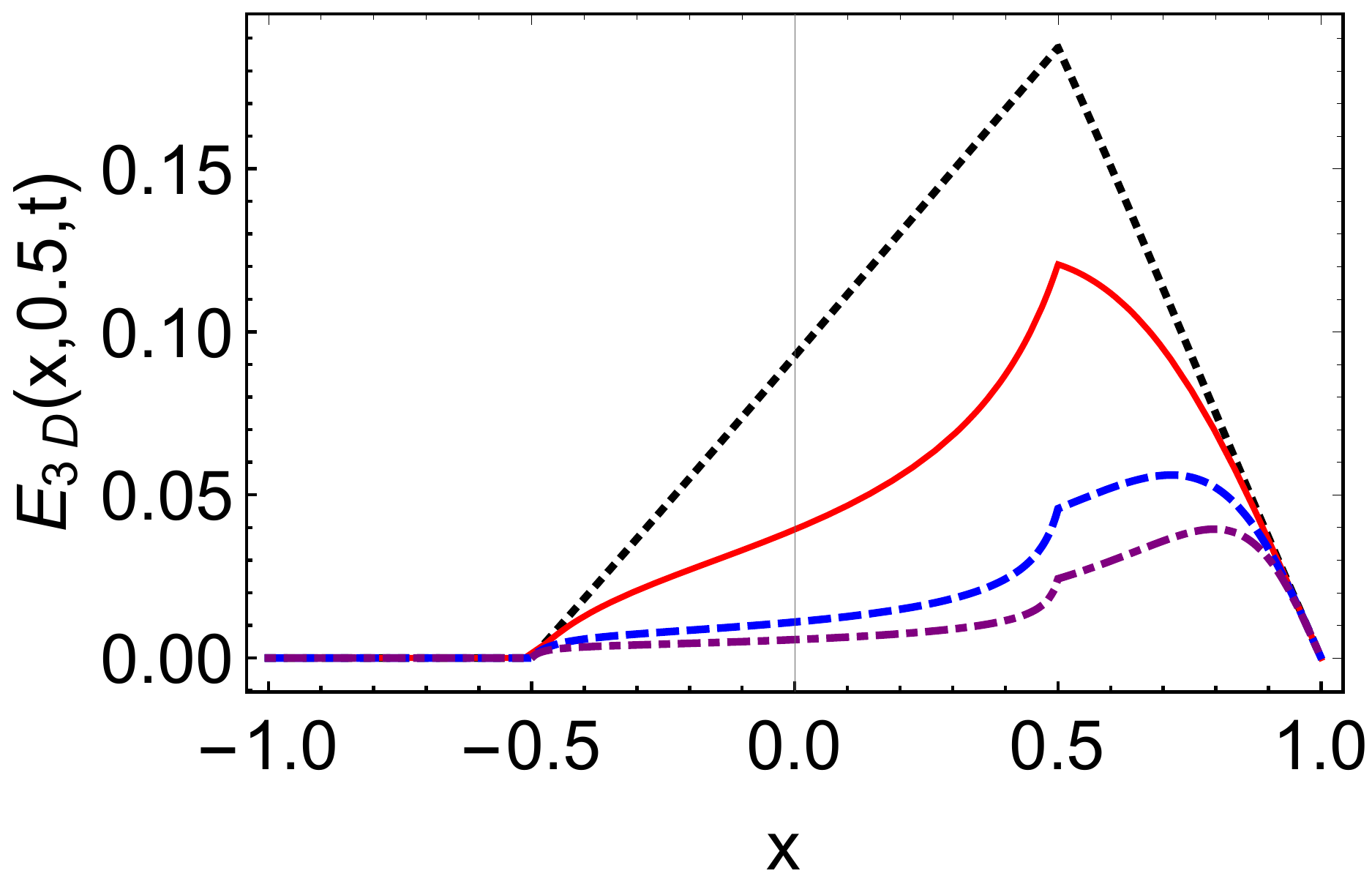}
\caption{Pion GPDs (left panel: vector GPD, right panel: tensor GPD) at $\xi=0.5$ with different $t$ using 3D regularization: black dotted line -- $t=0$ GeV$^2$, red line -- $t=-1$ GeV$^2$, blue dashed line -- $t=-5$ GeV$^2$, purple dotdashed line -- $t=-10$ GeV$^2$. }\label{3degpd}
\end{figure*}

\begin{figure*}
\centering
\includegraphics[width=0.47\textwidth]{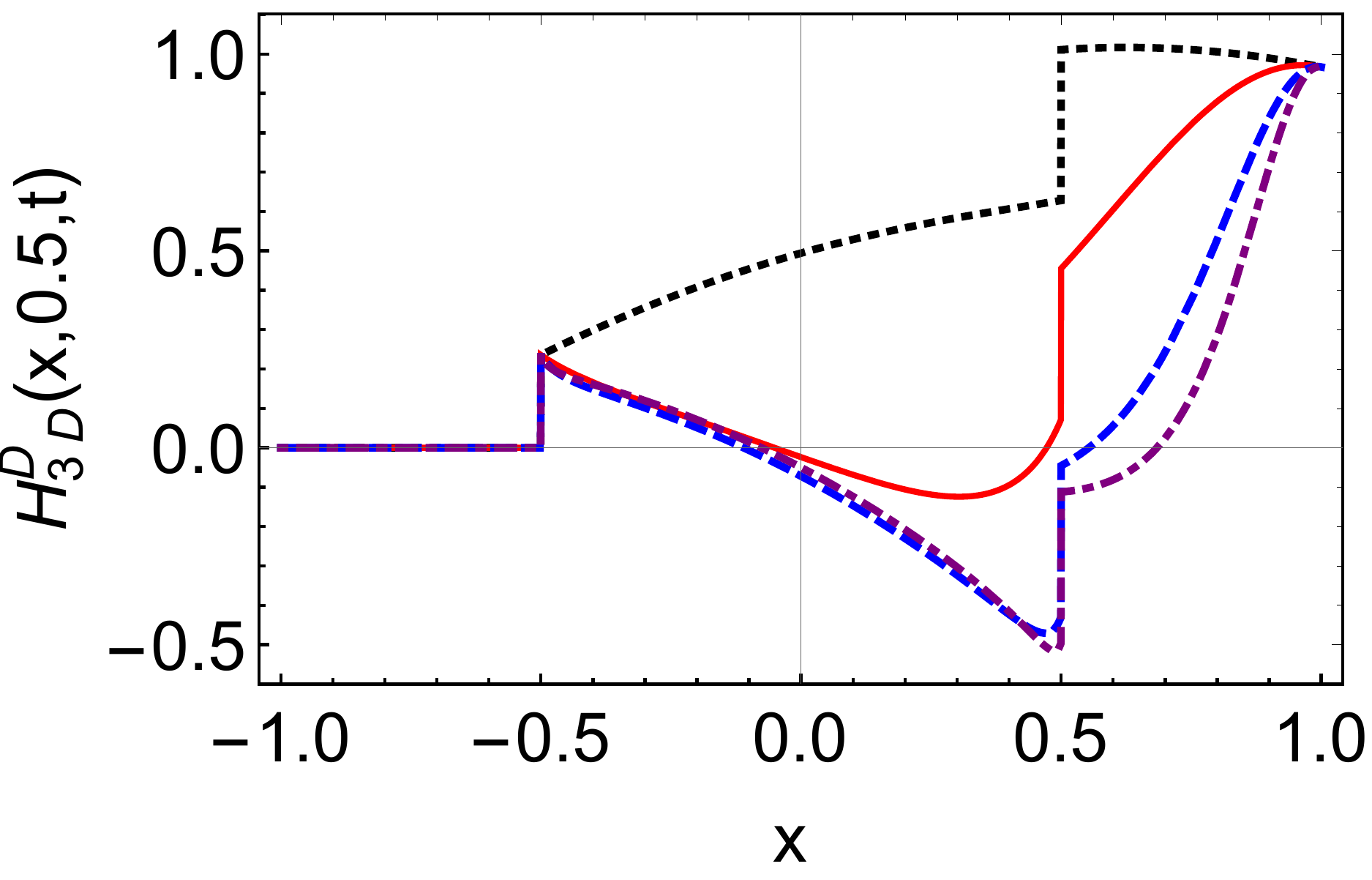}
\qquad
\includegraphics[width=0.47\textwidth]{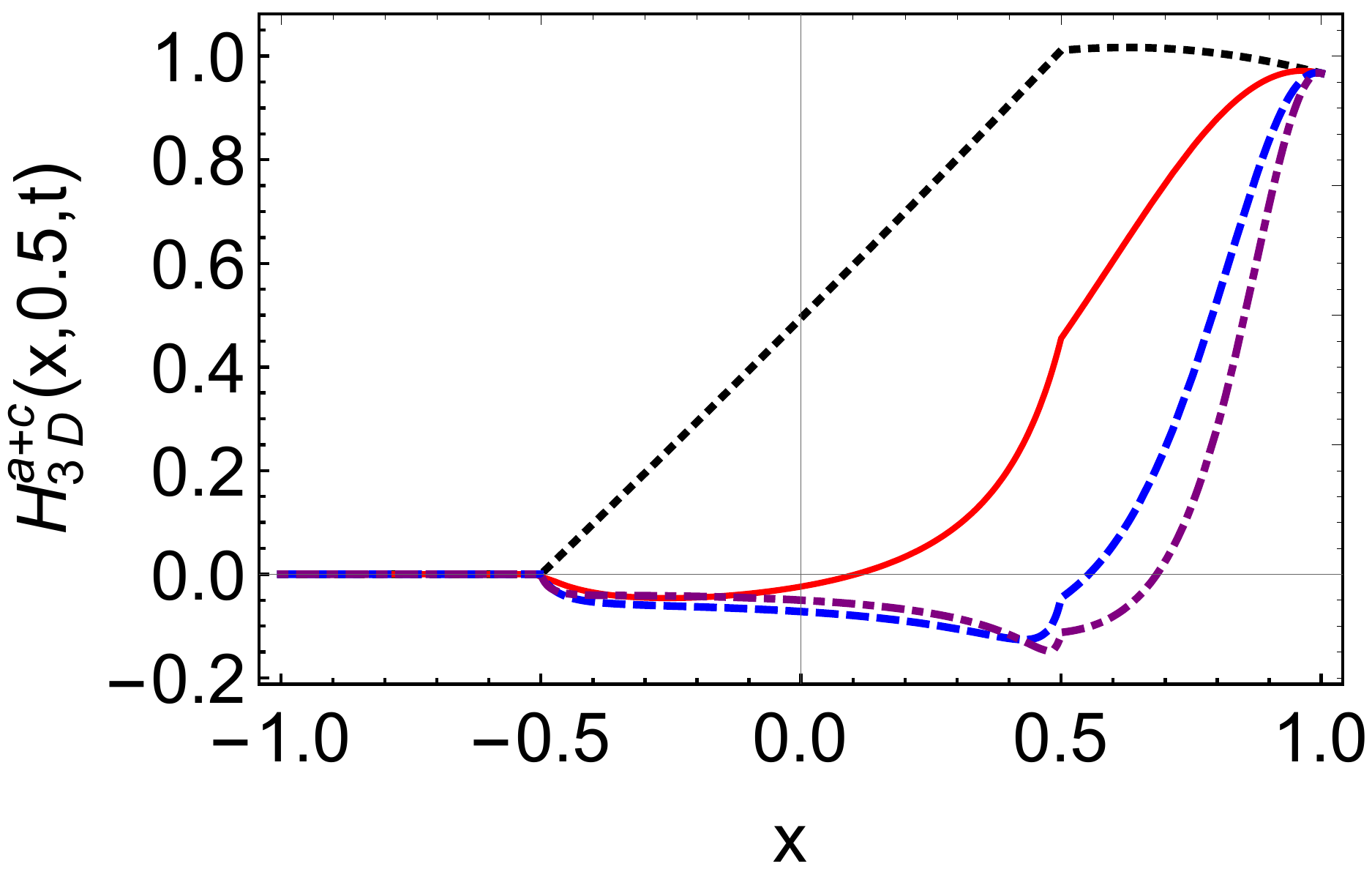}
\caption{Pion GPDs (left panel: $H_{\text{3D}}^D$, right panel: $H_{\text{3D}}^a+H_{\text{3D}}^c$) at $\xi=0.5$ with different $t$ using PT regularization: black dotted line -- $t=0$ GeV$^2$, red line -- $t=-1$ GeV$^2$, blue dashed line -- $t=-5$ GeV$^2$, purple dotdashed line -- $t=-10$ GeV$^2$. }\label{3dgpd1}
\end{figure*}

GPDs in 3D momentum cutoff scheme
\begin{align}\label{3dagpdf}
&\quad H_{\text{3D}}^a\left(x,\xi,t\right)\nonumber\\
&=\frac{N_cZ_{\pi }}{4\pi ^2} \left[\int_0^{\Lambda_{\text{3D}}}\mathrm{d}k\frac{\theta_{\bar{\xi} 1} k^2}{(k^2+\sigma_3)^{3/2}}+ \int_0^{\Lambda_{\text{3D}}}\mathrm{d}k \frac{\theta_{\xi 1} k^2}{(k^2+\sigma_4^{3/2})}\right]\nonumber\\
&+\frac{N_cZ_{\pi }}{4\pi ^2}\int_0^{\Lambda_{\text{3D}}}\mathrm{d}k \frac{\theta_{\bar{\xi} \xi} }{\xi}\frac{x k^2}{(k^2+\sigma_5)^{3/2}}\nonumber\\
&+\frac{N_c Z_{\pi } }{16\pi ^2}\int_0^{\Lambda_{\text{3D}}}\mathrm{d}k\int_0^1 \mathrm{d}\alpha \frac{\theta_{\alpha \xi}}{\xi} \frac{3k^2(2 x m_{\pi}^2+(1-x)t)}{(k^2+\sigma_6)^{5/2}} ,
\end{align}
\begin{align}\label{3dagtpdf}
E_{\text{3D}}\left(x,\xi,t\right)=\frac{N_c Z_{\pi } }{8\pi ^2}\int_0^{\Lambda_{\text{3D}}}\mathrm{d}k\int_0^1 \mathrm{d}\alpha \frac{\theta_{\alpha \xi}}{\xi} \frac{3k^2m_{\pi}M}{(k^2+\sigma_6)^{5/2}},
\end{align}
\begin{align}\label{3dagtpdf}
H_{\text{3D}}^b\left(x,\xi,t\right)=-\frac{N_cZ_{\pi }}{4\pi ^2} \int_0^{\Lambda_{\text{3D}}}\mathrm{d}k \frac{\theta_{\bar{\xi} \xi} }{\xi}\frac{2Mx k^2}{(k^2+\sigma_5)^{3/2}},
\end{align}
\begin{align}\label{3dagtpdf}
H_{\text{3D}}^c\left(x,\xi,t\right)&=F_{1\rho}^{\text{3D}}(t)A_{1,0}^{\text{3D}}(t)\frac{N_cG_{\rho }}{4\pi ^2} \int_0^{\Lambda_{\text{3D}}}\mathrm{d}k \frac{\theta_{\bar{\xi} \xi} }{\xi}\nonumber\\
&\times t\left(1-\frac{x^2}{\xi^2}\right)\frac{2k^2}{(k^2+\sigma_5)^{3/2}},
\end{align}
where the $\theta$ functions are the same as Eqs. (\ref{region1}). Pion GPDs in 3D regularization scheme are plotted in Figs. \ref{3degpd} and \ref{3dgpd1}.


\subsubsection{Forward limit}
\begin{figure}
\centering
\includegraphics[width=0.47\textwidth]{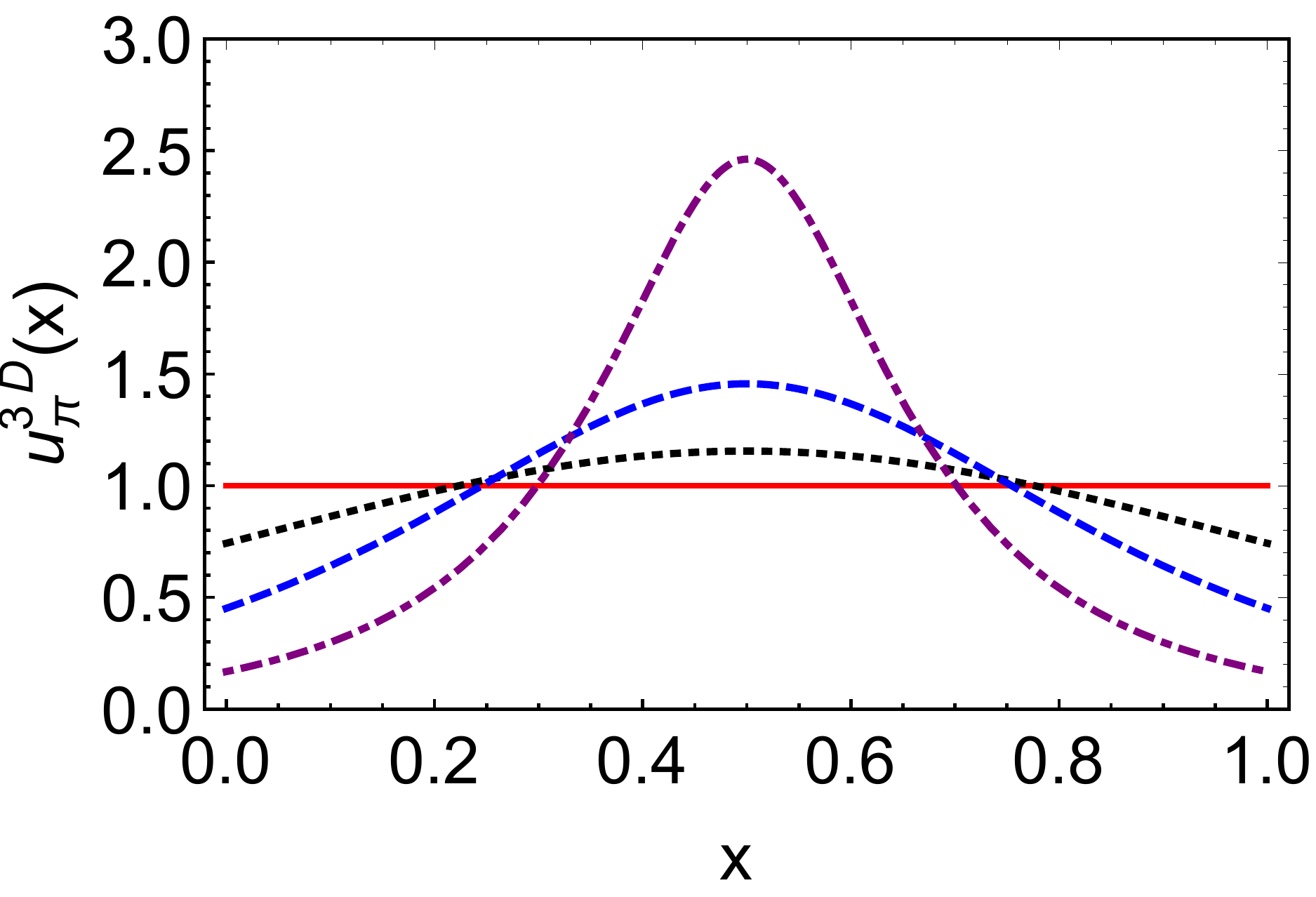}
\caption{Pion $u$ quark PDF with different values of $m_{\pi}$ using 3D regularization scheme: red line -- $m_{\pi}=0$ GeV, black dotted line -- $m_{\pi}=0.4$ GeV, blue dashed line -- $m_{\pi}=0.6$ GeV, purple dotdashed line -- $m_{\pi}=0.75$ GeV. }\label{3dpdf}
\end{figure}

In the forward limit
\begin{align}\label{3dhpdf}
u_{\text{3D}}(x)&=\frac{3Z_{\pi }}{2\pi ^2}\int_0^{\Lambda_{\text{3D}}}\mathrm{d}k \frac{ k^2}{(k^2+\sigma_1)^{3/2}} \nonumber\\
&+\frac{3Z_{\pi }}{4\pi ^2}  \int_0^{\Lambda_{\text{3D}}}\mathrm{d}k \frac{ 3k^2 x(1-x) m_{\pi}^2 }{(k^2+\sigma_1)^{5/2}} ,
\end{align}
we plot the diagrams of $u$ quark PDFs of pion with different values of $m_{\pi}$ in Fig. \ref{3dpdf}.

\subsubsection{Form Factors}
\begin{align}\label{3da1}
A_{1,0}^{\text{3D}}(t)&=\frac{N_cZ_{\pi } }{2\pi^2}\int_0^1 \mathrm{d}x\int_0^{\Lambda_{\text{3D}}}\mathrm{d}k \frac{ k^2}{(k^2+\sigma_1)^{3/2}}\nonumber\\
&+\frac{N_cZ_{\pi }}{8\pi^2} \int_0^1 \mathrm{d}x \int_0^{1-x}\mathrm{d}y \int_0^{\Lambda_{\text{3D}}}\mathrm{d}k  \nonumber\\
&\times  \frac{3k^2(2m_{\pi }^2-(x+y)(2m_{\pi }^2-t))}{(k^2+\sigma_7)^{5/2}},
\end{align}
\begin{align}\label{3db1}
B_{1,0}^{\text{3D}}(t)&=\frac{ N_c Z_{\pi } }{8\pi ^2}  \int _0^1\mathrm{d}x \int _0^{1-x}\mathrm{d}y \int_0^{\Lambda_{\text{3D}}}\mathrm{d}k\frac{3k^2m_{\pi}M}{(k^2+\sigma_7)^{5/2}},
\end{align}
\begin{align}\label{3da2}
A_{2,0}^{\text{3D}}(t)&=\frac{N_cZ_{\pi }}{4\pi ^2}\int_0^1 \mathrm{d}x\int_0^{\Lambda_{\text{3D}}}\mathrm{d}k\frac{ k^2}{(k^2+\sigma_1)^{3/2}}\nonumber\\
&+\frac{N_c Z_{\pi }}{8\pi ^2}  \int_0^1 \mathrm{d}x \int_0^{1-x} \mathrm{d}y\int_0^{\Lambda_{\text{3D}}}\mathrm{d}k (1-x-y)\nonumber\\
&\times \frac{3k^2(2 m_{\pi }^2(1-x-y)+t  (x+y))}{(k^2+\sigma_7)^{5/2}},
\end{align}
\begin{align}\label{a3da22}
A_{2,2}^{\text{3D}}(t)&=-\frac{N_cZ_{\pi }}{2\pi ^2}\int_0^1 \mathrm{d}x \int_0^{\Lambda_{\text{3D}}}\mathrm{d}k\,(1-x)\frac{ k^2}{(k^2+\sigma_1)^{3/2}}\nonumber\\
&-\frac{N_c Z_{\pi } }{\pi ^2}  \int_0^1 \mathrm{d}x \int_0^{\Lambda_{\text{3D}}}\mathrm{d}k\,x (1-2x) \frac{ k^2}{(k^2+\sigma_2)^{3/2}}\nonumber\\
&-\frac{N_c Z_{\pi } }{2\pi^2} \int_0^1  \mathrm{d}x \int_0^{\Lambda_{\text{3D}}}\mathrm{d}k \,\frac{ (1-x ) \left(2 m_{\pi }^2-t\right)k^2}{t(k^2+\sigma_1)^{3/2}}\nonumber\\
&+\frac{N_c Z_{\pi }}{8\pi^2}\int_0^1 \mathrm{d}x \int_0^{1-x} \mathrm{d}y \int_0^{\Lambda_{\text{3D}}}\mathrm{d}k\,\frac{3k^2\left(2 m_{\pi }^2-t\right)}{t(k^2+\sigma_7)^{5/2} },
\end{align}
\begin{eqnarray}\label{3db2}
B_{2,0}^{\text{3D}}(t)&=&\frac{N_c Z_{\pi}  }{8\pi ^2} \int_0^1 \mathrm{d}x \int_0^{1-x} \mathrm{d}y \int_0^{\Lambda_{\text{3D}}}\mathrm{d}k \nonumber\\
&\times & \frac{3k^2m_{\pi}M (1-x-y)}{(k^2+\sigma_7)^{5/2}},
\end{eqnarray}
\begin{align}
A_{2,2}^{b,\text{3D}}(t)=-\frac{N_c Z_{\pi } }{\pi ^2}  \int_0^1 \mathrm{d}x  \int_0^{\Lambda_{\text{3D}}}\mathrm{d}k\,  \frac{ 2M x (1-2x)k^2}{(k^2+\sigma_2)^{3/2}}.
\end{align}

\subsubsection{Impact parameter space PDFs}
\begin{figure}
\centering
\includegraphics[width=0.47\textwidth]{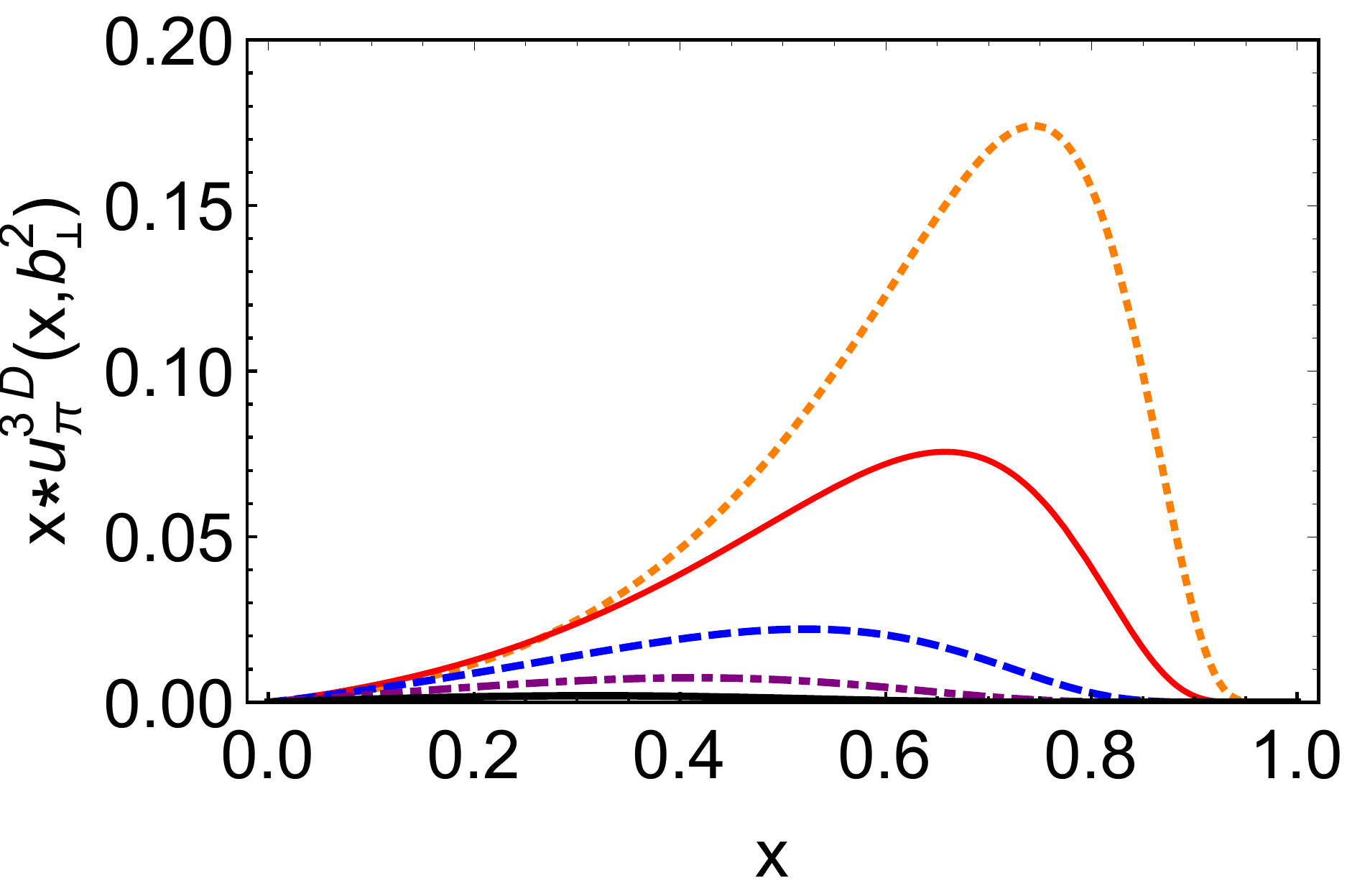}
\qquad
\includegraphics[width=0.47\textwidth]{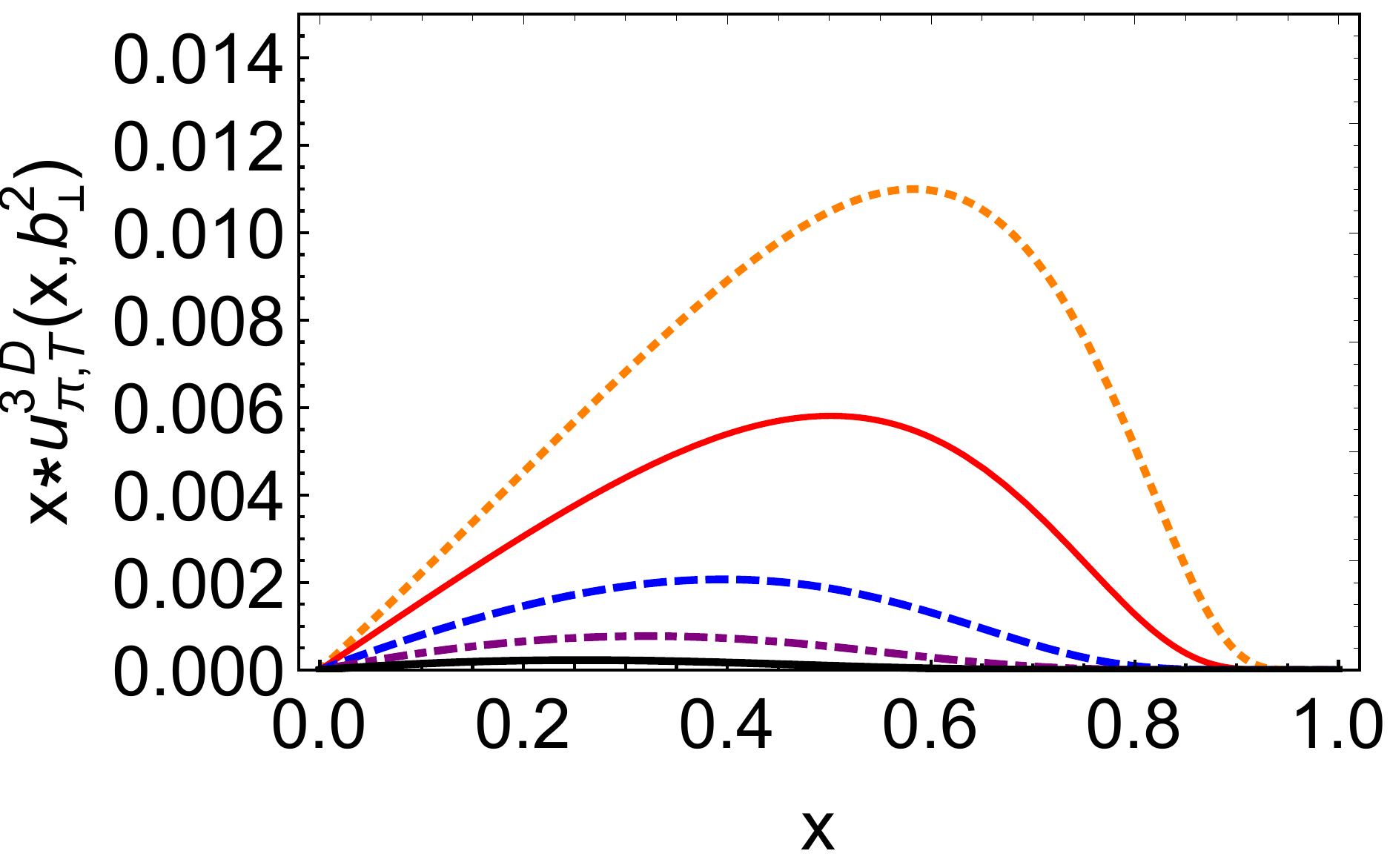}
\caption{Impact parameter space PDFs using 3D regularization scheme: left panel -- $x*u\left(x,\bm{b}_{\perp}^2\right)$, the $\delta^2(\bm{b}_{\perp})$ component first line of Eq. (\ref{3dpspdf}) --- is suppressed in the image, and right panel -- $x*u_T\left(x,\bm{b}_{\perp}^2\right)$ both panels with $\bm{b}_{\perp}^2=0.5$ GeV$^{-2}$ --- orange dotted curve, $\bm{b}_{\perp}^2=1$ GeV$^{-2}$ --- red solid curve, $\bm{b}_{\perp}^2=2.5$ GeV$^{-2}$ --- blue dashed curve, $\bm{b}_{\perp}^2=5$ GeV$^{-2}$ --- purple dot-dashed curve, $\bm{b}_{\perp}^2=10$ GeV$^{-2}$ --- thick black solid curve.}\label{3dqxb}
\end{figure}

In the 3D regularization scheme
\begin{align}\label{aG91}
&\quad H_{\text{3D}}\left(x,0,-\bm{q}_{\perp}^2\right)\nonumber\\
&=\frac{N_cZ_{\pi }}{4\pi ^2}  \int_0^{\Lambda_{\text{3D}}}\mathrm{d}k  \frac{2 k^2}{(k^2+\sigma_1)^{3/2}} \nonumber\\
&+\frac{N_cZ_{\pi }}{8\pi ^2}\int_0^{1-x} \mathrm{d}\alpha \int_0^{\Lambda_{\text{3D}}}\mathrm{d}k \frac{3k^2\left(2 x m_{\pi}^2+ x \bm{q}_{\perp}^2- \bm{q}_{\perp}^2\right)}{(k^2+\sigma_8)^{5/2}},
\end{align}
\begin{align}\label{aG911}
E_{\text{3D}}\left(x,0,-\bm{q}_{\perp}^2\right)=\frac{N_cZ_{\pi }}{4\pi ^2}\int_0^{1-x} \mathrm{d}\alpha \int_0^{\Lambda_{\text{3D}}}\mathrm{d}k \frac{3k^2m_{\pi} M }{(k^2+\sigma_8)^{5/2}},
\end{align}
%
\begin{widetext}
\begin{align}\label{3dpspdf}
&\quad u^{\text{3D}}\left(x,\bm{b}_{\perp}^2\right)\nonumber\\
&=\frac{N_cZ_{\pi }}{2\pi ^2} \int \frac{\mathrm{d}^2\bm{q}_{\perp}}{(2 \pi )^2}\int_0^{\Lambda_{\text{3D}}}e^{-i\bm{b}_{\perp}\cdot \bm{q}_{\perp}}\mathrm{d}k\frac{ k^2}{(k^2+\sigma_1)^{3/2}} \nonumber\\
&+\frac{N_cZ_{\pi }}{8\pi ^3}\int_0^{1-x} \mathrm{d}\alpha  \int_0^{\Lambda_{\text{3D}}}\mathrm{d}k \,e^{ -\frac{\sqrt{\bm{b}_{\perp}^2(k^2+\sigma_1)} }{\sqrt{\alpha  (1-\alpha -x)}} }\frac{k^2 \sqrt{\bm{b}_{\perp}^2} \left((1-x) \left(k^2+M^2\right)-xm_{\pi }^2  \left(2 \alpha ^2-2 \alpha +x^2+2 (\alpha -1) x+1\right)\right) }{2 \alpha ^{5/2} (1-\alpha -x)^{5/2} \left(k^2+\sigma_1\right)}\nonumber\\
&-\frac{N_cZ_{\pi }}{8\pi ^3}\int_0^{1-x} \mathrm{d}\alpha  \int_0^{\Lambda_{\text{3D}}}\mathrm{d}k\,e^{ -\frac{\sqrt{\bm{b}_{\perp}^2(k^2+\sigma_1)} }{\sqrt{\alpha  (1-\alpha -x)}} }\frac{k^2 \left((1-x) \left(k^2+M^2\right)-x m_{\pi }^2  \left(1-\alpha ^2+\alpha +x^2-(\alpha +2) x\right)\right) }{\alpha ^2 (1-\alpha -x)^2 \left(k^2+\sigma_1\right)^{3/2}},
\end{align}
\begin{align}\label{3dtpspdf}
u_{\text{T}}^{\text{3D}}\left(x,\bm{b}_{\perp}^2\right)
&=\frac{N_cZ_{\pi }}{8\pi^3}\int_0^{1-x}  \mathrm{d}\alpha \int_0^{\Lambda_{\text{3D}}}\mathrm{d}k\, e^{ -\frac{\sqrt{\bm{b}_{\perp}^2(k^2+\sigma_1)} }{\sqrt{\alpha  (1-\alpha -x)}}}\frac{k^2 m_{\pi } M \left(\sqrt{\bm{b}_{\perp}^2} \sqrt{k^2+\sigma_1}+\sqrt{\alpha  (1-\alpha -x)}\right) }{ \alpha^{3/2}  (1-\alpha -x)^{3/2}  \left(k^2+\sigma_1\right)^{3/2}} ,
\end{align}
\end{widetext}
for $u\left(x,\bm{b}_{\perp}^2\right)$, when integrating $\bm{b}_{\perp}$ one can get PDF $u(x)$ in Eq. (\ref{3dhpdf}). We plot the diagrams of $x *u\left(x,\bm{b}_{\perp}^2\right)$ and $x *u_{\text{T}}\left(x,\bm{b}_{\perp}^2\right)$ in Fig. \ref{3dqxb}.

\subsection{4D regularization}
The 4 momentum cut off in Euclidean space, $p_E^2=\mathbf{p}^2+p_4^2<\Lambda^2$, $p_0=ip_4$, Wick rotate the integrating in the gap equation (\ref{nocutoffgap}), we obtain

\begin{align}\label{4dgap}
M=m-\frac{6 G_{\pi}}{\pi^2}\int_0^{\Lambda_{\text{4D}}} \mathrm{d}k_E\frac{k_E^3 M}{(k_E^2+M^2)},
\end{align}
the pion decay constant becomes
\begin{align}\label{4dfpi}
f_{\pi}^{\text{4D}}=\frac{N_c\sqrt{Z_{\pi }}M}{2\pi ^2}\int_0^1 \mathrm{d}x\int_0^{\Lambda_{\text{4D}}}\mathrm{d}k\frac{k^3}{(k^2+\sigma_1)^2},
\end{align}
the vector bubble diagram $\Pi_{\text{VV}}$ is defined as
\begin{align}\label{4dvvbu}
\Pi_{\text{VV}}^{\text{4D}}(t)=-\frac{6 }{\pi ^2} \int_0^1\mathrm{d}x\int_0^{\Lambda_{\text{4D}}} \mathrm{d}k\,  \frac{x (1-x) tk^3}{(k^2+\sigma_2)^2}.
\end{align}
The parameters used in 3D momentum cutoff scheme are listed in Table \ref{tb4d}.
%
%
\begin{center}
\begin{table*}
\caption{Parameter set used for 4D regularization. The dressed quark mass and regularization parameters are in units of GeV, while coupling constant are in units of GeV$^{-2}$.}\label{tb4d}
\begin{tabular}{p{1.0cm} p{1.0cm} p{1.0cm}p{1.0cm}p{1.0cm}p{1.0cm}p{1.0cm}p{1.0cm}p{1.0cm}p{1.4cm}}
\hline\hline
$\Lambda _{\text{4D}}$&$M$&$G_{\pi}$&$G_{\rho}$&$G_{\omega}$&$m_{\pi}$&$f_{\pi}$&$m$&$Z_{\pi}$&$\langle\bar{u}u\rangle$\\
\hline
0.74&0.395&9.60&7.711&6.621&0.14&0.093&0.008&17.452&--(0.216)$^3$\\
\hline\hline
\end{tabular}
\end{table*}
\end{center}

\begin{figure*}
\centering
\includegraphics[width=0.47\textwidth]{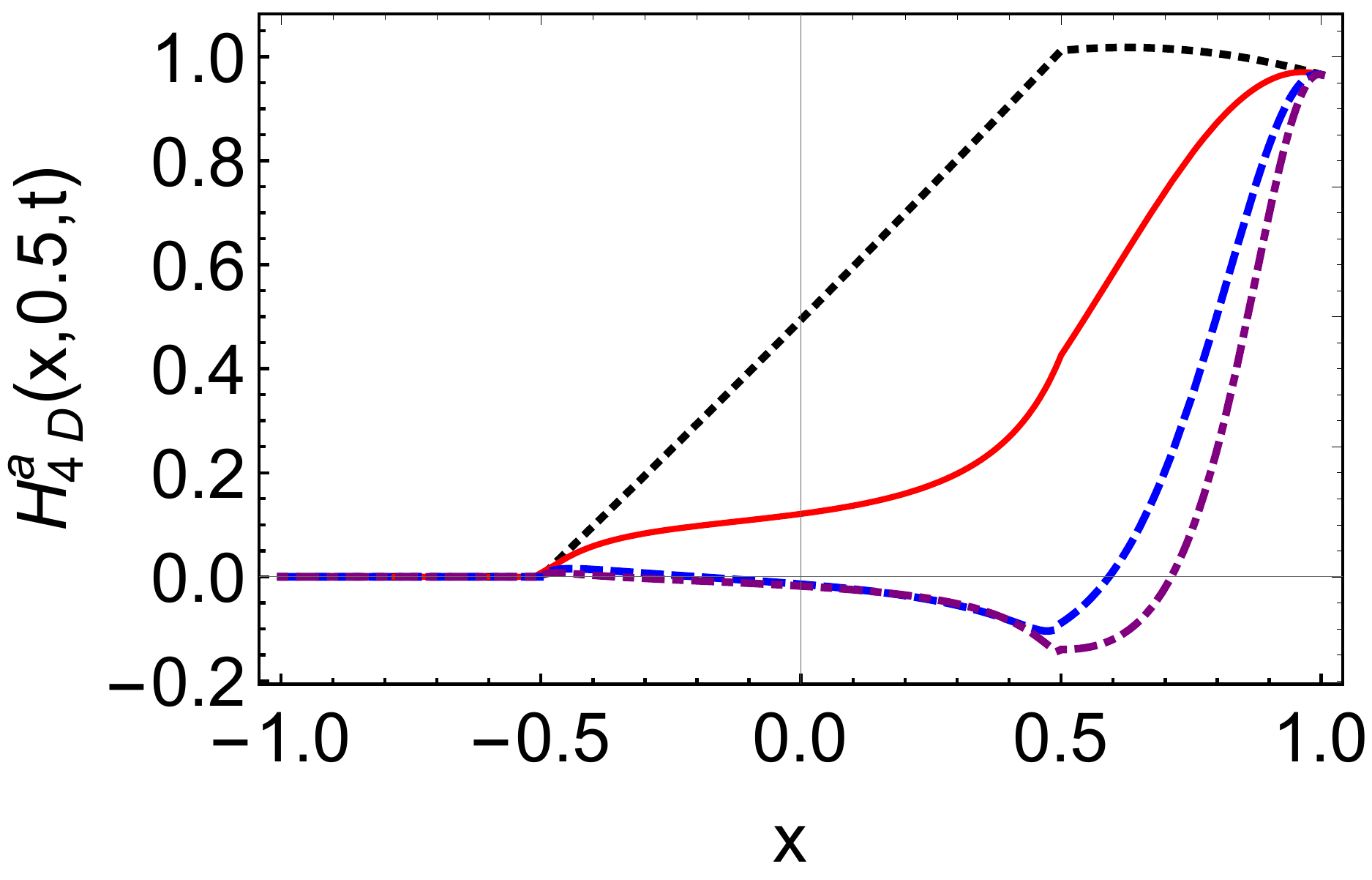}
\qquad
\includegraphics[width=0.47\textwidth]{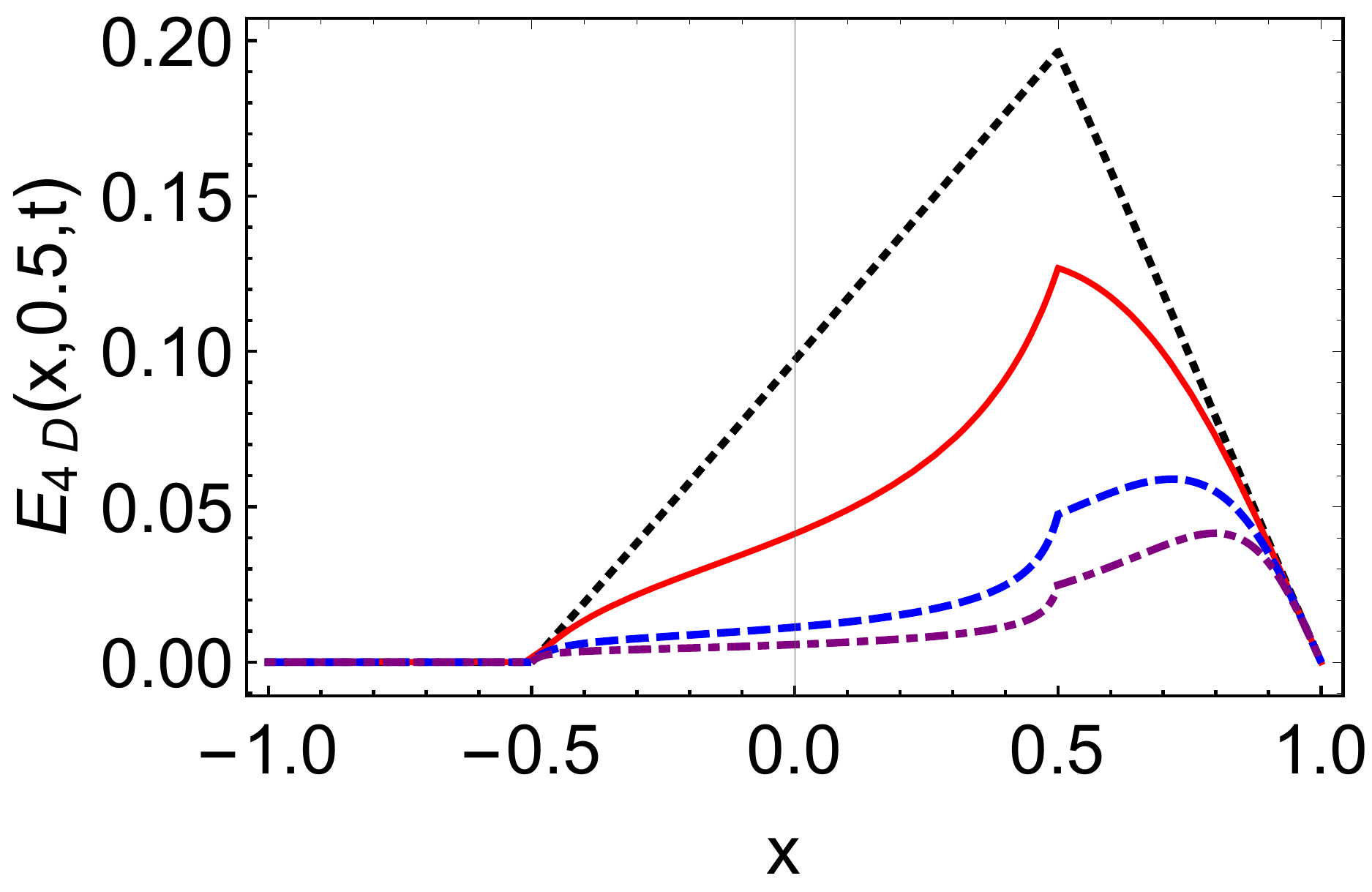}
\caption{Pion GPDs (left panel: vector GPD, right panel: tensor GPD) at $\xi=0.5$ with different $t$ using 4D regularization: black dotted line -- $t=0$ GeV$^2$, red line -- $t=-1$ GeV$^2$, blue dashed line -- $t=-5$ GeV$^2$, purple dotdashed line -- $t=-10$ GeV$^2$. }\label{4degpd}
\end{figure*}

\begin{figure*}
\centering
\includegraphics[width=0.47\textwidth]{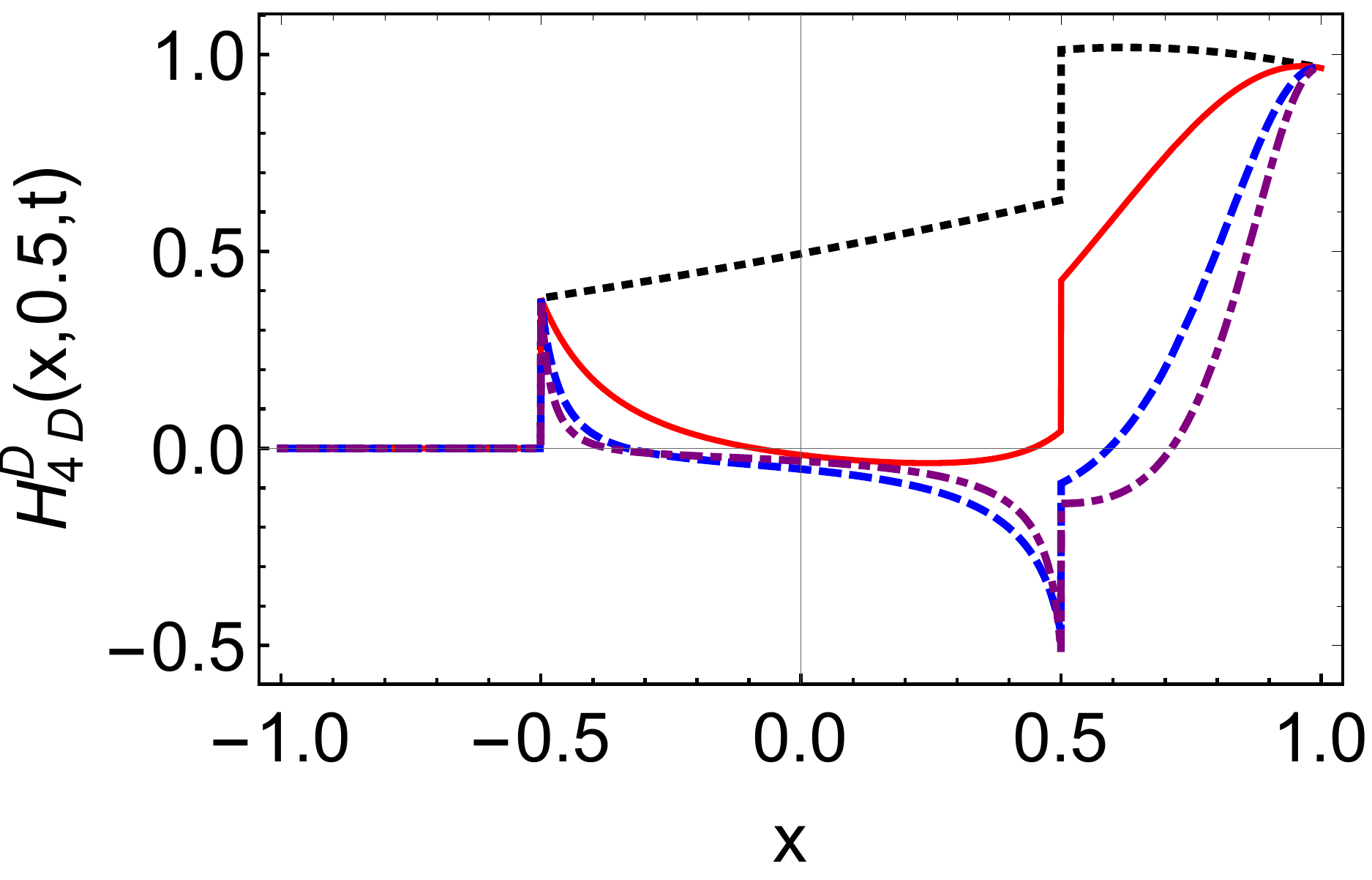}
\qquad
\includegraphics[width=0.47\textwidth]{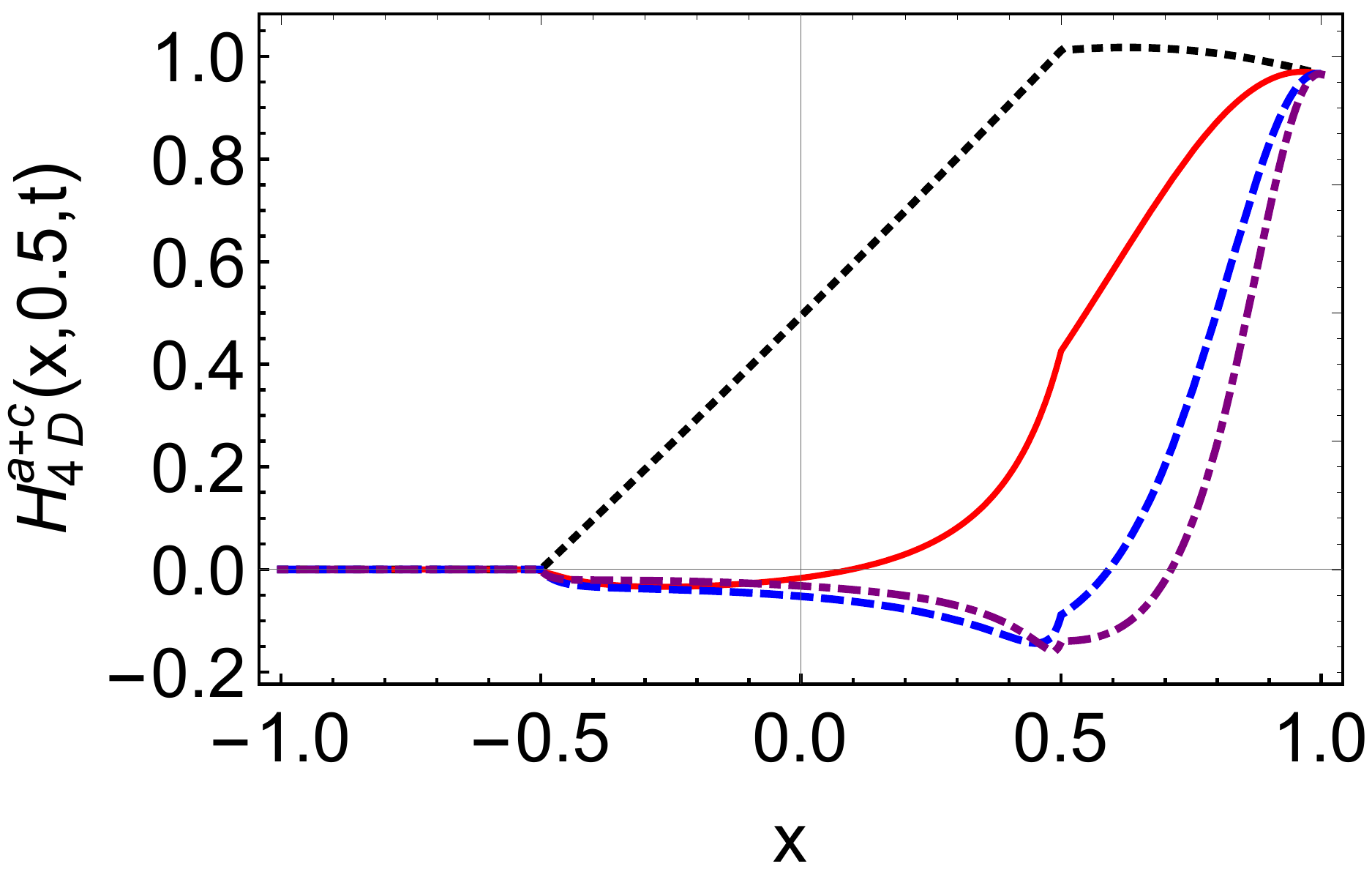}
\caption{Pion GPDs (left panel: $H_{\text{4D}}^D$, right panel: $H_{\text{4D}}^a+H_{\text{4D}}^c$) at $\xi=0.5$ with different $t$ using PT regularization: black dotted line -- $t=0$ GeV$^2$, red line -- $t=-1$ GeV$^2$, blue dashed line -- $t=-5$ GeV$^2$, purple dotdashed line -- $t=-10$ GeV$^2$. }\label{4dgpd1}
\end{figure*}

GPDs in 4D momentum cutoff scheme
\begin{align}\label{4dagpdf}
&\quad H_{\text{4D}}^a\left(x,\xi,t\right)\nonumber\\
&=\frac{N_cZ_{\pi }}{8\pi ^2} \left[\int_0^{\Lambda_{\text{4D}}}\mathrm{d}k\frac{2\theta_{\bar{\xi} 1} k^3}{(k^2+\sigma_3)^2}+ \int_0^{\Lambda_{\text{4D}}}\mathrm{d}k \frac{2\theta_{\xi 1} k^3}{(k^2+\sigma_4^2)}\right]\nonumber\\
&+\frac{N_cZ_{\pi }}{8\pi ^2}\int_0^{\Lambda_{\text{4D}}}\mathrm{d}k \frac{\theta_{\bar{\xi} \xi} }{\xi}\frac{2x k^3}{(k^2+\sigma_5)^2}\nonumber\\
&+\frac{N_c Z_{\pi } }{16\pi ^2}\int_0^{\Lambda_{\text{4D}}}\mathrm{d}k\int_0^1 \mathrm{d}\alpha \frac{\theta_{\alpha \xi}}{\xi} \frac{4k^3(2 x m_{\pi}^2+(1-x)t)}{(k^2+\sigma_6)^3} ,
\end{align}
\begin{align}\label{4dagtpdf}
E_{\text{4D}}\left(x,\xi,t\right)=\frac{N_c Z_{\pi } }{8\pi ^2}\int_0^{\Lambda_{\text{4D}}}\mathrm{d}k\int_0^1 \mathrm{d}\alpha \frac{\theta_{\alpha \xi}}{\xi} \frac{4k^3m_{\pi}M}{(k^2+\sigma_6)^3},
\end{align}
\begin{align}\label{3dagtpdf}
H_{\text{4D}}^b\left(x,\xi,t\right)=-\frac{N_cZ_{\pi }}{4\pi ^2} \int_0^{\Lambda_{\text{4D}}}\mathrm{d}k \frac{\theta_{\bar{\xi} \xi} }{\xi}\frac{2Mx k^3}{(k^2+\sigma_5)^2},
\end{align}
\begin{align}\label{3dagtpdf}
H_{\text{4D}}^c\left(x,\xi,t\right)&=F_{1\rho}^{\text{4D}}(t)A_{1,0}^{\text{4D}}(t)\frac{N_cG_{\rho }}{4\pi ^2} \int_0^{\Lambda_{\text{4D}}}\mathrm{d}k \frac{\theta_{\bar{\xi} \xi} }{\xi}\nonumber\\
&\times t\left(1-\frac{x^2}{\xi^2}\right)\frac{2k^3}{(k^2+\sigma_5)^2},
\end{align}
where the $\theta$ functions are defined in Eqs. (\ref{region1}). Pion GPDs in 4D regularization scheme are plotted in Figs. \ref{4degpd} and \ref{4dgpd1}.


\subsubsection{Forward limit}
\begin{figure}
\centering
\includegraphics[width=0.47\textwidth]{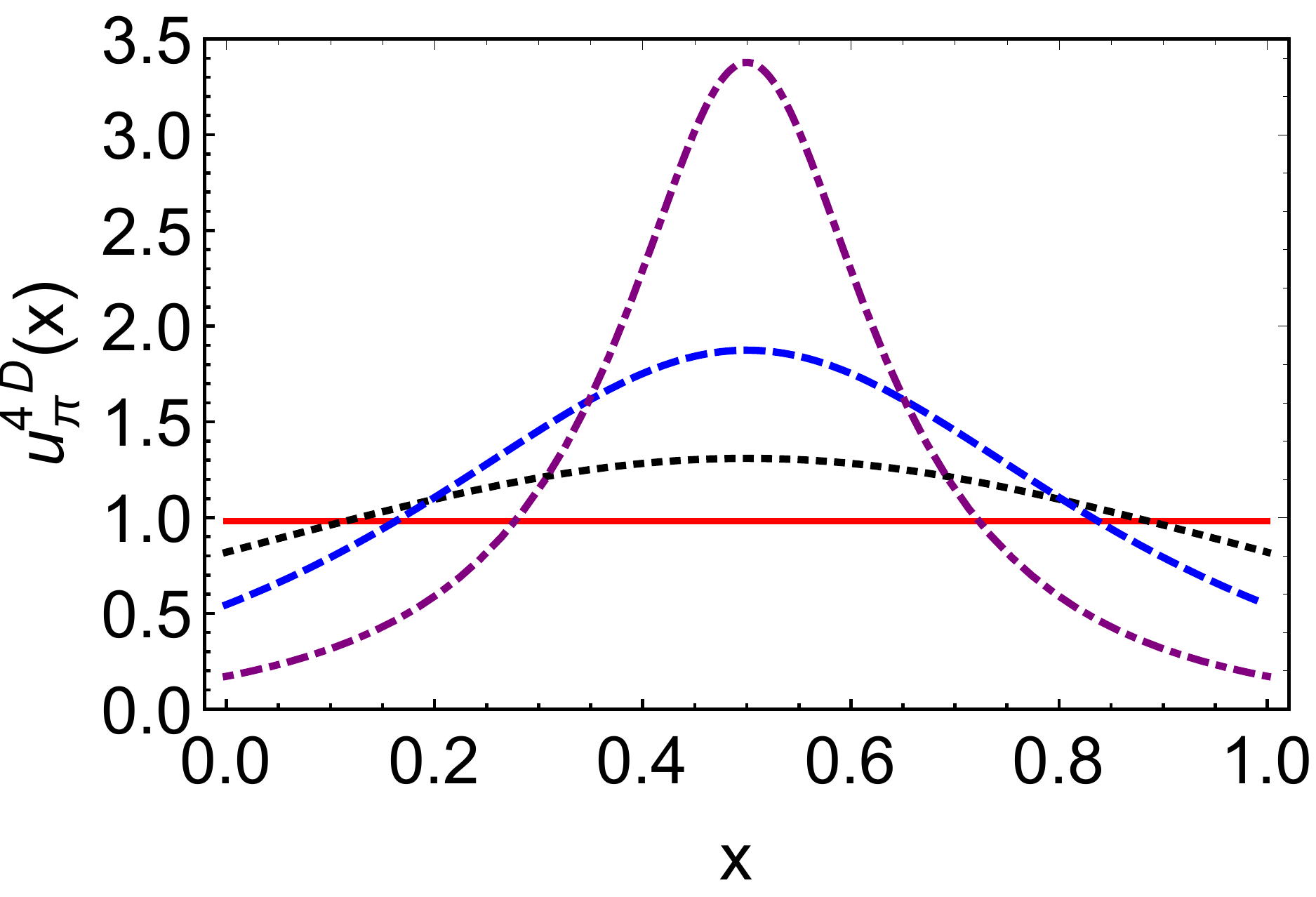}
\caption{Pion $u$ quark PDF with different values of $m_{\pi}$ using 4D regularization scheme: red line -- $m_{\pi}=0$ GeV, black dotted line -- $m_{\pi}=0.4$ GeV, blue dashed line -- $m_{\pi}=0.6$ GeV, purple dotdashed line -- $m_{\pi}=0.75$ GeV. }\label{4dpdf}
\end{figure}

In the forward limit
\begin{align}\label{4dhpdf}
u_{\text{4D}}(x)&=\frac{3Z_{\pi }}{4\pi ^2}\int_0^{\Lambda_{\text{4D}}}\mathrm{d}k \frac{ 2k^3}{(k^2+\sigma_1)^2} \nonumber\\
&+\frac{3Z_{\pi }}{4\pi ^2}  \int_0^{\Lambda_{\text{4D}}}\mathrm{d}k \frac{ 4k^3 x(1-x) m_{\pi}^2 }{(k^2+\sigma_1)^3} ,
\end{align}
we have plotted the PDF in 4D regularization scheme in Fig. \ref{4dpdf}.

\subsubsection{Form Factors}
\begin{align}\label{4da1}
A_{1,0}^{\text{4D}}(t)&=\frac{N_cZ_{\pi } }{4\pi^2}\int_0^1 \mathrm{d}x\int_0^{\Lambda_{\text{4D}}}\mathrm{d}k \frac{ 2k^3}{(k^2+\sigma_1)^2}\nonumber\\
&+\frac{N_cZ_{\pi }}{8\pi^2} \int_0^1 \mathrm{d}x \int_0^{1-x}\mathrm{d}y \int_0^{\Lambda_{\text{4D}}}\mathrm{d}k  \nonumber\\
&\times  \frac{4k^3(2m_{\pi }^2-(x+y)(2m_{\pi }^2-t))}{(k^2+\sigma_7)^3},
\end{align}
\begin{align}\label{4db1}
B_{1,0}^{\text{4D}}(t)&=\frac{ N_c Z_{\pi } }{8\pi ^2}  \int _0^1\mathrm{d}x \int _0^{1-x}\mathrm{d}y \int_0^{\Lambda_{\text{4D}}}\mathrm{d}k\frac{4k^3m_{\pi}M}{(k^2+\sigma_7)^3},
\end{align}
\begin{align}\label{4da2}
A_{2,0}^{\text{4D}}(t)&=\frac{N_cZ_{\pi }}{8\pi ^2}\int_0^1 \mathrm{d}x\int_0^{\Lambda_{\text{4D}}}\mathrm{d}k\frac{ 2k^3}{(k^2+\sigma_1)^2}\nonumber\\
&+\frac{N_c Z_{\pi }}{8\pi ^2}  \int_0^1 \mathrm{d}x \int_0^{1-x} \mathrm{d}y\int_0^{\Lambda_{\text{4D}}}\mathrm{d}k (1-x-y)\nonumber\\
&\times \frac{4k^3(2 m_{\pi }^2(1-x-y)+t  (x+y))}{(k^2+\sigma_7)^3},
\end{align}
\begin{align}\label{4da22}
A_{2,2}^{\text{4D}}(t)&=-\frac{N_cZ_{\pi }}{4\pi ^2}\int_0^1 \mathrm{d}x \int_0^{\Lambda_{\text{4D}}}\mathrm{d}k\,(1-x)\frac{ 2k^3}{(k^2+\sigma_1)^2}\nonumber\\
&-\frac{N_c Z_{\pi } }{2\pi ^2}  \int_0^1 \mathrm{d}x \int_0^{\Lambda_{\text{4D}}}\mathrm{d}k\,x (1-2x) \frac{ 2k^3}{(k^2+\sigma_2)^2}\nonumber\\
&-\frac{N_c Z_{\pi } }{4\pi^2} \int_0^1  \mathrm{d}x \int_0^{\Lambda_{\text{4D}}}\mathrm{d}k \,\frac{ (1-x ) \left(2 m_{\pi }^2-t\right)2k^3}{t(k^2+\sigma_1)^2}\nonumber\\
&+\frac{N_c Z_{\pi }}{8\pi^2}\int_0^1 \mathrm{d}x \int_0^{1-x} \mathrm{d}y \int_0^{\Lambda_{\text{4D}}}\mathrm{d}k\,\frac{4k^3\left(2 m_{\pi }^2-t\right)}{t(k^2+\sigma_7)^3},
\end{align}
\begin{eqnarray}\label{4db2}
B_{2,0}^{\text{4D}}(t)&=&\frac{N_c Z_{\pi}  }{8\pi ^2} \int_0^1 \mathrm{d}x \int_0^{1-x} \mathrm{d}y \int_0^{\Lambda_{\text{4D}}}\mathrm{d}k \nonumber\\
&\times & \frac{4k^3m_{\pi}M (1-x-y)}{(k^2+\sigma_7)^3},
\end{eqnarray}
\begin{align}\label{4db2da}
A_{2,2}^{b,\text{4D}}(t)=-\frac{N_c Z_{\pi } }{\pi ^2}  \int_0^1 \mathrm{d}x  \int_0^{\Lambda_{\text{4D}}}\mathrm{d}k\,  \frac{ 2M x (1-2x)k^3}{(k^2+\sigma_2)^2}.
\end{align}

\begin{figure}
\centering
\includegraphics[width=0.47\textwidth]{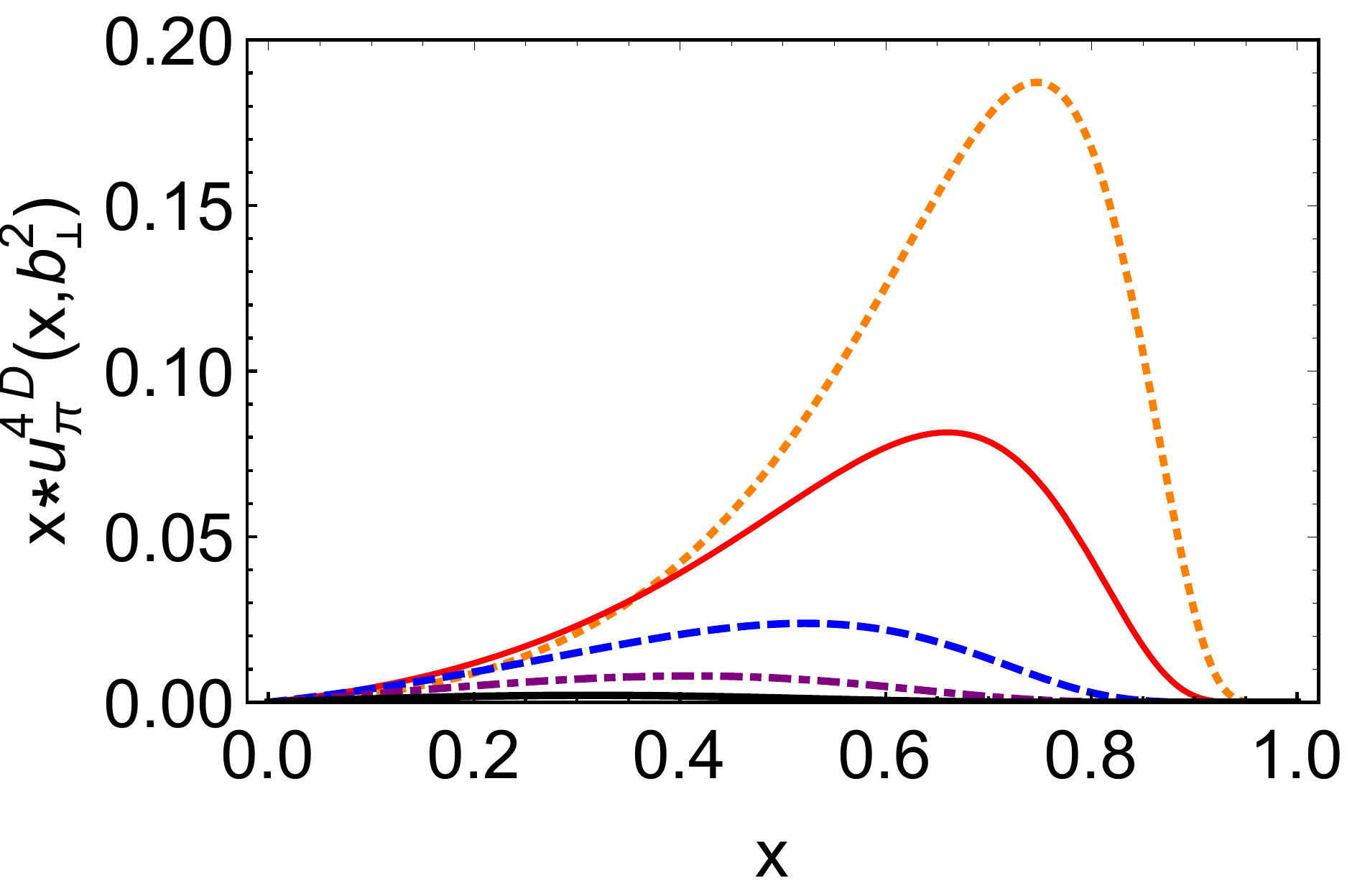}
\qquad
\includegraphics[width=0.47\textwidth]{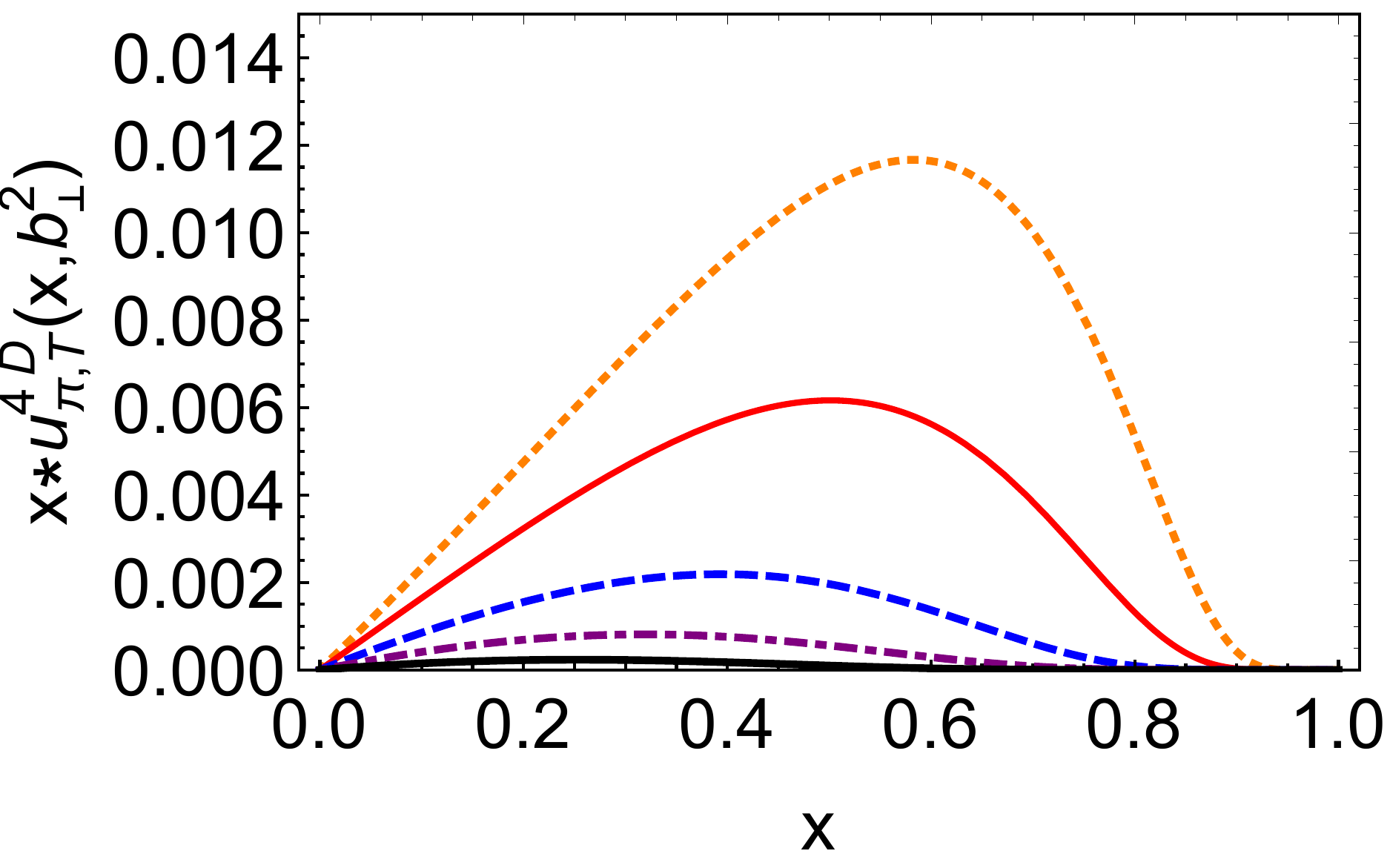}
\caption{Impact parameter space PDFs using 4D regularization scheme: left panel -- $x*u\left(x,\bm{b}_{\perp}^2\right)$, the $\delta^2(\bm{b}_{\perp})$ component first line of Eq. (\ref{4dpspdf}) -- is suppressed in the image, and right panel -- $x*u_T\left(x,\bm{b}_{\perp}^2\right)$ both panels with $\bm{b}_{\perp}^2=0.5$ GeV$^{-2}$ --- orange dotted curve, $\bm{b}_{\perp}^2=1$ GeV$^{-2}$ --- red solid curve, $\bm{b}_{\perp}^2=2.5$ GeV$^{-2}$ --- blue dashed curve, $\bm{b}_{\perp}^2=5$ GeV$^{-2}$ --- purple dot-dashed curve, $\bm{b}_{\perp}^2=10$ GeV$^{-2}$ --- thick black solid curve.}\label{4dqxb}
\end{figure}

\subsubsection{Impact parameter space PDFs}

\begin{align}\label{aG91}
&\quad H_{\text{4D}}\left(x,0,-\bm{q}_{\perp}^2\right)\nonumber\\
&=\frac{N_cZ_{\pi }}{4\pi ^2}  \int_0^{\Lambda_{\text{4D}}}\mathrm{d}k  \frac{2 k^3}{(k^2+\sigma_1)^2} \nonumber\\
&+\frac{N_cZ_{\pi }}{8\pi ^2}\int_0^{1-x} \mathrm{d}\alpha \int_0^{\Lambda_{\text{4D}}}\mathrm{d}k \frac{4k^3\left(2 x m_{\pi}^2+ x \bm{q}_{\perp}^2- \bm{q}_{\perp}^2\right)}{(k^2+\sigma_8)^3},
\end{align}
\begin{align}\label{aG911}
E_{\text{4D}}\left(x,0,-\bm{q}_{\perp}^2\right)=\frac{N_cZ_{\pi }}{4\pi ^2}\int_0^{1-x} \mathrm{d}\alpha \int_0^{\Lambda_{\text{4D}}}\mathrm{d}k \frac{4k^3m_{\pi} M }{(k^2+\sigma_8)^3},
\end{align}
\begin{widetext}
\begin{align}\label{4dpspdf}
&\quad u^{\text{4D}}\left(x,\bm{b}_{\perp}^2\right)\nonumber\\
&=\frac{N_cZ_{\pi }}{4\pi ^2} \int \frac{\mathrm{d}^2\bm{q}_{\perp}}{(2 \pi )^2}\int_0^{\Lambda_{\text{4D}}}e^{-i\bm{b}_{\perp}\cdot \bm{q}_{\perp}}\mathrm{d}k\frac{ k^2}{(k^2+\sigma_1)^{3/2}} \nonumber\\
&+\frac{N_cZ_{\pi }}{16\pi ^3}\int_0^{1-x} \mathrm{d}\alpha \int_0^{\Lambda_{\text{4D}}}\mathrm{d}k\,   \mathrm{K}_1\left[\frac{\sqrt{\bm{b}_{\perp}^2} \sqrt{k^2+\sigma_1}}{\sqrt{\alpha  (1-x-\alpha )}}\right] \frac{2  k^3 (x-1) \sqrt{\bm{b}_{\perp}^2} }{ \alpha ^{5/2} (1-\alpha -x)^{5/2} \sqrt{k^2+\sigma_1}}\nonumber\\
&+\frac{N_cZ_{\pi }}{16\pi ^3}\int_0^{1-x} \mathrm{d}\alpha \int_0^{\Lambda_{\text{4D}}}\mathrm{d}k\, \bm{b}_{\perp}^2 k^3 \, \mathrm{K}_2\left[\frac{\sqrt{\bm{b}_{\perp}^2} \sqrt{k^2+\sigma_1}}{\sqrt{\alpha  (1-x-\alpha)}}\right]\frac{\left((1-x) \left(k^2+M^2\right)-xm_{\pi }^2 \left(2 \alpha ^2-2 \alpha +x^2+2 (\alpha -1) x+1\right)\right) }{2 \alpha ^3 (1-\alpha -x)^3 \left(k^2+\sigma_1\right)},
\end{align}
\begin{align}\label{4dtpspdf}
&u_\text{T}^{\text{4D}}\left(x,\bm{b}_{\perp}^2\right)=\frac{N_cZ_{\pi }}{16\pi^3}\int_0^{1-x}  \mathrm{d}\alpha \int_0^{\Lambda_{\text{4D}}}\mathrm{d}k \frac{k^3 m_{\pi } M \bm{b}_{\perp}^2 }{ \alpha ^2 (1-\alpha-x)^2 \left(k^2+\sigma_1\right)}\mathrm{K}_2\left[\frac{\sqrt{\bm{b}_{\perp}^2} \sqrt{k^2+\sigma_1}}{\sqrt{\alpha  (1-x-\alpha )}}\right],
\end{align}
\end{widetext}
where $\mathrm{K}_1$ and $\mathrm{K}_2$ are the Bessel function of the second kind $\mathrm{K}_n(z)$, for $u^{\text{4D}}\left(x,\bm{b}_{\perp}^2\right)$, when integrating $\bm{b}_{\perp}$ one can get PDF $u(x)$ in Eq. (\ref{4dhpdf}). We plot the diagrams of $x *u^{\text{4D}}\left(x,\bm{b}_{\perp}^2\right)$ and $x *u_\text{T}^{\text{4D}}\left(x,\bm{b}_{\perp}^2\right)$ in Fig. \ref{4dqxb}.

\subsection{PV regularization}
The Pauli-Villars scheme has an attractive feature, which is it keeps the gauge invariance. In Pauli-Villars regularization, in order to get rid of the divergences from loop integrals, people introduce the virtually heavy particles~\cite{Klevansky:1992qe}
\begin{align}\label{3}
\frac{1}{k^2-M^2}\rightarrow \sum_i c_i \frac{1}{k^2-M_i^2},
\end{align}
where the labels run from $i=0$ to $i=2$
\begin{align}\label{3}
M_i^2=M^2+\alpha_i\Lambda_i^2,
\end{align}
where
\begin{align}\label{3}
c_0&=1,\quad c_1=1,\quad c_2=-2;\nonumber\\
\alpha_0&=0,\quad\alpha_1=2,\quad\alpha_2=1;
\end{align}
then the gap equation becomes
\begin{align}\label{3}
M
&=m-\frac{3G_{\pi}M}{\pi^2}\sum_ic_iM_i^2\log\left[M_i^2/M^2\right],
\end{align}
the pion decay constant becomes
\begin{align}\label{pvfpi}
f_{\pi}^{\text{PV}}=\frac{N_c\sqrt{Z_{\pi }}M}{4\pi ^2}\int_0^1 \mathrm{d}x \log \left[\frac{\left(\Lambda_{\text{PV}}^2+\sigma_1\right)^2}{\sigma_1 \left(2 \Lambda_{\text{PV}}^2+\sigma_1\right)}\right],
\end{align}
the bubble diagram $\Pi_{\text{VV}}$ is
\begin{align}\label{pvvvbu}
\Pi_{\text{VV}}^{\text{PV}}(t)=-\frac{3 }{\pi ^2} \int_0^1\mathrm{d}x\, x (1-x) t\log \left[\frac{\left(\Lambda_{\text{PV}}^2+\sigma_2\right)^2}{\sigma_2\left(2 \Lambda_{\text{PV}}^2+\sigma_2\right)}\right].
\end{align}
The parameters used in PV regularization scheme are listed in Table \ref{tbpv}.
%
%
%
%
%
%

%

\begin{center}
\begin{table*}
\caption{Parameter set used for PV regularization. The dressed quark mass and regularization parameters are in units of GeV, while coupling constant are in units of GeV$^{-2}$.}\label{tbpv}
\begin{tabular}{p{1.0cm} p{1.0cm}p{1.0cm}p{1.0cm} p{1.0cm}p{1.0cm}p{1.0cm}p{1.0cm}p{1.0cm}p{1.4cm}}
\hline\hline
$\Lambda _{\text{PV}}$&$M$&$G_{\pi}$&$G_{\rho}$&$G_{\omega}$&$m_{\pi}$&$f_{\pi}$&$m$&$Z_{\pi}$&$\langle\bar{u}u\rangle$\\
\hline
0.63&0.4&9.724&8.407&7.451&0.14&0.093&0.008&18.17&--(0.216)$^3$\\
\hline\hline
\end{tabular}
\end{table*}
\end{center}

\begin{figure*}
\centering
\includegraphics[width=0.47\textwidth]{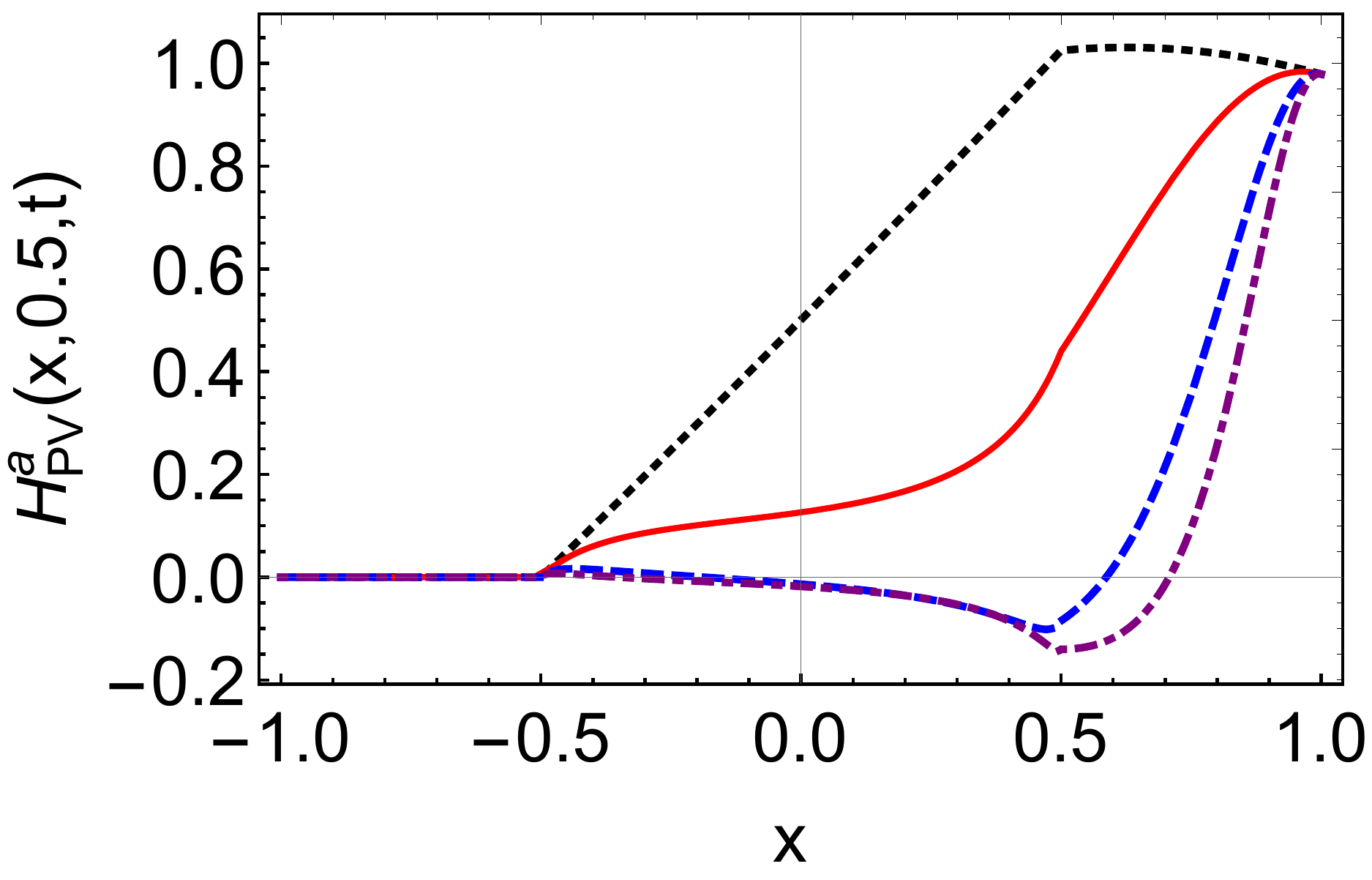}
\qquad
\includegraphics[width=0.47\textwidth]{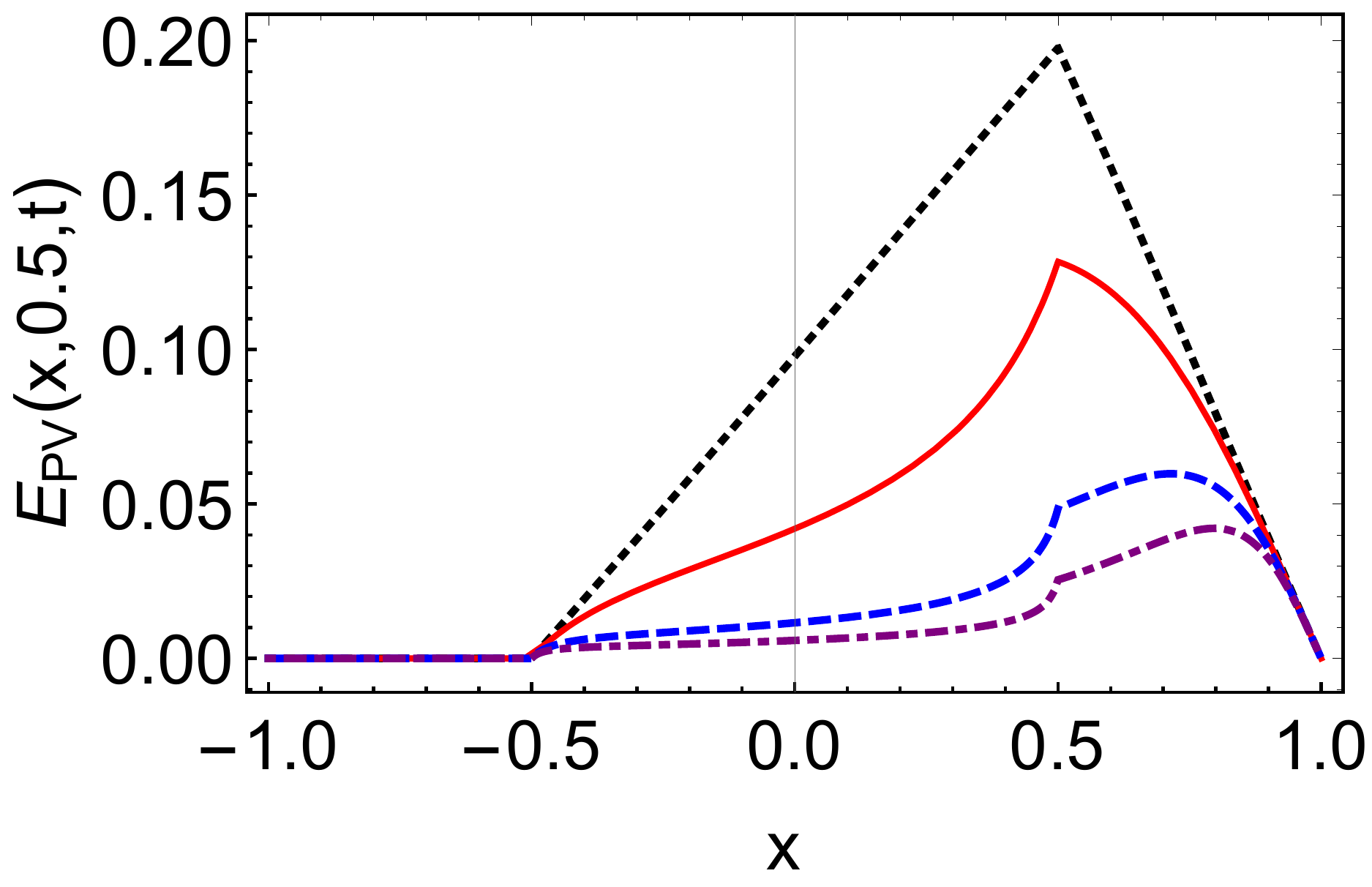}
\caption{Pion GPDs (left panel: vector GPD, right panel: tensor GPD) at $\xi=0.5$ with different $t$ using PV regularization: black dotted line -- $t=0$ GeV$^2$, red line -- $t=-1$ GeV$^2$, blue dashed line -- $t=-5$ GeV$^2$, purple dotdashed line -- $t=-10$ GeV$^2$. }\label{pvegpd}
\end{figure*}

\begin{figure*}
\centering
\includegraphics[width=0.47\textwidth]{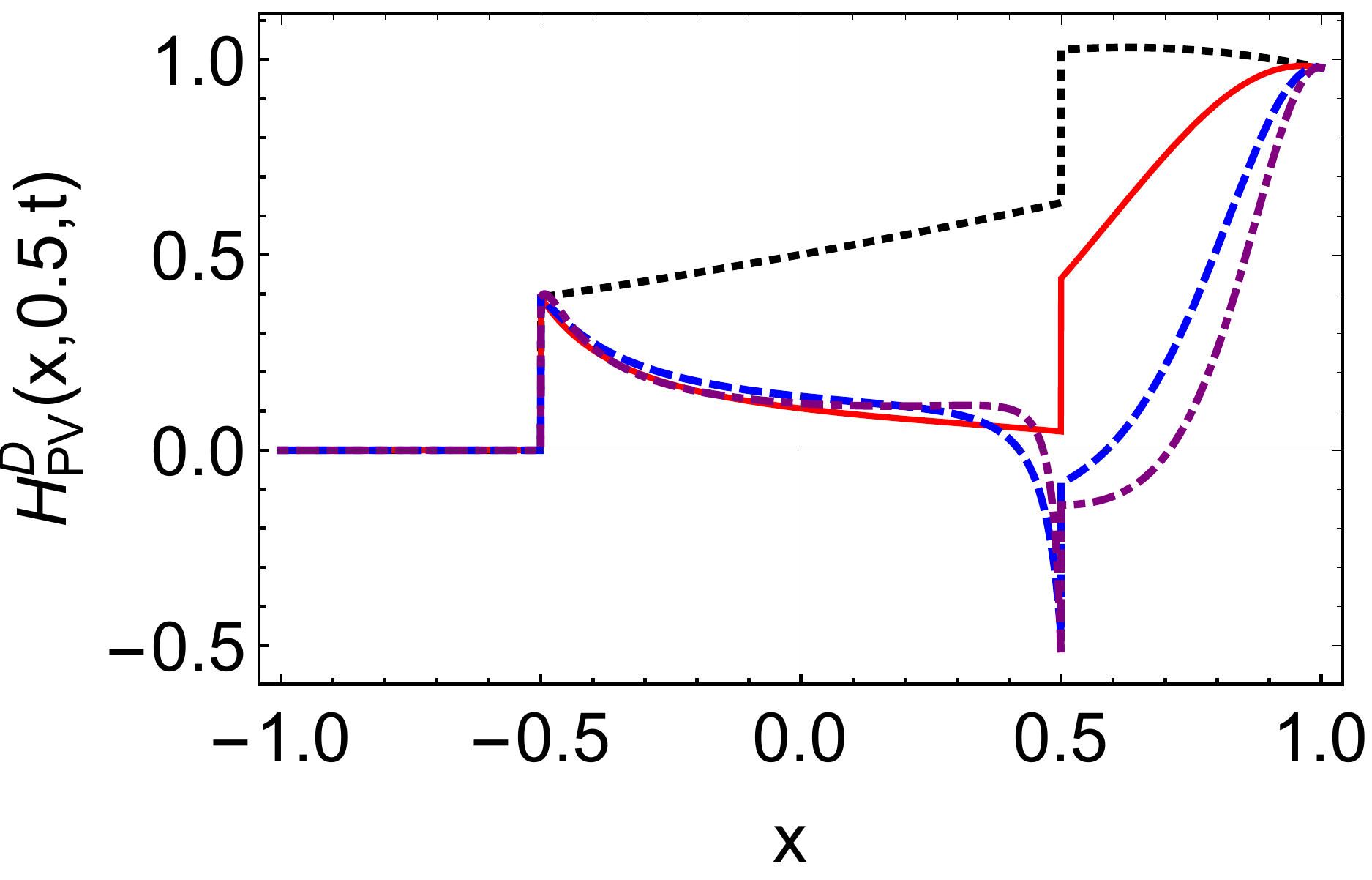}
\qquad
\includegraphics[width=0.47\textwidth]{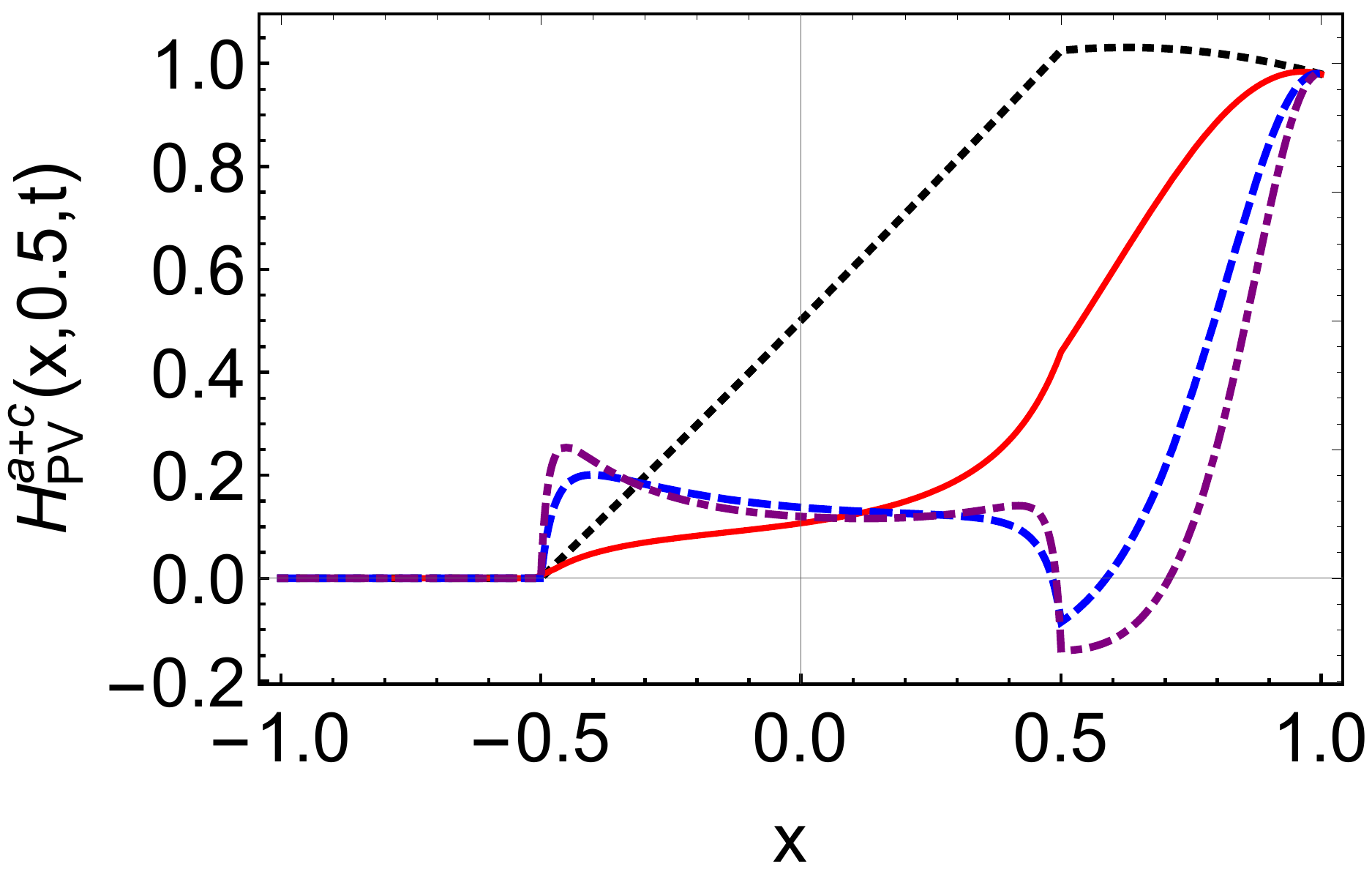}
\caption{Pion GPDs (left panel: $H_{\text{PV}}^D$, right panel: $H_{\text{PV}}^a+H_{\text{PV}}^c$) at $\xi=0.5$ with different $t$ using PT regularization: black dotted line -- $t=0$ GeV$^2$, red line -- $t=-1$ GeV$^2$, blue dashed line -- $t=-5$ GeV$^2$, purple dotdashed line -- $t=-10$ GeV$^2$. }\label{pvgpd1}
\end{figure*}

The we obtain the pion GPDs in Pauli-Villars scheme
\begin{align}\label{pvagpdf}
&\quad H_{\text{PV}}^a\left(x,\xi,t\right)\nonumber\\
&= \frac{N_cZ_{\pi }}{8\pi ^2}\theta_{\bar{\xi} 1} \log \left[\frac{\left(\Lambda_{\text{PV}}^2+\sigma_3\right)^2}{\sigma_3 \left(2 \Lambda_{\text{PV}}^2+\sigma_3\right)}\right]\nonumber\\
&+ \frac{N_cZ_{\pi }}{8\pi ^2}\theta_{\bar{\xi} 1} \log \left[\frac{\left(\Lambda_{\text{PV}}^2+\sigma_4\right)^2}{\sigma_4 \left(2 \Lambda_{\text{PV}}^2+\sigma_4\right)}\right]\nonumber\\
&+\frac{N_cZ_{\pi }}{8\pi ^2}\frac{\theta_{\bar{\xi} \xi} }{\xi} x\log \left[\frac{\left(\Lambda_{\text{PV}}^2+\sigma_5\right)^2}{\sigma_5 \left(2 \Lambda_{\text{PV}}^2+\sigma_5\right)}\right]\nonumber\\
&+\frac{N_c Z_{\pi } }{16\pi ^2}\int_0^1 \mathrm{d}\alpha \frac{\theta_{\alpha \xi}}{\xi} (2 x m_{\pi}^2+(1-x)t)\nonumber\\
&\times \left(\frac{1}{\sigma_6+2\Lambda_{\text{PV}}^2}-\frac{2}{\sigma_6+\Lambda_{\text{PV}}^2}-\frac{1}{\sigma_6} \right) ,
\end{align}
\begin{align}\label{pvagtpdf}
E_{\text{PV}}\left(x,\xi,t\right)&=\frac{N_c Z_{\pi } }{8\pi ^2}\int_0^1 \mathrm{d}\alpha \frac{\theta_{\alpha \xi}}{\xi} m_{\pi}M\nonumber\\
&\times \left(\frac{1}{\sigma_6+2\Lambda_{\text{PV}}^2}-\frac{2}{\sigma_6+\Lambda_{\text{PV}}^2}-\frac{1}{\sigma_6} \right),
\end{align}
\begin{align}\label{3dagtpdf}
H_{\text{PV}}^b\left(x,\xi,t\right)=-\frac{N_cZ_{\pi }}{4\pi ^2} \frac{\theta_{\bar{\xi} \xi} }{\xi} Mx \log \left[\frac{\left(\Lambda_{\text{PV}}^2+\sigma_5\right)^2}{\sigma_5 \left(2 \Lambda_{\text{PV}}^2+\sigma_5\right)}\right],
\end{align}
\begin{align}\label{3dagtpdf}
H_{\text{PV}}^c\left(x,\xi,t\right)&=F_{1\rho}^{\text{PV}}(t)A_{1,0}^{\text{PV}}(t)\frac{N_cG_{\rho }}{4\pi ^2} \frac{\theta_{\bar{\xi} \xi} }{\xi}t\left(1-\frac{x^2}{\xi^2}\right)\nonumber\\
&\times \log \left[\frac{\left(\Lambda_{\text{PV}}^2+\sigma_5\right)^2}{\sigma_5 \left(2 \Lambda_{\text{PV}}^2+\sigma_5\right)}\right],
\end{align}
The pion GPDs in PV regularization scheme are plotted in Figs. \ref{pvegpd} and \ref{pvgpd1}.


\subsubsection{Forward limit}
\begin{figure}
\centering
\includegraphics[width=0.47\textwidth]{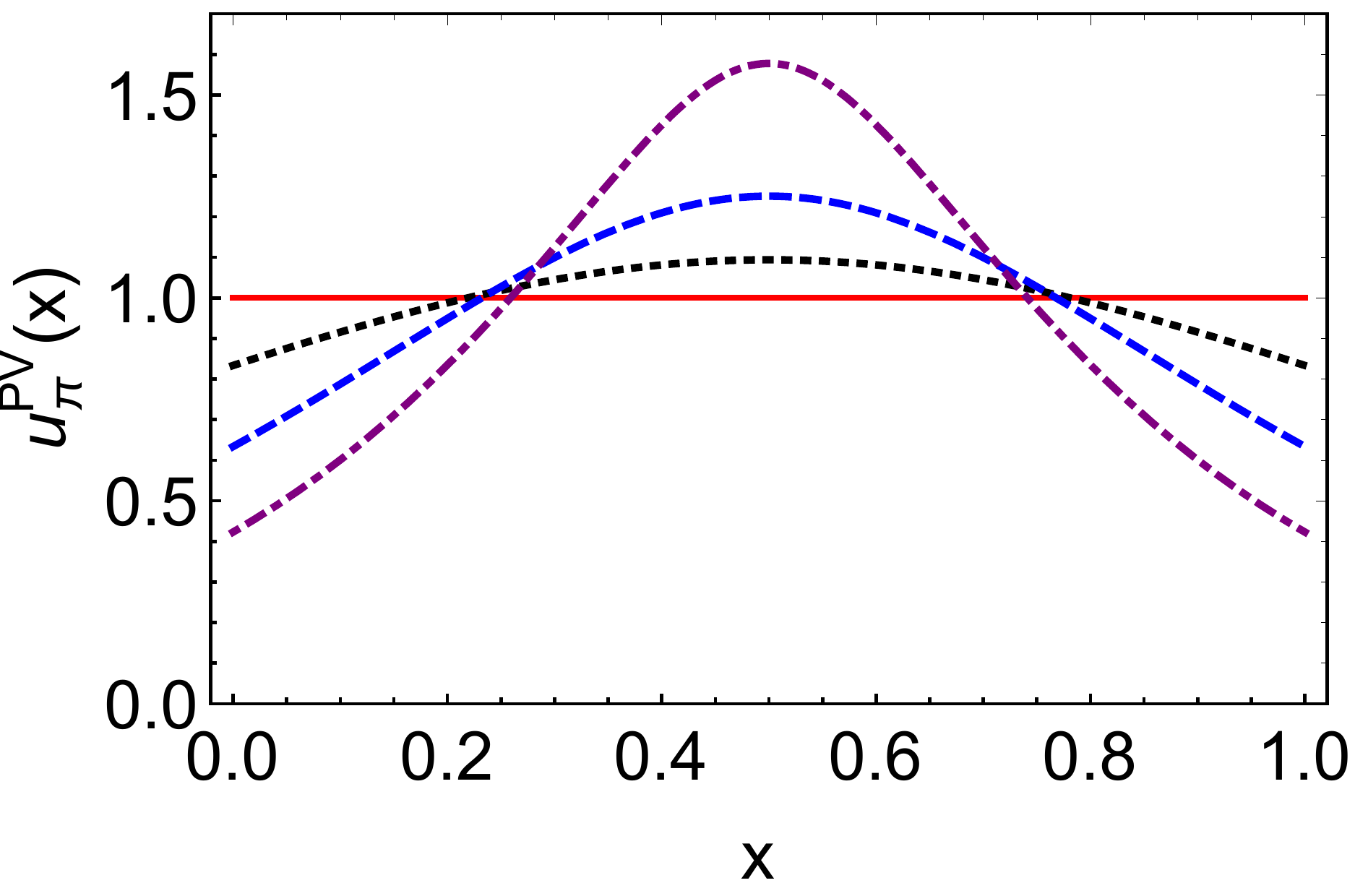}
\caption{Pion $u$ quark PDF with different values of $m_{\pi}$ using PV regularization: red line -- $m_{\pi}=0$ GeV, black dotted line -- $m_{\pi}=0.4$ GeV, blue dashed line -- $m_{\pi}=0.6$ GeV, purple dotdashed line -- $m_{\pi}=0.75$ GeV. }\label{pvpdf}
\end{figure}

In the forward limit
\begin{align}\label{pvhpdf}
&u_{\text{PV}}(x)\nonumber\\
=&\frac{3Z_{\pi }}{4\pi ^2}  \log \left[\frac{\left(\Lambda_{\text{PV}} ^2+\sigma_1\right)^2}{\sigma_1\left(2 \Lambda_{\text{PV}} ^2+\sigma_1\right)}\right]\nonumber\\
+&\frac{3Z_{\pi }}{4\pi ^2} x(1-x) m_{\pi}^2 \left(\frac{1}{\sigma_1+2\Lambda_{\text{PV}}^2}-\frac{2}{\sigma_1+\Lambda_{\text{PV}}^2}+\frac{1}{\sigma_1} \right) ,
\end{align}
we plot the PDF in PV regularization scheme in Fig. \ref{pvpdf}. From Figs. \ref{ptpdf}, \ref{3dpdf}, \ref{4dpdf} and \ref{pvpdf} we can see that in PT and PV regularization schemes, pion PDFs are softer, in the 3D and 4D regularization schemes the peaks are sharper.

\subsubsection{Form Factors}
\begin{align}\label{pva1}
A_{1,0}^{\text{PV}}(t)&=\frac{N_cZ_{\pi } }{4\pi^2}\int_0^1 \mathrm{d}x\log \left[\frac{\left(\Lambda_{\text{PV}} ^2+\sigma_1\right)^2}{\sigma_1\left(2 \Lambda_{\text{PV}} ^2+\sigma_1\right)}\right]\nonumber\\
&+\frac{N_cZ_{\pi }}{8\pi^2} \int_0^1 \mathrm{d}x \int_0^{1-x}\mathrm{d}y\, (2m_{\pi }^2-(x+y)(2m_{\pi }^2-t))\nonumber\\
&\times \left(\frac{1}{\sigma_7+2\Lambda_{\text{PV}}^2}-\frac{2}{\sigma_7+\Lambda_{\text{PV}}^2}+\frac{1}{\sigma_7} \right),
\end{align}
\begin{align}\label{pvb1}
B_{1,0}^{\text{PV}}(t)&=\frac{ N_c Z_{\pi } }{8\pi ^2}  \int _0^1\mathrm{d}x \int _0^{1-x}\mathrm{d}y \,m_{\pi}M\nonumber\\
&\times \left(\frac{1}{\sigma_7+2\Lambda_{\text{PV}}^2}-\frac{2}{\sigma_7+\Lambda_{\text{PV}}^2}+\frac{1}{\sigma_7} \right),
\end{align}
\begin{align}\label{pva2}
A_{2,0}^{\text{PV}}(t)&=\frac{N_cZ_{\pi }}{8\pi ^2}\int_0^1 \mathrm{d}x \log \left[\frac{\left(\Lambda_{\text{PV}} ^2+\sigma_1\right)^2}{\sigma_1\left(2 \Lambda_{\text{PV}} ^2+\sigma_1\right)}\right]\nonumber\\
&+\frac{N_c Z_{\pi }}{8\pi ^2}  \int_0^1 \mathrm{d}x \int_0^{1-x} \mathrm{d}y \,(1-x-y)\nonumber\\
&\times (2 m_{\pi }^2(1-x-y)+t  (x+y))\nonumber\\
&\times \left(\frac{1}{\sigma_7+2\Lambda_{\text{PV}}^2}-\frac{2}{\sigma_7+\Lambda_{\text{PV}}^2}+\frac{1}{\sigma_7} \right),
\end{align}
\begin{align}\label{pvda22}
A_{2,2}^{\text{PV}}(t)&=-\frac{N_cZ_{\pi }}{4\pi ^2}\int_0^1 \mathrm{d}x \,(1-x)\log \left[\frac{\left(\Lambda_{\text{PV}} ^2+\sigma_1\right)^2}{\sigma_1\left(2 \Lambda_{\text{PV}} ^2+\sigma_1\right)}\right]\nonumber\\
&-\frac{N_c Z_{\pi } }{2\pi ^2}  \int_0^1 \mathrm{d}x \,x (1-2x) \log \left[\frac{\left(\Lambda_{\text{PV}} ^2+\sigma_2\right)^2}{\sigma_2\left(2 \Lambda_{\text{PV}} ^2+\sigma_2\right)}\right]\nonumber\\
&-\frac{N_c Z_{\pi } }{4\pi^2} \int_0^1  \mathrm{d}x \,\frac{ (1-x ) \left(2 m_{\pi }^2-t\right)}{t}\nonumber\\
&\times \log \left[\frac{\left(\Lambda_{\text{PV}} ^2+\sigma_1\right)^2}{\sigma_1\left(2 \Lambda_{\text{PV}} ^2+\sigma_1\right)}\right]\nonumber\\
&+\frac{N_c Z_{\pi }}{8\pi^2}\int_0^1 \mathrm{d}x \int_0^{1-x} \mathrm{d}y \,\frac{\left(2 m_{\pi }^2-t\right)}{t}\nonumber\\
&\times \left(\frac{1}{\sigma_7+2\Lambda_{\text{PV}}^2}-\frac{2}{\sigma_7+\Lambda_{\text{PV}}^2}+\frac{1}{\sigma_7} \right),
\end{align}
\begin{align}\label{pvb2}
B_{2,0}^{\text{PV}}(t)&=\frac{N_c Z_{\pi}  }{8\pi ^2} \int_0^1 \mathrm{d}x \int_0^{1-x} \mathrm{d}y \,m_{\pi}M (1-x-y)\nonumber\\
&\times \left(\frac{1}{\sigma_7+2\Lambda_{\text{PV}}^2}-\frac{2}{\sigma_7+\Lambda_{\text{PV}}^2}+\frac{1}{\sigma_7} \right),
\end{align}
\begin{align}\label{4db2da}
A_{2,2}^{b,\text{PV}}(t)&=-\frac{N_c Z_{\pi } }{\pi ^2}  \int_0^1 \mathrm{d}x \,M x (1-2x)\nonumber\\
&\times\log \left[\frac{\left(\Lambda_{\text{PV}} ^2+\sigma_2\right)^2}{\sigma_2\left(2 \Lambda_{\text{PV}} ^2+\sigma_2\right)}\right].
\end{align}

We have listed charge radius and the values of GFFs at $t=0$ in Table \ref{tbrg}. The bare charge radius in PT regularization regularization is the biggest, in the 3D momentum cut-off is the smallest, but do not have too much difference. For the dressed pion charge radius, the value in PT regularization is obviously bigger and more approximates the experiment data $r_{\pi}=0.66 $ fm. The values of the quark mass distribution $A_{2,0}(0)$ and the quark pressure distribution $A_{2,2}^D(0)$, in the PV regularization are the biggest, but for the anomalous magnetic moment $B_{1,0}(0)$ and $B_{2,0}(0)$ in the PT regularization do not have too much difference with the results in PV regularization. The low-energy theorem requires $\theta_1(0)-\theta_2(0)=\mathcal{O}(m_{\pi}^2)$, if we do not consider the contribution of D-term, we can see from Table. \ref{tbrg} that $A_{2,0}(0)$ and $|A_{2,2}(0)|$ is quite different. When considering the D-term, the value of $|A_{2,2}^D(0)|$ is close to $A_{2,0}(0)$.

\begin{figure*}
\centering
\includegraphics[width=0.47\textwidth]{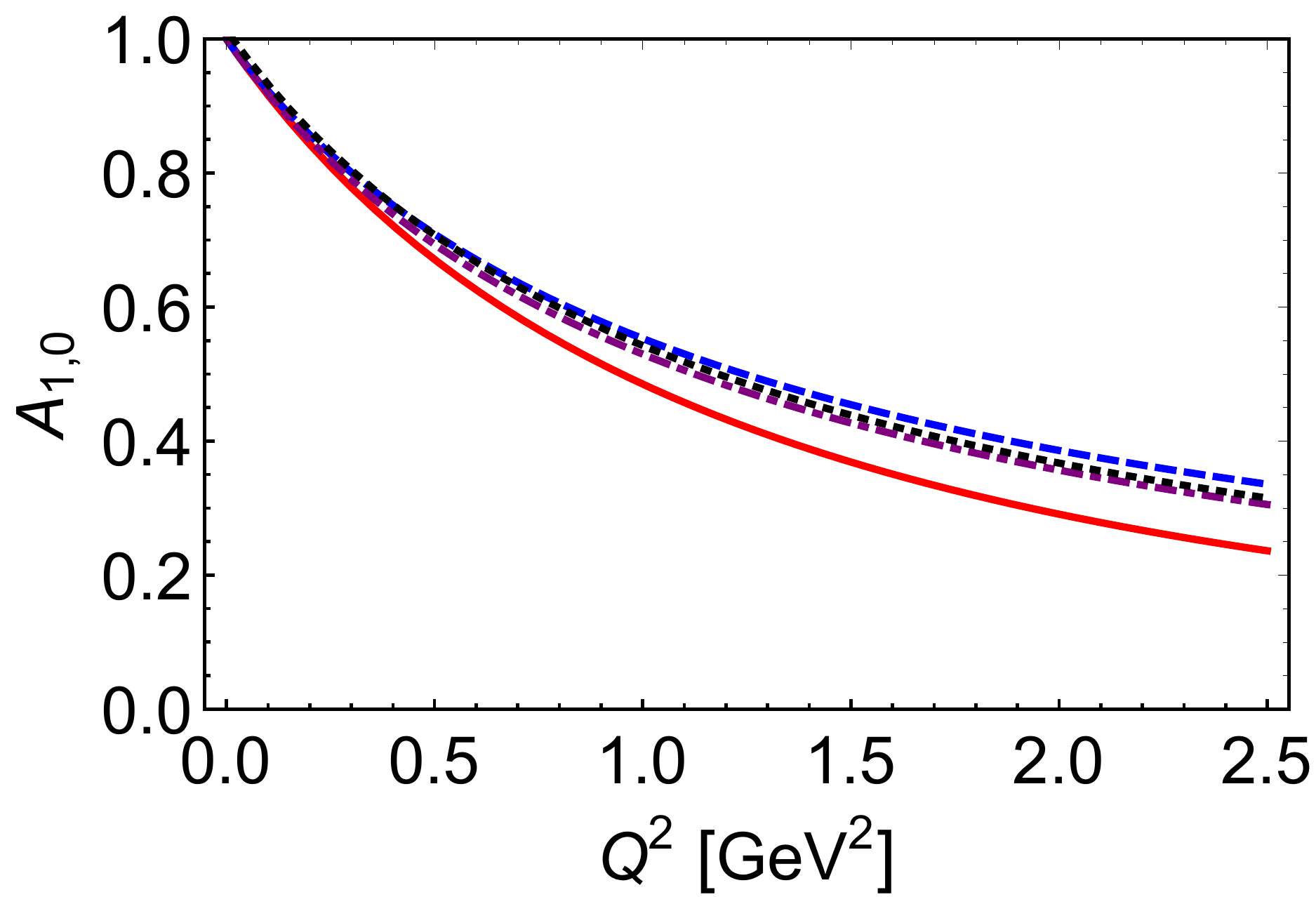}
\qquad
\includegraphics[width=0.47\textwidth]{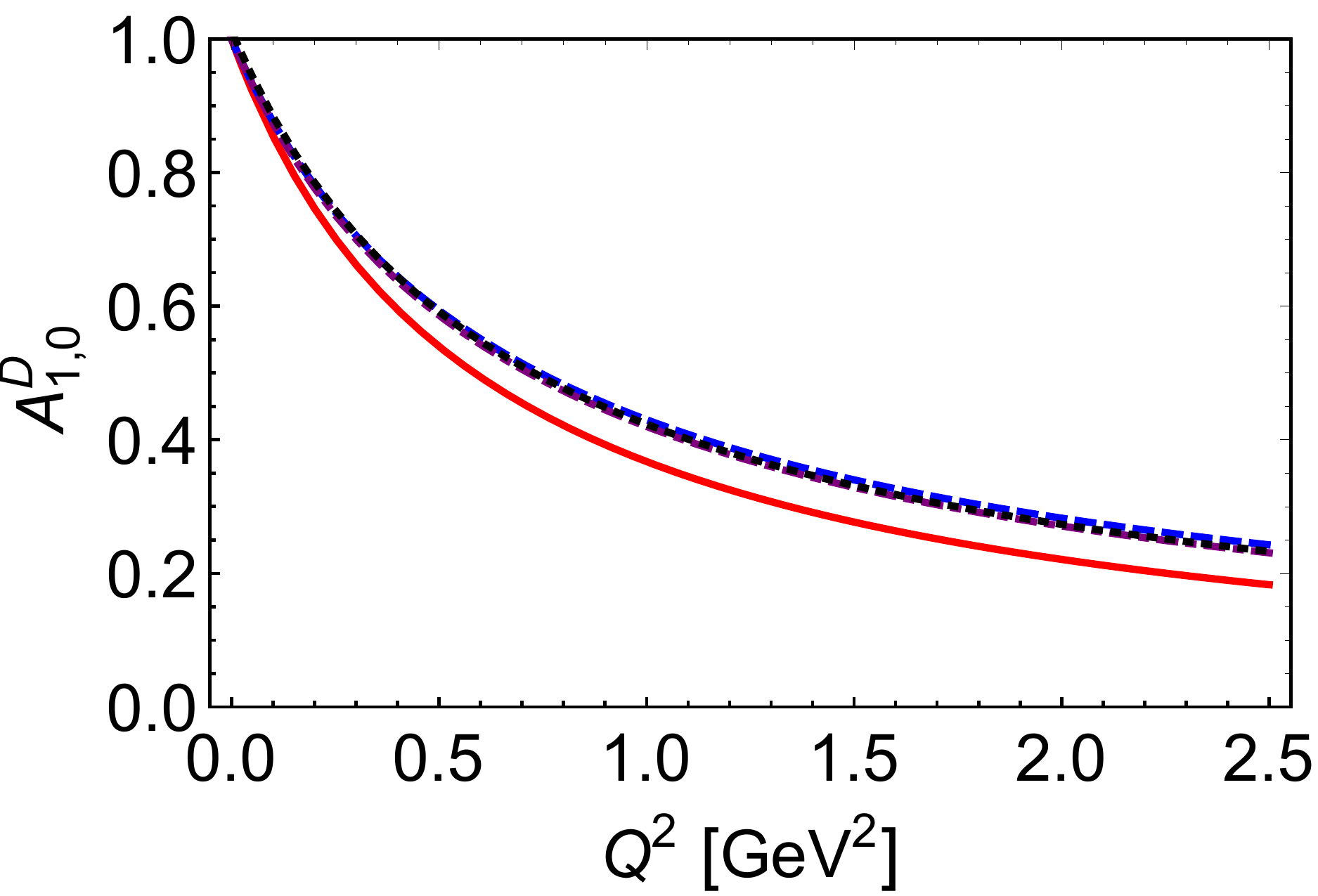}
\caption{Pion electromagnetic FFs (left panel: bare FF, right panel: dressed FF) using different regularization schemes: red line -- PT regularization, blue dashed line -- 3D regularization, purple dotdashed line -- 4D regularization, black dotted line -- PV regularization. }\label{a10}
\end{figure*}

\begin{figure*}
\centering
\includegraphics[width=0.47\textwidth]{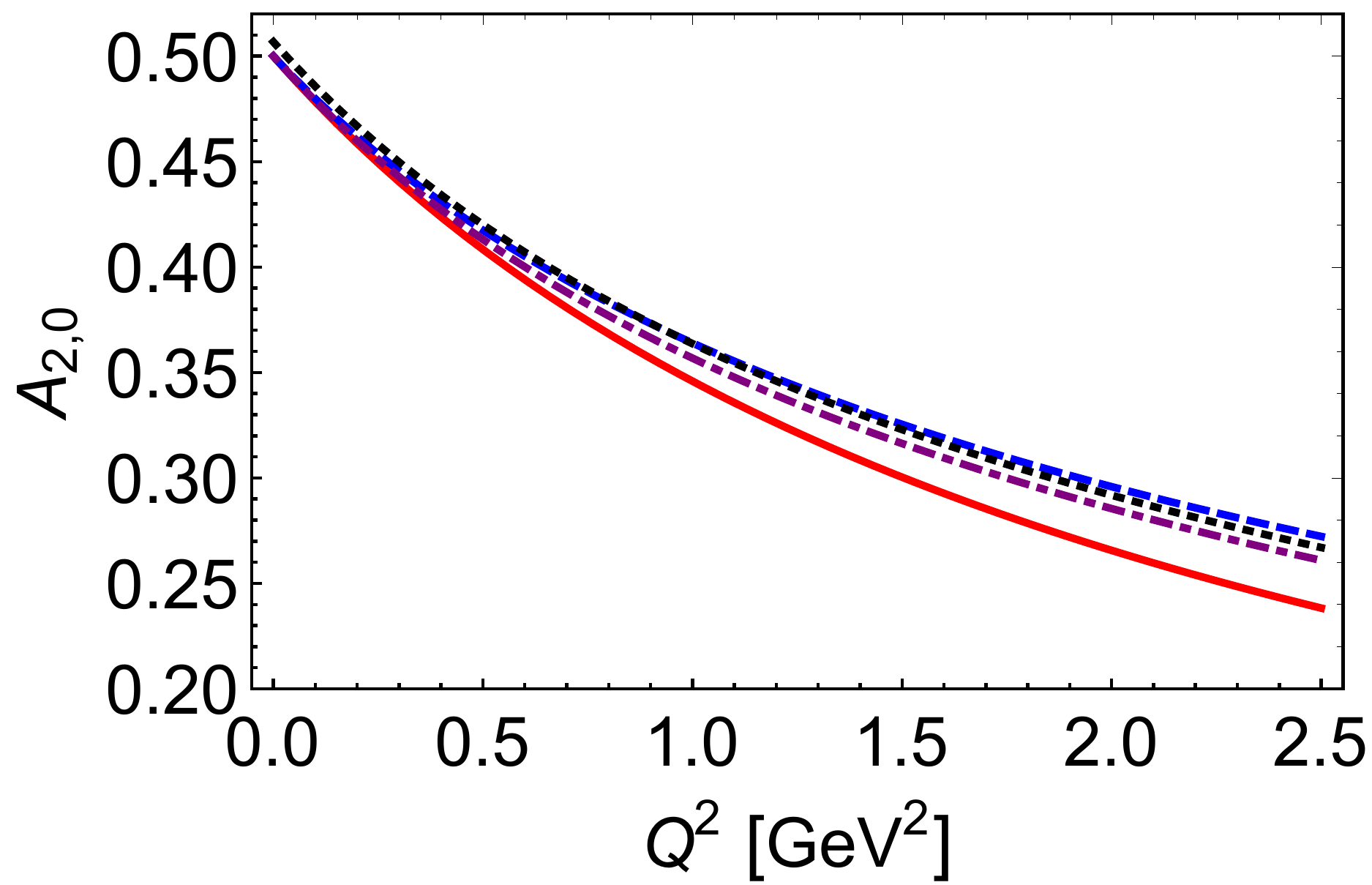}
\qquad
\includegraphics[width=0.47\textwidth]{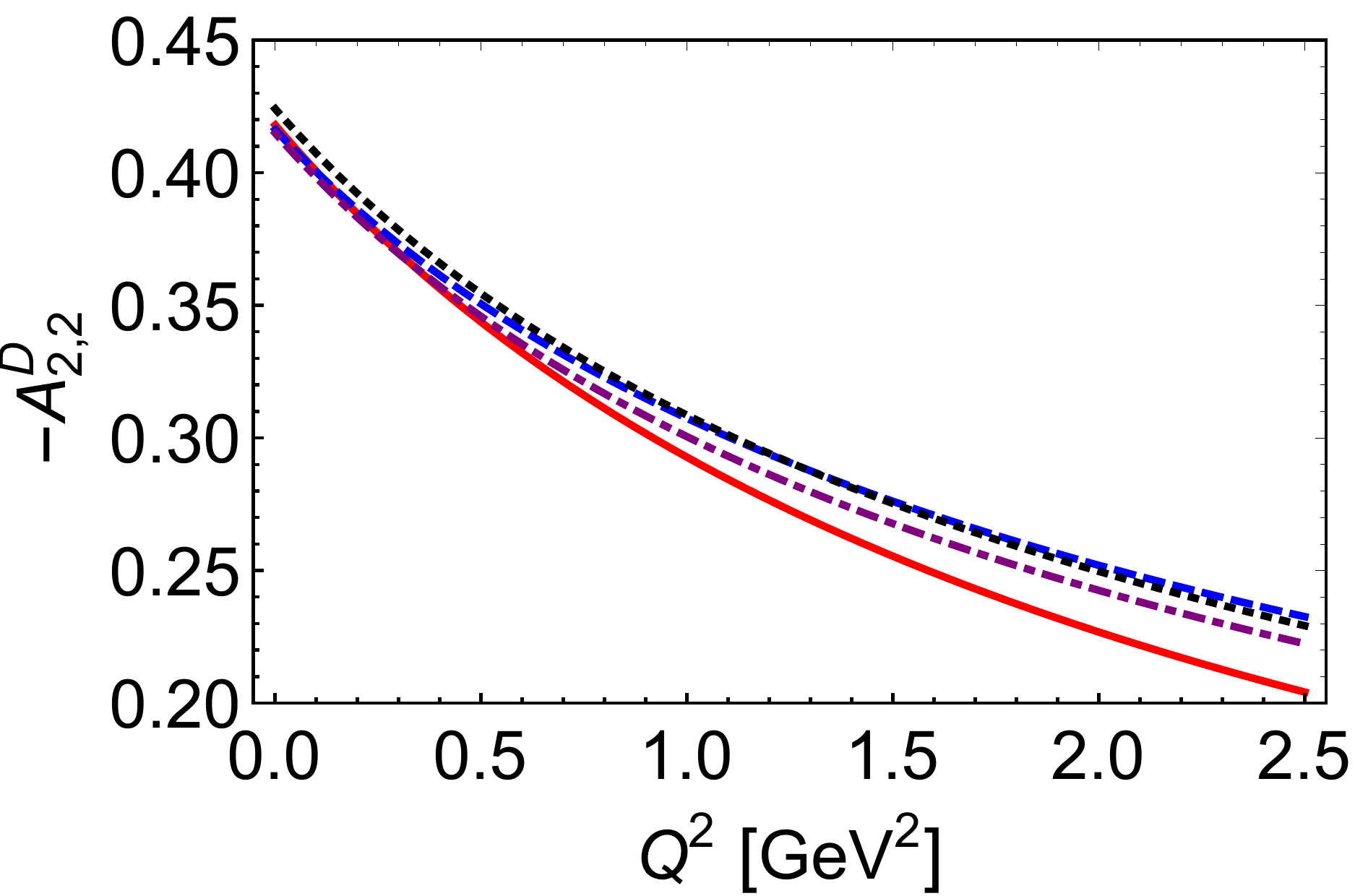}
\caption{Pion generalised vector FFs (left panel: $A_{2,0}(t)$, right panel: $A_{2,2}^D(t)$) using different regularization schemes: red line -- PT regularization, blue dashed line -- 3D regularization, purple dotdashed line -- 4D regularization, black dotted line -- PV regularization. }\label{a20}
\end{figure*}

\begin{figure*}
\centering
\includegraphics[width=0.47\textwidth]{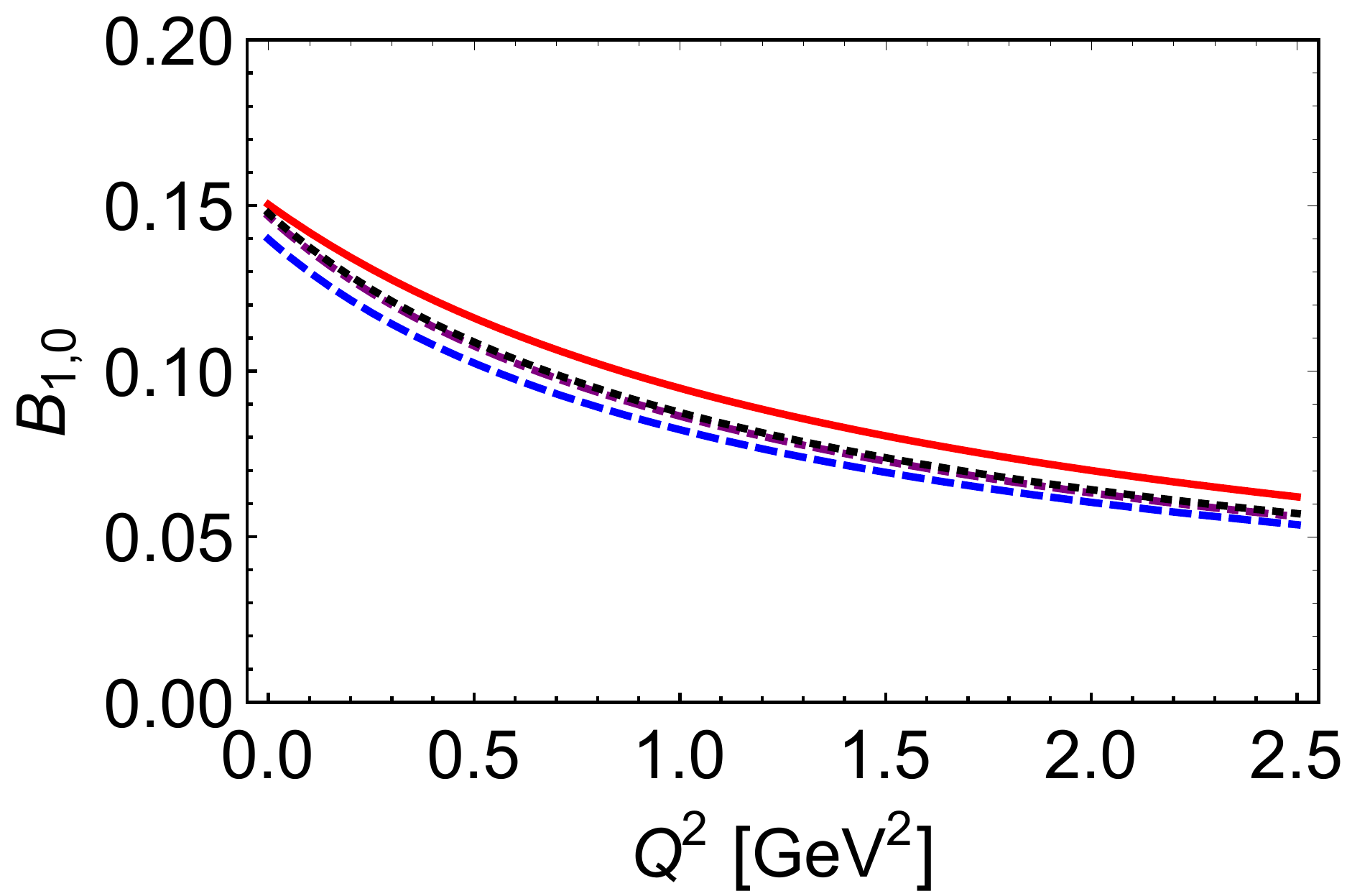}
\qquad
\includegraphics[width=0.47\textwidth]{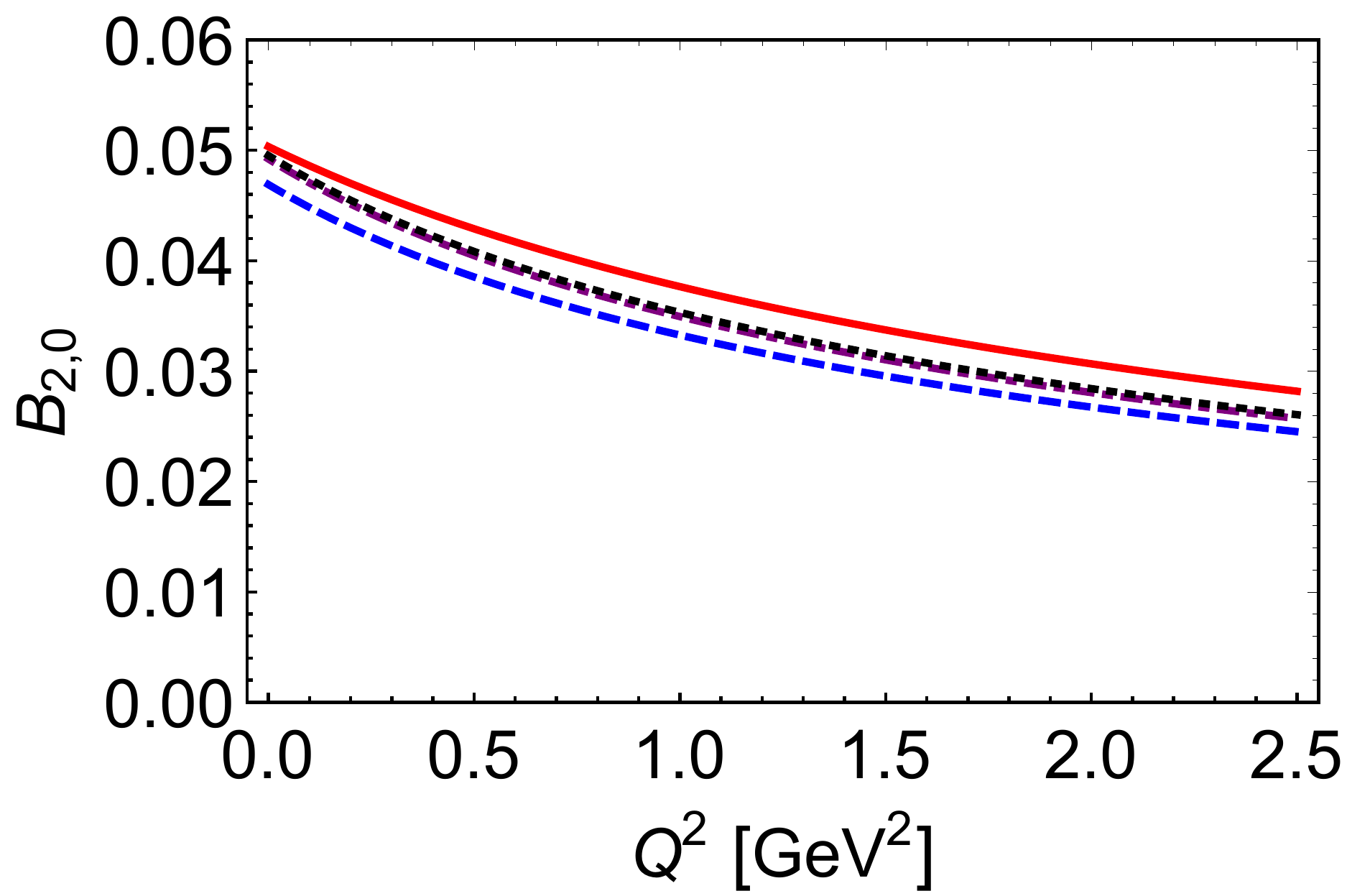}
\caption{Pion tensor FFs (left panel: $B_{1,0}(t)$, right panel: $B_{2,0}(t)$) using different regularization schemes: red line -- PT regularization, blue dashed line -- 3D regularization, purple dotdashed line -- 4D regularization, black dotted line -- PV regularization. }\label{b10}
\end{figure*}

In Figs. \ref{a10} -- \ref{b10} we plot the diagrams of FFs. Fig. \ref{a10} shows the electromagnetic FFs, the bare FF in the left panel and the dressed FF in the right panel. Fig. \ref{a20} plots the quark mass distribution $A_{2,0}(t)$ in the left panel and the quark pressure distribution $A_{2,2}^D(t)$ in the right panel, the gravitational FFs in PT regularization are the softest. Fig. \ref{b10} gives the tensor FFs $B_{1,0}(t)$ and $B_{2,0}(t)$, the tensor FFs are almost have the same slope in different regularization schemes, the four lines are almost parallel, but the values in PT regularization are the biggest.

\begin{center}
\begin{table}
\caption{Charge radius and GFFs at $t = 0$, charge radius are in units of fm.}\label{tbrg}
\begin{tabular}{p{0.6cm} p{0.9cm} p{0.9cm}p{1.0cm}p{1.1cm}p{1.1cm}p{1.1cm}p{1.1cm}}
\hline\hline
 &$r_{\pi}$&$r_{\pi}^D$&$A_{2,0}(0)$&$A_{2,2}(0)$&$A_{2,2}^D(0)$&$B_{1,0}(0)$&$B_{2,0}(0)$\\
\hline
PT&0.457&0.628&0.500&--0.161&--0.418&0.150&0.050\\
\hline
3D&0.444&0.578&0.500&--0.161&--0.415&0.140&0.047\\
\hline
4D&0.455&0.583&0.500&--0.161&--0.415&0.147&0.049\\
\hline
PV&0.454&0.590&0.507&--0.163&--0.424&0.148&0.050\\
\hline\hline
\end{tabular}
\end{table}
\end{center}

\begin{figure}
\centering
\includegraphics[width=0.47\textwidth]{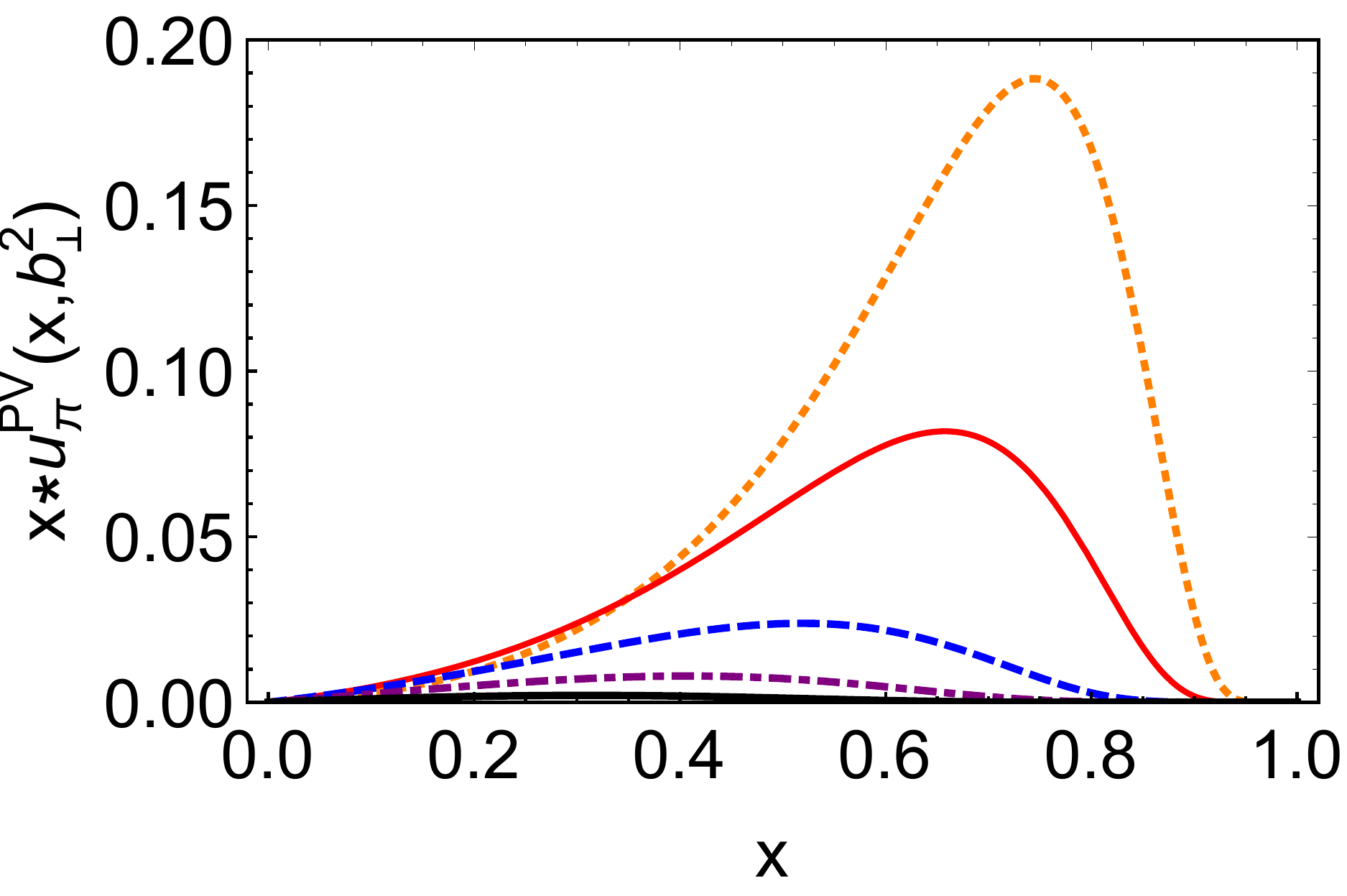}
\qquad
\includegraphics[width=0.47\textwidth]{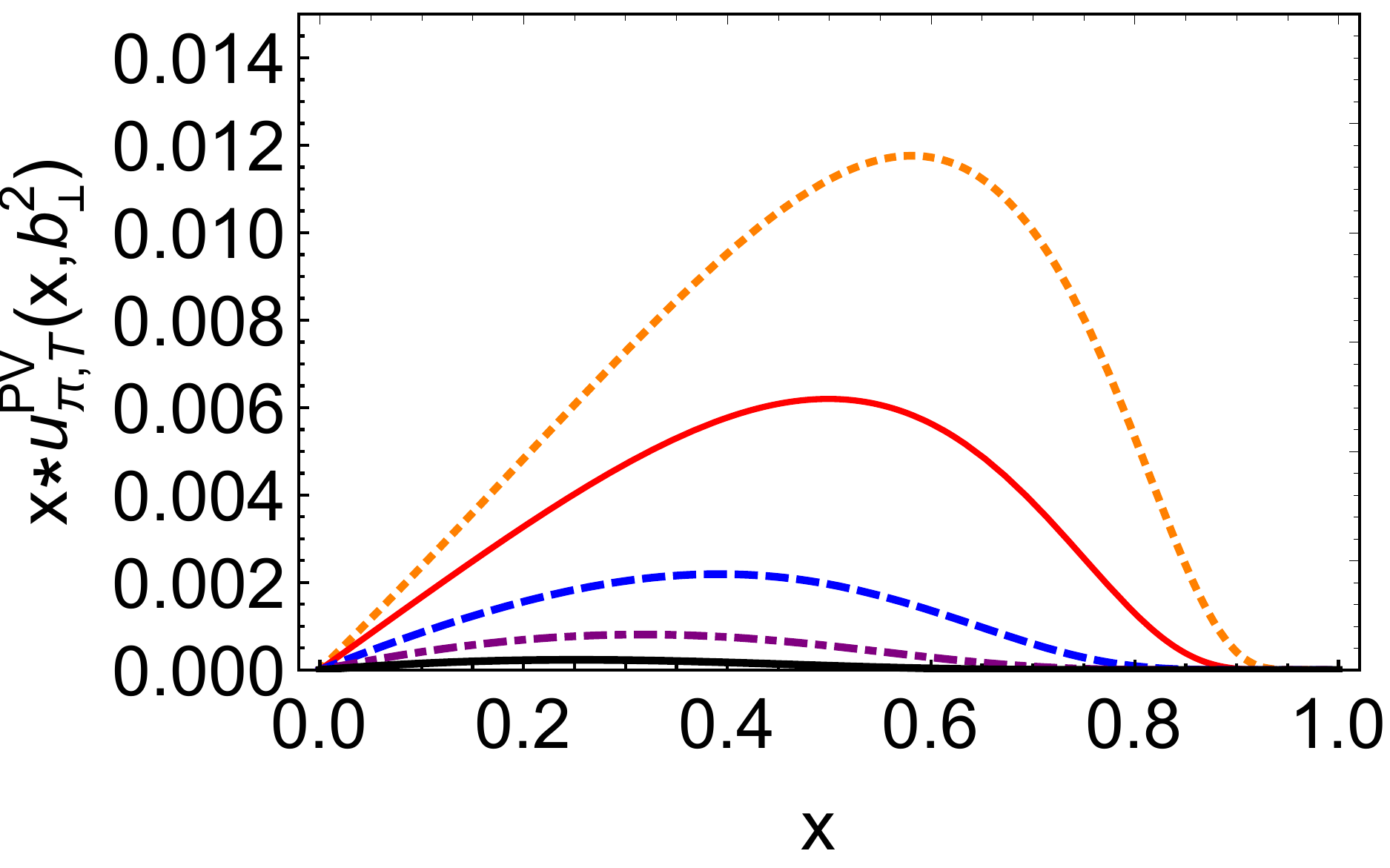}
\caption{Impact parameter space PDFs using PV regularization scheme: left panel -- $x*u\left(x,\bm{b}_{\perp}^2\right)$, the $\delta^2(\bm{b}_{\perp})$ component first line of Eq. (\ref{pvpspdf}) -- is suppressed in the image, and right panel -- $x*u_T\left(x,\bm{b}_{\perp}^2\right)$ both panels with $\bm{b}_{\perp}^2=0.5$ GeV$^{-2}$ --- orange dotted curve, $\bm{b}_{\perp}^2=1$ GeV$^{-2}$ --- red solid curve, $\bm{b}_{\perp}^2=2.5$ GeV$^{-2}$ --- blue dashed curve, $\bm{b}_{\perp}^2=5$ GeV$^{-2}$ --- purple dot-dashed curve, $\bm{b}_{\perp}^2=10$ GeV$^{-2}$ --- thick black solid curve.}\label{pvqxb}
\end{figure}

\subsubsection{Impact parameter space PDFs}

\begin{align}\label{aG91}
&\quad H_{\text{PV}}\left(x,0,-\bm{q}_{\perp}^2\right)\nonumber\\
&=\frac{N_cZ_{\pi }}{4\pi ^2}  \log \left[\frac{\left(\Lambda_{\text{PV}} ^2+\sigma_1\right)^2}{\sigma_1\left(2 \Lambda_{\text{PV}} ^2+\sigma_1\right)}\right] \nonumber\\
&+\frac{N_cZ_{\pi }}{8\pi ^2}\int_0^{1-x} \mathrm{d}\alpha  \left(2 x m_{\pi}^2+ x \bm{q}_{\perp}^2- \bm{q}_{\perp}^2\right)\nonumber\\
&\times \left(\frac{1}{\sigma_8+2\Lambda_{\text{PV}}^2}-\frac{2}{\sigma_8+\Lambda_{\text{PV}}^2}+\frac{1}{\sigma_8} \right),
\end{align}
\begin{align}\label{aG911}
E_{\text{PV}}\left(x,0,-\bm{q}_{\perp}^2\right)&=\frac{N_cZ_{\pi }}{4\pi ^2}\int_0^{1-x} \mathrm{d}\alpha\,m_{\pi} M \nonumber\\
&\times \left(\frac{1}{\sigma_8+2\Lambda_{\text{PV}}^2}-\frac{2}{\sigma_8+\Lambda_{\text{PV}}^2}+\frac{1}{\sigma_8} \right),
\end{align}
\begin{widetext}
\begin{align}\label{pvpspdf}
&\quad u^{\text{PV}}\left(x,\bm{b}_{\perp}^2\right)\nonumber\\
&=\frac{N_cZ_{\pi }}{4\pi ^2} \int \frac{\mathrm{d}^2\bm{q}_{\perp}}{(2 \pi )^2}e^{-i\bm{b}_{\perp}\cdot \bm{q}_{\perp}}\log \left[\frac{\left(\Lambda_{ \text{PV}}^2+\sigma_1\right)^2}{\sigma_1 \left(2 \Lambda_{ \text{PV}}^2+\sigma_1\right)}\right] \nonumber\\
&+\frac{N_cZ_{\pi }}{16\pi ^3}\int_0^{1-x} \mathrm{d}\alpha \,   \mathrm{K}_0\left[\frac{\sqrt{\bm{b}_{\perp}^2(2 \Lambda_{ \text{PV}}^2+\sigma_1)} }{\sqrt{\alpha  (1-x-\alpha )}}\right] \frac{\left((1-x) \left(2\Lambda_{ \text{PV}}^2+M^2\right)-m_{\pi }^2 x \left(2 \alpha ^2-2 \alpha +x^2+2 (\alpha -1) x+1\right)\right) }{\alpha ^2 (\alpha +x-1)^2} \nonumber\\
&-\frac{N_cZ_{\pi }}{8\pi ^3}\int_0^{1-x} \mathrm{d}\alpha \,   \mathrm{K}_0\left[\frac{\sqrt{\bm{b}_{\perp}^2(\Lambda_{ \text{PV}}^2+\sigma_1)}}{\sqrt{\alpha  (1-x-\alpha )}}\right] \frac{\left((1-x) \left(\Lambda_{ \text{PV}}^2+M^2\right)-m_{\pi }^2 x \left(2 \alpha ^2-2 \alpha +x^2+2 (\alpha -1) x+1\right)\right) }{\alpha ^2 (\alpha +x-1)^2} \nonumber\\
&+\frac{N_cZ_{\pi }}{16\pi ^3}\int_0^{1-x} \mathrm{d}\alpha \,   \mathrm{K}_0\left[\frac{\sqrt{\bm{b}_{\perp}^2\sigma_1}}{\sqrt{\alpha  (1-x-\alpha )}}\right] \frac{\left((1-x)M^2-m_{\pi }^2 x \left(2 \alpha ^2-2 \alpha +x^2+2 (\alpha -1) x+1\right)\right) }{\alpha ^2 (\alpha +x-1)^2},
\end{align}
\begin{align}\label{pvtpspdf}
&u_T^{\text{PV}}\left(x,\bm{b}_{\perp}^2\right)\nonumber\\
=&\frac{N_cZ_{\pi }}{8\pi^3}\int_0^{1-x}  \mathrm{d}\alpha\frac{m_{\pi } M }{\alpha  (1-\alpha -x)}\left(\mathrm{K}_0\left[\frac{\sqrt{\bm{b}_{\perp}^2(\sigma_1+2 \Lambda_{ \text{PV}}^2)}}{\sqrt{\alpha  (1-x-\alpha )}}\right]-2\mathrm{K}_0\left[\frac{\sqrt{\bm{b}_{\perp}^2(\sigma_1+\Lambda_{ \text{PV}}^2)} }{\sqrt{\alpha  (1-x-\alpha )}}\right]+\mathrm{K}_0\left[\frac{\sqrt{\bm{b}_{\perp}^2\sigma_1} }{\sqrt{\alpha  (1-x-\alpha )}}\right]\right),
\end{align}
\end{widetext}
where $\mathrm{K}_0$ is Bessel function of the second kind, for $u\left(x,\bm{b}_{\perp}^2\right)$, when integrating $\bm{b}_{\perp}$ one can get PDF $u(x)$ in Eq. (\ref{pvhpdf}). We plot the diagrams of $x *u\left(x,\bm{b}_{\perp}^2\right)$ and $x *u_\text{T}\left(x,\bm{b}_{\perp}^2\right)$ in Fig. \ref{pvqxb}.

\begin{figure}
\centering
\includegraphics[width=0.47\textwidth]{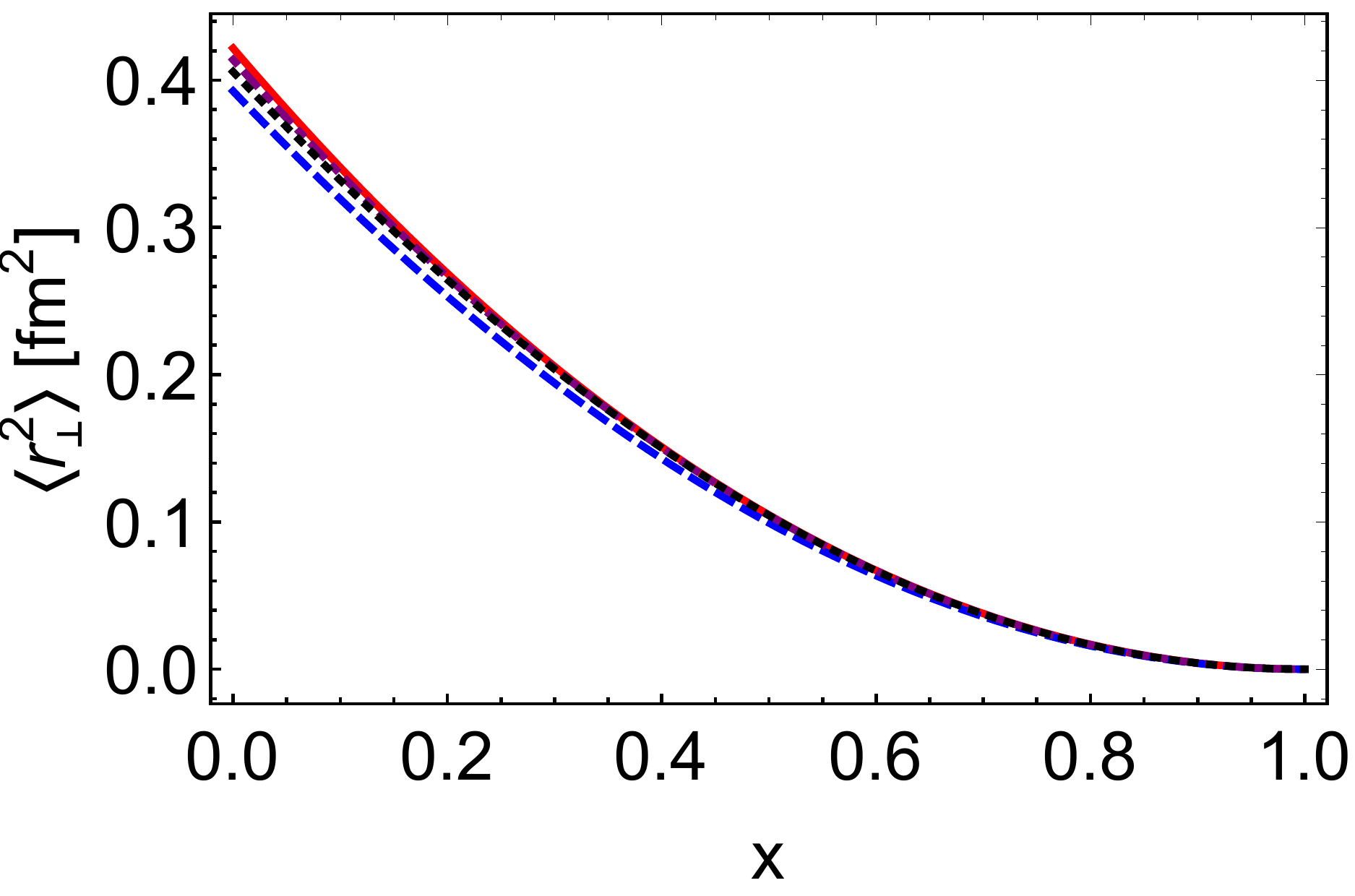}
\caption{The width distribution of $u$ quark in the pion for a given momentum fraction $x$ defined in Eq. (\ref{tmmf}) using different regularization schemes: red line -- PT regularization, blue dashed line -- 3D regularization, purple dotdashed line -- 4D regularization, black dotted line -- PV regularization.}\label{witf}
\end{figure}

From the diagrams in Figs. \ref{ptqxb}, \ref{3dqxb}, \ref{4dqxb} and \ref{pvqxb} we can see that PDFs in impact parameter space are similar, there is no big difference. In Fig. \ref{witf}, we plot the diagrams of the width distribution of $u$ quark in the pion for a given momentum fraction $x$. From the diagram we can see the width distributions in the four regularization schemes all satisfy the condition that when $x\rightarrow 1$ they become zero. The mean-squared $\langle \bm{b}_{\bot}^2\rangle_{\pi}^u$ are listed in Table \ref{tbr}, in PT regularization scheme it is the largest, in 3D momentum cut-off it is the smallest. The case of average transverse shift $\langle b_{\bot}^y\rangle_1^u$ and $\langle b_{\bot}^y\rangle_2^u$ are similar as the mean-squared $\langle \bm{b}_{\bot}^2\rangle_{\pi}^u$. The light-cone energy radius $r_{E,LC}$ and the light-cone charge radius $r_{c,LC}$ are also listed in Table \ref{tbr}.
\begin{center}
\begin{table}
\caption{The mean-squared $\langle \bm{b}_{\bot}^2\rangle_{\pi}^u$, the average transverse shift $\langle b_{\bot}^y\rangle_n^u$, the light-cone energy radius and the light-cone charge radius. The mean-squared $\langle \bm{b}_{\bot}^2\rangle_{\pi}^u$ are in units of fm$^2$, the left quantities are in units of fm.}\label{tbr}
\begin{tabular}{p{0.8cm} p{1.1cm} p{1.1cm}p{1.1cm}p{1.1cm}p{1.1cm}p{1.1cm}}
\hline\hline
 &$\langle \bm{b}_{\bot}^2\rangle_{\pi}^u$&$\langle b_{\bot}^y\rangle_1^u$&$\langle b_{\bot}^y\rangle_2^u$&$r_{E,LC}$&$r_{c,LC}$&$r_{c,LC}^D$\\
\hline
PT&0.140&0.106&0.071&0.188&0.374&0.513\\
\hline
3D&0.131&0.098&0.066&0.182&0.362&0.472\\
\hline
4D&0.138&0.103&0.069&0.187&0.371&0.476\\
\hline
PV&0.137&0.103&0.069&0.186&0.371&0.482\\
\hline\hline
\end{tabular}
\end{table}
\end{center}

\section{Summary and conclusion}\label{excellent}
In this paper, we calculated the pion GPDs in the NJL model using different regularization schemes, including the proper time regularization scheme, three dimensional momentum cut off, four dimensional momentum cut off, and Pauli-Villars regularization scheme.

Firstly, we plot the diagrams of GPDs, both vector and tensor GPDs are continuous but not differentiable at $x=\pm \,\xi$ except $\xi=0$, at this point, GPDs are not continuous. When considering the dressed quark photon vertex and D-term, vector GPD becomes not continuous because of the D-term. The contribution of D-term is not only the discontinuous of vector GPD, but also contributes to $A_{2,2}$, it makes the value of $|A_{2,2}(0)|$ approximate $A_{2,0}(0)$, if we don't consider D-term, the two values are quite different, when considering the D-term, they are approximate.

Secondly, we calculate the pion PDFs and plot the diagrams of PDFs from vector GPDs with different $m_{\pi}$, in the chiral limit $m_{\pi}=0$, pion PDF is constant $1$, when $m_{\pi}=0.14\,\text{GeV}$, in the NJL model PDFs float around $1$, when we set $m_{\pi}= 0.4, 0.6, 0.75$ GeV, in 4D and 3D regularization schemes, the peaks are sharper than PT and PV regularization schemes.

Thirdly, we calculate the form factors from vector GPDs and tensor GPDs, and plot the diagrams. For the electromagnetic factors, the bare charge radius from different regularization schemes are very close, but for the dressed charge radius in PT regularization scheme is closer to the experiment value $r_{\pi}=0.66$ fm. This can be seen from the Fig. \ref{a10}, $A_{1,0}^{D,\text{PT}}$ is softer than form factors in other regularization schemes. The gravitational form factors $A_{2,0}^{\text{PT}}$ and $A_{2,2}^{\text{PT}}$ are also softer than in other regularization schemes. The tensor form factors $B_{1,0}$ and $B_{2,0}$ are different from the vector case, they are almost parallel lines, the tensor form factors in 3D regularization scheme are the smallest.

Fourthly, we study the PDFs in impact parameter space, which are the Fourier transform of GPDs. We plot the diagrams of $x*u(x,\bm{b}_{\perp}^2)$ and $x*u_\text{T}(x,\bm{b}_{\perp}^2)$, the diagrams in different regularization schemes are similar, there's very small difference in values. The width distribution of u quark in the pion are plotted in Fig. \ref{witf}, they all satisfy the requirement that when $x\rightarrow 1$, $H(x,0,-\bm{q}_{\bot}^2)$ should be independent of $-\bm{q}_{\bot}^2$, the struck quark becomes closer to the centre of momentum since its weight increases, the average impact parameter become zero. The mean-squared $\langle \bm{b}_{\bot}^2\rangle_{\pi}^u$ are calculated. When quark polarized in the light-front-transverse $+\,x$ direction, the transverse-spin density is not symmetric around $\bm{b}_{\bot}=(b_x=0,b_y=0)$ anymore, the peaks shift to $(b_x=0,b_y>0)$. The average transverse shift $\langle b_{\bot}^y\rangle_1^u$ and $\langle b_{\bot}^y\rangle_2^u$ in PT regularization scheme are the biggest, in the 3D momentum cutoff are the smallest, the mean-squared $\langle \bm{b}_{\bot}^2\rangle_{\pi}^u$ are similar.

Now we arrive at the conclusion, pion GPDs in the NJL model using different regularization schemes are similar, when considering D-term, vector GPD is not continuous at $x=\pm \,\xi$, the form factors in impact parameter space are analogous. The average transverse shift in PT regularization scheme are the biggest, the 3D momentum cut-off are the smallest. The light-cone energy radius and the light-cone charge radius are similar.

\acknowledgments

We are grateful for constructive comments and technical assistance from Zhu-Fang Cui. Work supported by: National Natural Science Foundation of China (under grant no. 11775118).

\appendix
\section{Appendix 1: useful formulas}\label{AppendixT1}
Here we use the gamma-functions ($n\in \mathbb{Z}$, $n\geq 0$)
\begin{subequations}\label{cfun}
\begin{align}
\mathcal{C}_0(z)&:=\int_0^{\infty}\frac{s}{s+z} ds=\int_{\tau_{uv}^2}^{\tau_{ir}^2} d\tau \frac{1}{\tau^2} e^{-\tau z}\,, \\
\mathcal{C}_n(z)&:=(-)^n\frac{\sigma^n}{n!}\frac{d^n}{d\sigma^n}\mathcal{C}_0(\sigma)\,, \\
\bar{\mathcal{C}}_i(z)&:=\frac{1}{z}\mathcal{C}_i(z).
\end{align}
\end{subequations}
The $\sigma$ functions are define as
\begin{subequations}\label{cfun1}
\begin{align}
\sigma_1&=M^2-x(1-x)m_{\pi}^2\,, \\
\sigma_2&=M^2-x(1-x)t=M^2+x(1-x)Q^2\,, \\
\sigma_3&=M^2-\frac{x+\xi}{1+\xi} \frac{1-x}{1+\xi} m_{\pi}^2\,, \\
\sigma_4&=M^2-\frac{x-\xi}{1-\xi}\frac{1-x}{1-\xi} m_{\pi}^2\,, \\
\sigma_5&=M^2-\frac{1}{4}(1+\frac{x}{ \xi })(1-\frac{x}{\xi }) t\,, \\
\sigma_6&=M^2-\alpha \left(1-\alpha \right)  m_{\pi}^2\nonumber\\
&-\left(\frac{\xi+x}{2\xi}-\alpha \frac{1+\xi}{2\xi}\right) (\frac{\xi-x}{2\xi}+\alpha \frac{1-\xi}{2\xi}) t\,, \\
\sigma_7&=(x+y)(x+y-1)m_{\pi }^2-xyt+M^2\,, \\
\sigma_8&=M^2+(1-\alpha-x) \alpha \bm{q}_{\perp}^2-x \left(1-x\right) m_{\pi}^2.
\end{align}
\end{subequations}

\bibliographystyle{apsrev4-1}
\bibliography{zhang}


\end{document}